\documentclass[aps,prd,amsmath, amssymb,twocolumn,superscriptaddress,nofootinbib]{revtex4}

\usepackage{orcidlink}
\usepackage{amsfonts}
\usepackage{color}
\usepackage{enumerate} 
\usepackage{xcolor}
\usepackage{multirow}
\usepackage{dcolumn}
\usepackage[export]{adjustbox}
\usepackage{tikz}
\usepackage{hyperref}
\usepackage[normalem]{ulem}
\usepackage{soul}
\usepackage{aasmacros}

\newcommand{\Omb}{$\Omega_{\rm b}$} 
\newcommand{\Omc}{$\Omega_{\rm c}$} 
\newcommand{\h}{$h$} 
\newcommand{\Omk}{$\Omega_{k}$} 
\newcommand{\mnu}{$m_{\nu}$} 
\newcommand{\wO}{$w_{0}$} 
\newcommand{\wa}{$w_{\rm a}$} 
\newcommand{\ns}{$n_{\rm s}$} 
\newcommand{\As}{$A_{\rm s}$}
\newcommand{\gam}{$\gamma$}
\newcommand{\bfthetabase}{$\boldsymbol{\theta}_{\rm base}$}
\newcommand{\bfthetaDE}{$\boldsymbol{\theta}_{\rm DE}$}
\newcommand{\FoMbase}{${\rm FoM}_{\boldsymbol{\theta}_{\rm base}}$}
\newcommand{\FoMDE}{${\rm FoM}_{\boldsymbol{\theta}_{\rm DE}}$}
\newcommand{\FoMent}{${\rm FoM}_{\boldsymbol{\theta}_{\rm ent}}$}
\newcommand{\basemodel}{$w_{0}{\rm CDM}$}
\newcommand{\baseplusmodel}{$w_{0}w_{\rm a}{\rm CDM}$}

\newcommand{\be}{\begin{equation}}
\newcommand{\ee}{\end{equation}}
\newcommand{\bey}{\begin{eqnarray}}
\newcommand{\eey}{\end{eqnarray}}
\newcommand{\nn}{\nonumber}
\newcommand{\obs}{{\rm obs}}

\newcommand{\fid}{{\rm fid}}
\newcommand{\bfk}{\boldsymbol{k}}
\newcommand{\bfr}{\boldsymbol{r}}
\newcommand{\bfq}{\boldsymbol{q}}
\newcommand{\bfx}{\boldsymbol{x}}

\newcommand{\bfF}{\boldsymbol{F}}

\newcommand{\bfC}{\boldsymbol{C}}

\newcommand{\DA}{D_{\rm A}}

\newcommand{\Omegam}{\Omega_{\rm m}}
\newcommand{\Omegak}{\Omega_k}
\newcommand{\OmegaDE}{\Omega_{\rm DE}}

\newcommand{\Omegab}{\Omega_{\rm b}}
\newcommand{\Omegac}{\Omega_{\rm c}}
\newcommand{\AIA}{A_{\rm IA}}

\newcommand{\wt}{\widetilde}

\newcommand{\Pgg}{P_{\rm gg}}
\newcommand{\PEE}{P_{\rm EE}}
\newcommand{\PgE}{P_{\rm gE}}
\newcommand{\wtPgg}{\wt{P}_{\rm gg}}
\newcommand{\wtPEE}{\wt{P}_{\rm EE}}
\newcommand{\FoM}{{\rm FoM}}
\newcommand{\Cov}{{\rm Cov}}

\begin{document}
\title{Improving cosmological constraints via galaxy intrinsic alignment in full-shape analysis}

\author{Junsup Shim \orcidlink{0000-0001-7352-6175}}\email{jshim@asiaa.sinica.edu.tw} 
\affiliation{Academia Sinica Institute of Astronomy and Astrophysics (ASIAA), No. 1, Section 4, Roosevelt Road, Taipei 106216, Taiwan}
\author{Teppei Okumura \orcidlink{0000-0002-8942-9772}}
\affiliation{Academia Sinica Institute of Astronomy and Astrophysics (ASIAA), No. 1, Section 4, Roosevelt Road, Taipei 106216, Taiwan}
\affiliation{Kavli IPMU (WPI), UTIAS, The University of Tokyo, Kashiwa, Chiba 277-8583, Japan}

\author{Atsushi Taruya \orcidlink{0000-0002-4016-1955
}}
\affiliation{Yukawa Institute for Theoretical Physics, Kyoto University, Kyoto 606-8502, Japan}
\affiliation{Kavli IPMU (WPI), UTIAS, The University of Tokyo, Kashiwa, Chiba 277-8583, Japan}

\date{\today}

\begin{abstract}
The intrinsic alignment (IA) of galaxy shapes probes the underlying gravitational tidal field, thus offering cosmological information complementary to galaxy clustering. In this paper, we perform a Fisher forecast to assess the benefit of IA in improving cosmological parameter constraints, for the first time, leveraging the full-shape (FS) information of IA statistics.
Our forecast is based on PFS-like and Euclid-like surveys as examples of deep and wide galaxy surveys, respectively. We explore various cosmological models, with the most comprehensive one simultaneously including dynamical dark energy, curvature, massive neutrinos, and modified gravity (MG). We find that adding FS IA information significantly tightens cosmological constraints relative to the FS clustering-only cases, particularly for dynamical dark energy and nonflat-MG models. For a deep galaxy survey, the Figure-of-Merit for the dark energy equation of state parameters is improved by at least more than $40\%$ in all dynamical dark energy models investigated. For nonflat-MG models, parameter constraints are tightened by $6-28\%$, except for the dark matter density and spectral index parameters. For a wide galaxy survey, improvements with IA become milder, although its joint constraints are tighter than those from the deep survey. Our findings highlight the efficacy of the galaxy IA as a complementary cosmological probe to galaxy clustering.

\end{abstract}
\maketitle
\flushbottom


\section{Introduction}\label{sec:introduction}
Large-scale matter distribution inferred from galaxy redshift surveys has been one of the richest sources of cosmological information for revealing the underlying physics governing the evolution of the Universe. Utilizing characteristic clustering features in the matter/galaxy distribution, i.e., the baryon acoustic oscillation (BAO) \cite{peebles&yu70,eisenstein&hu98, cole+05,eisenstein+05} and redshift-space distortion (RSD) \cite{jackson72,sargent&turner77, kaiser87,hamilton92}, the measurements of the expansion and growth rates of the Universe have served as effective tools for constraining the nature of dark energy and the modification of gravity theories \cite{peacock01,seo&eisenstein03,tegmark+04,okumura+08,guzzo+08,beutler+12,blake+12,reid+12,samushia+13,beutler+14,aubourg+15,okumura+16,alam+17,beutler+17,gilmarin+17,hawken+17,hou+21,aubert+22}.

In the traditional clustering analyses, cosmological constraints have been routinely extracted from a small set of parameters that compresses the information on BAO and RSD \cite{seo&eisenstein03,anderson+14,beutler+14,beutler+17,gilmarin+17,nadathur+19,zhao+22,desi_BAO+24}. While still focusing on galaxy clustering, one approach adopted to tighten cosmological constraints is the full-shape or full-modeling analysis \cite{sanchez+09,montesano+10,montesano+12,sanchez+13, ivanov+20, nunes+22, philcox&ivanov22, simon+23, Gsponer+24, ramirez+24, desi_FS+24}.
In addition to the BAO and RSD information, the full-shape analysis further leverages the broadband shape of the matter/galaxy power spectrum -- e.g., slope, amplitude, and turnover peak -- where information on various physical processes are encoded over a wide range of scales. 
Thus, the full-shape analysis can extract more information from a given observed galaxy power spectrum and tighten cosmological parameter constraints, improving upon the BAO/RSD-limited ones \cite{ivanov+20,philcox+20,brieden+21a,brieden+21b, desi_FS+24,DESI_FS_MG+24}.
Furthermore, it potentially allows capturing the signatures of scale-dependent physics in more extended cosmological models, e.g., massive neutrinos \cite{boyle&komatsu+18,kumar+22,moretti+23} and modified gravity theories \cite{moretti+23,rodriguez-meza+24} as they can suppress or enhance the growth of matter perturbation differently depending on scales \cite{carroll+04,lesgourgues&pastor06,hu&sawicki07}.

On the other hand, a joint analysis combining galaxy clustering with other complementary probes has become another effective and promising strategy to improve cosmological parameter constraints.
A cosmological probe recently gaining more attention is the intrinsic alignment (IA) of galaxy shapes. Unlike the apparent shape alignments due to the weak gravitational lensing, i.e., cosmic shear \cite{kaiser92,bartelmann&schneider01}, the intrinsic shapes of galaxies are oriented to align in particular directions under the influence of the gravitational tidal field \cite{catelan+01,hirata&seljak04}, producing observable IA signals \cite{brown+02, mandelbaum+06, hirata+07, okumura+09, tonegawa+22, tsaprazi+22, OT23, zhou+23}.

Through the tidal field induced by surrounding large-scale structures, the IA of galaxies reflects valuable information, and thus, can be utilized as cosmological probes for various physics \cite{schmidt&jeong12, faltenbacher+12,chisari&dvorkin13,chisari+16,kogai+18,biagetti&orland020,OT20b,okumura+20,chuang+22,akitsu+23,shiraishi+23,philcox+24,saga+24}.
In particular, it has been demonstrated that combining IA information with galaxy clustering significantly improves the cosmological parameter constraints in the conventional geometric and dynamic analyses (without full-shape information) \cite{TO20,OT22,OT23,xu+23} and dark energy constraints with full-shape IA information \cite{shim+25}, implying that the galaxy IA serves an effective and complementary statistics to further constrain cosmological models.
Considering the capabilities of ongoing/forthcoming surveys -- e.g., the Dark Energy Spectroscopic Instrument (DESI) \cite{desi+16}, Subaru Prime Focus Spectrograph (PFS) \cite{takada+14}, Euclid space telescope \cite{euclid+11,euclid_prep+20}, Nancy Grace Roman Space Telescope \cite{roman+13}, and Rubin Observatory Legacy Survey of Space and Time (LSST) \cite{lsst+18} -- it would be optimal to exploit the full-shape information of the power spectrum and combine as many complementary probes to more precisely constrain cosmological models.

In this paper, we perform a Fisher forecast jointly utilizing galaxy clustering and IA within the framework of a full-shape analysis. We assess the cosmological benefit of combining IA information in the full-shape framework. Our forecast is based on the two types of surveys; the PFS-like deep galaxy survey which can be aided with high-quality imaging from Hyper Suprime-Cam (HSC) \cite{hsc+18,hsc2+18} and the Euclid-like wide survey, expecting different IA contributions due to different survey setups. We consider general extensions of the standard $\Lambda$CDM models, spanning a wide range of parameter space, including the evolving dark energy, massive neutrinos, and modified gravity. We show that the intrinsic alignment can considerably contribute to further improving parameter constraints even though the full-shape clustering-only information alone already provides tight constraints \cite[also see Ref.][ for full-shape IA contribution particularly to dark energy constraints]{shim+25}. 

The rest of the paper is organized as follows. In section \ref{sec:pre}, we describe the formulation of the geometric and dynamical quantities measuring expansion and growth rates in terms of cosmological parameters. Section \ref{sec:statistics} presents the galaxy clustering and IA statistics, while section~\ref{sec:fisher} describes the Fisher matrix formalism including the treatment of prior. We then present Fisher forecast constraints in section \ref{sec:results}, with some details further discussed in section \ref{sec:discussion}, before we conclude in section \ref{sec:conclusion}.


\section{Preliminaries}\label{sec:pre}
\subsection{Expansion history}\label{sec:expansion}

We first begin by relating the observed redshift, $z$, of a galaxy and the
comoving distance to it, $\chi(z)$.
It is given by
\be\label{eq:comoving}
\chi(z)=\int^z_0 \frac{cdz'}{H(z')}\ ,
\ee
with $c$ being the speed of light. 
The Hubble parameter, $H(z)$, is, typically decomposed into $H(z)=H_{0}E(z)$ where $H_{0}\equiv100h$ is the present-day Hubble parameter with the dimensionless Hubble constant, $h$.
Considering a nonflat universe with dynamical dark energy and massive neutrinos, the redshift-dependent part, $E(z)$, can be written as
\begin{multline*}
    E^2(z) = \Omega_{\gamma} (1+z)^{4}+ (\Omegab+\Omegac) (1+z)^{3}  \nn \\ + F\left[m_{\rm rel}(1+z)/m_{\nu}\right] \Omega_{\nu}(1+z)^{3}+ \Omegak(1+z)^2  \nn \\ +\OmegaDE (1+z)^{3(1+w_0+w_a)} \exp{\left[ -3w_a\frac{z}{1+z}\right]}\label{eq:friedmann}\ ,
\end{multline*}
where $\Omega_{i}$ with $i=\gamma,{\rm b,c},\nu,k,{\rm DE}$ represent the dimensionless density parameters for the photon, baryon, cold dark matter, massive neutrino, curvature,  and dark energy, satisfying $\sum_{i}\Omega_{i}=1$. 

The time-evolution of dark energy is captured in the equation of state (EOS) of dark energy, $w\equiv p_{\rm DE}/\rho_{\rm DE}$, adopting the CPL-parametrization \cite{chevallier&polarski01,linder03},
\be
w(z) =w_0+w_a\frac{z}{1+z} = w_0 + w_a(1-a),
\ee
where $w_0$ corresponds to the present-day dark energy EOS, whereas $w_a$ represents the slope of EOS variation with the scale factor $a=1/(1+z)$. This parametrization reduces to the Cosmological Constant $\Lambda$ in the concordance cosmology when setting $w_0=-1$ and $w_a=0$.

The transition of massive neutrinos from the relativistic to non-relativistic limit is described by the modulation function, $F\left[m_{\rm rel}(1+z)/m_{\nu}\right]$. 
We adopt the fit presented in \cite{wright06},
\be
F(y)\approx(1+y^{\beta})^{1/\beta}\ ,
\ee
as a function of the ratio between the mass of a neutrino, $m_{\nu}$, and the rest mass of a thermalized neutrino, $m_{\rm rel}=0.000531{\rm eV}$ at the present neutrino temperature, $T_{\nu,0}=(4/11)^{1/3}T_{0}=1.95{\rm K}$ with $\beta=1.842$ \cite[c.f. see Ref.][for the approximation normalized to the relativistic limit]{komatsu+11}. We assume one massive and two massless neutrinos.
\footnote{The choice of neutrino mass hierarchy has negligible impact on the power spectrum shape \cite{jimenez+10,zhao+18}. In particular, their mass ordering is strongly degenerate in dynamical dark energy models \cite{yang+17,li+18}. Thus, a mass hierarchy-insensitive conclusion is expected.}
Their total mass and the neutrino density parameter are related via $\Omega_{\nu}h^{2}=\sum m_{\nu}/93.14{\rm eV}$ \cite{mangano+05}.
At low redshifts, i.e., $y\ll1$, one can find that the redshift evolution of neutrino density parameter asymptotes to $(1+z)^{3}$ as if the massive neutrinos are non-relativistic matter. In contrast, in high redshifts, it evolves as $(1+z)^{4}$ similar to the radiation component.

Another crucial distance measure is the angular diameter distance, $D_{\rm A}(z)$, and it is defined as
\begin{align}
D_{\rm A}(z) = \frac{1}{1+z}\frac{c}{H_{0}}\left\{ 
\begin{array}{lll}
\sinh{\left( \sqrt{\Omegak}\frac{H_{0}\chi}{c} \right)} / \sqrt{\Omegak}  & & \Omegak>0,\\
\frac{H_{0}\chi}{c} & & \Omegak=0, \\
\sin{\left( \sqrt{-\Omegak}\frac{H_{0}\chi}{c} \right)} / \sqrt{-\Omegak} &  & \Omegak<0,
\end{array}
\right.
\end{align}
in relation with the comoving distance, depending on the curvature of the Universe.

\subsection{Growth history}\label{sec:growth}

The matter density fluctuation around its cosmic mean, $\bar\rho_{\rm m}$, is defined as
\be
\delta_{\rm m}(\bfx,z) \equiv \frac{\rho_{\rm m}(\bfx,z)-\bar{\rho}_{\rm m}(z)}{ \bar{\rho}_{\rm m}(z)}\ . \label{eq:delta}
\ee
For a statistical description of the fluctuation, we decompose the fluctuation into plane waves in the Fourier space,
\be
\delta_{\rm m}(\bfx,z)=\int \frac{d^{3}k}{(2\pi)^3}\delta_{\rm m}(\bfk,z){\rm exp}(-\textit{i}\bfk\cdot\bfx).
\ee
The matter density power spectrum, $P_{\rm m}(k,z)$, is then defined as
\be
\left< \delta_{\rm m}(\bfk,z) \delta_{\rm m}(\bfk',z)\right>=(2\pi)^{3}\delta_{\rm D}(\bfk+\bfk')P_{\rm m}(k,z)\,
\ee
where $\delta_{\rm D}$ denotes the Dirac delta function and we assume the statistical homogeneity and isotropy so that the power spectrum only depends on $k=|\bfk|$.

The time-evolution of the matter density fluctuation can be described with the linear growth factor,
$D_{\rm GR}(k,z)\equiv\delta_{\rm m}(\bfk,z) / \delta_{\rm m}(\bfk,0)$, where we assume general relativity (GR) for the fiducial gravity model and explicitly indicate in the subscript as `GR'. Then, the rate at which cosmic structures grow can be measured by the linear growth rate parameter, $f_{\rm GR}(k,z)$, defined as
\be
f_{\rm GR}(k,z) = -\frac{d\ln D_{\rm GR}(k,z)}{d \ln (1+z)} =  \frac{d\ln D_{\rm GR}(k,a)}{d \ln a}.
\label{eq:linF}
\ee
We note that we explicitly consider the $k$-dependence in the growth factor and growth rate because massive neutrinos introduce scale-dependent features \cite{kiakotou+08,boyle&komatsu+18}, through which we can extract information about their total mass. This is because their free-streaming motion suppresses the density fluctuation and hinders structure growth in a scale-dependent manner depending on their total mass \cite[see Ref.][and references therein]{lesgourgues&pastor06}.

As modified gravity (MG) alters the perturbation growth from GR, we incorporate such changes into the linear growth rate modeling by adopting the \gam-parametrization \cite{wang&steinhardt98, linder05}. In this parametrization, the scale-independent linear growth rate is given
\be
f_{\rm mod}(z) = \left[ \Omegam(z) \right]^{\gamma_{\rm mod}}, \label{eq:gamma}
\ee
with the time-dependent matter density parameter, $\Omegam(z)=\Omegam (1+z)^3/ E^{2}(z)$ and the power index $\gamma_{\rm mod}$ with the subscript `mod' representing a gravity model-dependent parameter,
e.g., $\gamma_{\rm GR} \approx 6/11$ for GR \cite{peebles80,lahav+91}.
Using this scale-independent parametrization, we obtain the scale-dependent linear growth rate in MG models by only modulating its amplitude through
\be
f_{\rm MG}(k,z) = f_{\rm GR}(k,z) \frac{\Omega_{\rm m}^{\gamma_{\rm MG}}(z)}{\Omega_{\rm m}^{\gamma_{\rm GR}}(z)},
\ee
assuming that MG only rescales structure growth identically on all scales, i.e., scale-independent modification.
Accordingly, we model the growth factor in MG cases in the same manner,
\be
D_{\rm MG}(k,z)=D_{\rm GR}(k,z)\frac{D_{\rm MG}(z)}{D_{\rm GR}(z)},
\ee
where scale-independent $D_{\rm MG}(z)$ and $D_{\rm GR}(z)$ can be calculated by integrating Eq.~(\ref{eq:linF}) using Eq.~(\ref{eq:gamma}),
\be
    D_{\rm mod}(z)\simeq {\rm exp}\biggl\{-\int_0^{z} \frac{[\Omega_{\rm m}(z')]^{\gamma_{\rm mod}}}{1+z'}dz'
    \biggl\}\ .
\ee
Our scale-dependent treatment of structure growth aligns with the approaches in Ref.~\cite{euclid_prep+20, moretti+23} in that the scale-dependent suppression is imprinted by massive neutrinos, while MG only alters their overall amplitudes. However, unlike Ref.~\cite{euclid_prep+20}, which uses a fitting formula derived for a flat universe \cite{kiakotou+08}, we numerically compute the linear growth rate in the presence of massive neutrinos directly from power spectra using CLASS code, without the flat geometry assumption.

\section{Clustering and Intrinsic Alignment statistics}\label{sec:statistics}

We now describe how the galaxy clustering and intrinsic alignment statistics are modeled. We utilize 2-point statistics in redshift space; auto-power spectra of density and ellipticity and cross-power spectrum between density and ellipticity. In the following, we illustrate how the galaxy density and ellipticity fields observed in redshift space are related to the real space matter density in linear theory.

\subsection{Density and ellipticity fields}

The observed galaxy density fluctuation in Fourier space, $\delta_{\rm g}$, can be related to the underlying matter density fluctuation, $\delta_{\rm m}$, through
\be
    \delta_{\rm g}(\bfk,z) = (b_{\rm g}(z)+f(k,z)\mu^{2})\delta_{\rm m}(\bfk,z)\ ,
\ee
where we adopt the plane-parallel approximation assuming the line-of-sight is fixed direction along the z-axis.
The galaxy bias, $b_{\rm g}$, assumes that galaxies are linearly biased tracers of the underlying matter density field on large scales \cite{kaiser84}, while $f(k,z)\mu^{2}$ describes the anisotropy induced in the redshift space matter distribution due to the peculiar motion of galaxies \cite{kaiser87}. The directional cosine is defined as $\mu\equiv\bfk\cdot\bfr/|\bfk||\bfr|$ with $\bfk$ and $\bfr$ being the wave and line-of-sight vectors.
Our choice for the bias parameters will be given in Sec.~\ref{subsec:setup}.

On the other hand, the ellipticity field of observed galaxies is defined as
\be
\gamma_{(+,\times)}(\bfx,z) = \frac{1-q^2}{1+q^2} \left( \cos{2\theta},\sin{2\theta} \right),
\ee
where $\theta$ measures the angle between the major axis of the shape of a galaxy and a reference axis both on the celestial plane perpendicular to the line-of-sight, and $q$ is the minor-to-major axis ratio of the galaxy shape. Adopting the linear alignment (LA) model \cite{catelan+01,hirata&seljak04}, the ellipticity field is modeled to be linearly related to the gravitational tidal field.
In the Fourier space, the relation follows
\be
\gamma_{(+,\times)}(\bfk,z) = b_K(z) \left( k_x^2-k_y^2, 2k_xk_y \right)\frac{\delta_{\rm m}(\bfk,z)}{k^2},
\label{eq:gamma_+x}
\ee
where the gravitational potential is replaced with the matter density via the Poisson equation. A more useful characterization of the ellipticity field can be made from its rotation-invariant form by decomposing into its E-/B-modes \cite{stebbins+96,kamionkowski+98,crittenden+02},
\be
\gamma_{\rm E}(\bfk,z) + i\gamma_{\rm B}(\bfk,z) = e^{-2i\phi_k}\left\{\gamma_{+}(\bfk,z)+i\gamma_{\times}(\bfk,z) \right\},
\label{eq:gamma_EB}
\ee
with the angle $\phi_k=\arctan(k_{y}/k_{x})$. Combining Eqs.~(\ref{eq:gamma_+x}) and (\ref{eq:gamma_EB}), we obtain
\be
\gamma_{\rm E}(\bfk,z)=b_K(z) (1-\mu^2) \delta_{\rm m}(\bfk,z)\ ,
\label{eq:gamma_E}
\ee
with vanishing $\gamma_{B}$, implying that the linear tidal alignment does not generate the B-mode component in the IA of galaxies.
We note that the galaxy ellipticity field does not depend on RSD in the linear limit, unlike galaxy density. This helps break the degeneracy between the linear growth rate and geometric distances, contributing to tightening cosmological constraints \cite{TO20, OT22}.
The shape bias, $b_K$, quantifies the responsivity of galaxy shapes to the tidal field, and can be given as,
\be
b_K(z) = - 0.01344 \AIA{}\Omegam/D(z),
\ee
where we introduce a dimensionless parameter $\AIA$ that quantifies the amplitude of IA, following the convention adopted in \cite{joachimi+11, kurita+21, shi+21a, shi+21b, OT22, inoue+24}.
We treat $\AIA$ as a constant throughout this paper because simulations suggest that it remains nearly constant over redshifts
for fixed galaxy/halo properties, e.g., halo \cite{kurita+21} and stellar masses \cite{shi+21b}.

\subsection{Density and ellipticity power spectra}

Utilizing the explicit relations of the galaxy density and ellipticity fields to the underlying matter density field, we compute their auto- and cross-power spectra; i.e., density-density (GG), ellipticity-ellipticity (II), and density-ellipticity (GI) power spectra. The three power spectra are related to the linear matter power spectrum, $P_{\rm m}$, and are expressed as
\be
    P_{\rm gg}(k,\mu,z) =(b_{\rm g}+f(k,z)\mu^{2})^{2}P_{\rm m}(k,z),\label{eq:pgg} 
\ee
\be
    P_{\rm gE}(k,\mu,z) = b_K(1-\mu^{2})(b_{\rm g}+f(k,z)\mu^{2})P_{\rm m}(k,z),\label{eq:pgE} 
\ee
\be
    P_{\rm EE}(k,\mu,z) = b_K^{2}(1-\mu^{2})^{2}P_{\rm m}(k,z).\label{eq:pEE} 
\ee

Given the model power spectra above, the observed power spectra, which account for additional anisotropies known as the Alcock-Paczynski (AP) effect \cite{alcock&paczynski79}, are expressed as follows:
\be
P_{i}^{\obs}\left(k_\perp^\fid,k_\parallel^{\fid},z \right) = 
\frac{H(z)}{H^\fid(z)}\left\{ \frac{D_{\rm A}^\fid(z)}{\DA(z)} \right\}^2
P_{i}\left(k_\perp,k_\parallel, z \right), 
\label{eq:Pi_AP}
\ee
where $i=\rm (gg, gE, EE)$ and the quantities with a superscript `fid' indicate that they are evaluated by adopting the fiducial cosmological parameter values. The prefactor $\frac{H(z)}{H^\fid(z)}\left\{ \frac{D_{\rm A}^\fid(z)}{\DA(z)} \right\}^2$ measures the volume change when adopting the fiducial cosmology relative to the underlying true cosmology. The wavenumber $k$ is decomposed into components perpendicular and parallel to the line-of-sight, $(k_\perp,k_\parallel) = k(\sqrt{1-\mu^2},\mu)$, and their counterparts in the fiducial cosmology can be calculated as 
\be
    k_\perp^\fid=k_\perp\,\frac{\DA(z)}{D_{\rm A}^\fid(z)},
\ee
\be
k_\parallel^\fid=k_\parallel \frac{H^\fid(z)}{H (z)}.
\ee

These capture the geometric distortions caused by the AP effect arising from a mismatch between the reference and true cosmology for inferring the line-of-sight and transverse distances.

\section{Fisher matrix formalism}\label{sec:fisher}

\subsection{Cosmological and nuisance parameters}\label{sec:parameters}

We first introduce our parameter space; cosmological parameters and redshift- and survey-dependent nuisance parameters.
The simplest cosmological model we consider is \basemodel{} model, a minimal extension of the standard $\Lambda$CDM model, which assumes non-evolving dark energy other than the Cosmological Constant $\Lambda$. 
Thus, the minimal cosmological parameter space consists of $\boldsymbol{\theta}_{\rm min}=(\Omega_{\rm b}, \Omega_{\rm c}, h, A_{\rm s}, n_{\rm s}, w_{0})$. We then consider more extended cosmological models by additionally including parameters for evolving dark energy, \wa, the existence of massive neutrinos, \mnu, modified gravity theories, \gam, and nonflat geometries, \Omk. Thus, the most extensive model contains 10 cosmological parameters in total. Our cosmological parameters and their fiducial values are summarized in Table.~\ref{tab:para_fid}.
\begin{table}[bt!]
\caption{Cosmological parameters and their fiducial values.}
\begin{center}
\begin{tabular}{ccc}
\hline \hline 
parameters & description & fiducial value \\
\hline
\Omb & baryon density parameter & $0.0492$ \\  
\Omc & DM density parameter & $0.264$ \\  
\h   & dimensionless Hubble constant & $0.674$ \\  
${\rm ln}(10^{10}A_{s})$ & power spectrum amplitude & $3.044$ \\  
\ns  & spectral index & $0.965$ \\  
\wO  & time-independent dark energy EOS &$-1.0$ \\  
\wa  & time-dependent dark energy EOS & $0.0$ \\
\mnu{} (eV) & total neutrino mass & $0.06$ \\  
\gam & gravity parameter & $0.545$ \\ 
\Omk & curvature density parameter & $0.0$ \\  
\hline \hline
\end{tabular}
\label{tab:para_fid}
\end{center}
\end{table}

Furthermore, there are reshift-dependent and survey-dependent parameters with varying fiducial values, i.e., $\boldsymbol{\theta}_{\rm n}(z)=(b_{\rm g}(z), b_{K}(z))$. We treat these two parameters in one redshift bin as distinct nuisance parameters from those in other redshift bins. Thus, the total number of parameters involved in the Fisher analysis becomes $n_{\rm tot}=n_{\rm p}+n_{\rm n}\times n_{z}$, where $n_{\rm p}$, $n_{\rm n}$, and $n_{z}$ represent the numbers of cosmological parameters, nuisance parameters per redshift bin, and redshift bins, respectively.

\subsection{Fisher matrix from galaxy surveys}\label{sec:matrices}

We now describe how the cosmological gain expected from IA is quantified in the full-shape framework.
We use the Fisher matrix formalism \cite{tegmark97,seo&eisenstein03} and compare the full-shape constraints on the cosmological parameters from the clustering-only information and joint information combined with IA. The clustering-only analysis only uses the observed galaxy density power spectrum, whereas the joint analysis utilizes all three power spectra.

Given the observed power spectra, the Fisher matrix for the cosmological and nuisance parameters, $\boldsymbol{\theta}=(\boldsymbol{\theta}_{\rm p},\boldsymbol{\theta}_{\rm n}(z_1),\cdot\cdot\cdot,\boldsymbol{\theta}_{\rm n}(z_{n_{z}}))=(\theta_{1},\cdot\cdot\cdot,\theta_{n_{\rm tot}}$), can be constructed following 
\begin{align}
F_{\alpha\beta}(z) = & \frac{V_s}{4\pi^2} \int^{k_{\rm max}}_{k_{\rm min}} dk k^2 \int ^{1}_{-1}d\mu \nn \\
& \times \sum_{i,j} \frac{\partial P_i(k,\mu,z)}{\partial\theta_\alpha}
\left[ \Cov^{-1}\right]_{ij}\frac{\partial P_j(k,\mu,z)}{\partial\theta_\beta}.
\label{eq:Fisher_matrix}
\end{align}
We omitted the superscript from $P_{i}^{\rm obs}$ for brevity so that $P_{i}$ denotes the observed power spectra from here on.
Here, $V_s$ is the survey volume spanning the redshift range $z_{\min} \leq z \leq z_{\max}$, and $k_{\rm min}$ and $k_{\rm max}$ correspond to the minimum and maximum wavenumbers, respectively. The survey volume determines the minimum wavenumber, $k_{\rm min}=2\pi/V_s^{1/3}$. We set $k_{\rm max}=0.2\ h{\rm Mpc}^{-1}$ for our analysis, while the impact of changing $k_{\rm max}$ to a more conservative limit will be discussed in section~\ref{sec:discussion} and Appendix~\ref{app:setup}.

The observed power spectra derivatives and the linear growth rate are numerically calculated using the finite difference method,
\be
\frac{\partial Y}{\partial x}=\frac{Y(x+\Delta x) -Y(x-\Delta x)}{2\Delta x}\ .
\ee
For such numerical differentiations, we use the \textit{CLASS} code \cite{blas+11} to compute the linear matter power spectrum for different cosmological parameter values. Note that the observed power spectra at one redshift, $z_1$, are independent of the redshift-specific nuisance parameters at different redshifts, $z_2$, e.g.,
\be
\frac{\partial P_i(k,\mu,z_1)}{\partial b_{\rm g}(z_2)} = \frac{\partial P_i(k,\mu,z_1)}{\partial b_{K}(z_2)} = 0\ .\label{eq:nuisance_zindep}
\ee

The Gaussian covariance matrix is defined as ${\rm Cov}_{ij}(k,\mu,z)=\langle P_{i}P_{j}\rangle-\langle P_{i}\rangle\langle P_{j}\rangle$ for a given wavevector and redshift bin.
For example, the covariance matrix for the joint analysis is a $3\times3$ matrix and reads
\be
\Cov_{ij} =
\left[
\begin{array}{ccc}
2 \{ \wtPgg\}^2 & 2\wtPgg\PgE & 2 \{ \PgE\}^2 \\
2\wtPgg\PgE & \wtPgg \wtPEE + \{ \PgE\}^2 & 2\PgE\wtPEE \\
2 \{ \PgE\}^2 & 2\PgE\wtPEE & 2 \{ \wtPEE\}^2 \\
\end{array}
\right]\ ,
\label{eq:covariance}
\ee
with $\wt{P}_{\rm gg}$ and $\wt{P}_{\rm EE}$ representing auto-power spectra with the Poisson shot noise,
\begin{align}
&\wtPgg = \Pgg + \frac{1}{n_{\rm g}} ,
\\
&\wtPEE = \PEE + \frac{\sigma_\gamma^2}{n_{\rm g}},
\end{align}
where $n_{\rm g}$ denotes the mean galaxy number density, and $\sigma_{\gamma}$ is the shape-noise quantifying the scatter in the intrinsic shape and measurement uncertainty.

The full Fisher matrix from galaxy surveys, $F^{\rm LSS}$, is then calculated as the summation of individual Fisher matrices at different redshifts,
\be
{\bfF}^{\rm LSS}=F^{\rm LSS}_{\alpha\beta}=\sum^{n_z}_{a=1}F_{\alpha\beta}(z_a)\ .
\ee
Then, $\bfF^{\rm LSS}$ schematically reads
\be
\left[
\begin{array}{ccccc}
\sum\limits_{a=1}^{n_{z}}(n_{\rm p}\small{\times} n_{\rm p})_{a} & (n_{\rm p}\small{\times} n_{\rm n})_{1} & (n_{\rm p}\small{\times} n_{\rm n})_{2} & \cdots & (n_{\rm p}\small{\times} n_{\rm n})_{n_{z}}  \\
(n_{\rm n}\small{\times} n_{\rm p})_{1} & (n_{\rm n}\small{\times} n_{\rm n})_{1} & 0 & \cdots & 0\\
(n_{\rm n}\small{\times} n_{\rm p})_{2} & 0 & (n_{\rm n}\small{\times} n_{\rm n})_{2}  & \cdots & 0\\
\vdots & \vdots &\vdots & \ddots& \vdots\\
(n_{\rm n}\small{\times} n_{\rm p})_{n_{z}} & 0 & 0 & \cdots & (n_{\rm n}\small{\times} n_{\rm n})_{n_{z}}\\
\end{array}
\right].
\ee
The only non-zero components in $\bfF^{\rm LSS}$ are the diagonal and the first column and row. The rest of the components are zero (Eq.~\ref{eq:nuisance_zindep}).
The Fisher components for the cosmological parameters in different redshift bins are summed up into the top-left $(n_{\rm p}\times n_{\rm p})$ submatrix, and the diagonal $(n_{\rm n}\times n_{\rm n})$ submatrices represent those for the nuisance parameters in the same redshift bin.
This is the identical Fisher matrix construction to that adopted in the Euclid forecast analysis \cite{euclid_prep+20}.
Then, we marginalize over all nuisance parameters from $\bfF^{\rm LSS}$ to obtain a smaller $n_{\rm p}\times n_{\rm p}$ marginalized Fisher matrix, $\tilde{\bfF}^{\rm LSS}$. The marginalized Fisher matrix contains components only for cosmological parameters and is computed as,
\be
 \tilde{\bfF}^{\rm LSS}\equiv[\bfC^{\rm LSS}]^{-1},
\ee
where $C_{\alpha\beta}^{\rm LSS}=[\bfF^{\rm LSS}]^{-1}_{\alpha\beta}$ with the indices $\alpha,\beta \leq n_{\rm p}$.

\subsection{CMB prior}\label{sec:prior}
In addition to the information from galaxy surveys, we include CMB information in the final Fisher forecast.
We adopt the Planck-15 compressed likelihood (Table 4 in \cite{planck+15}) as our CMB prior, which is relatively insensitive to the assumptions on the curvature and dark energy EOS \cite{wang&mukherjee07, mukherjee+08,zhai+20}.
This prior effectively summarizes CMB information into four parameters, $\mathbf{q}=(R, l_{\rm A}, \omega_{\rm b}, n_{\rm s})$ -- the rescaled distance to the last scattering surface $R\equiv\sqrt{\Omega_{m}H_0^{2}}D_{\rm A}(z_{*})/c$ \cite{efstathiou&bond99}, the angular size of the sound horizon at the last scattering $l_{\rm A}\equiv\pi D_{A}(z_{*})/r_{s}(z_*)$ \cite{wang&mukherjee07, mukherjee+08}, $\omega_{\rm b}\equiv\Omega_{\rm b}h^{2}$, and \ns, where the former two parameters are CMB shift parameters. 

By inverting the error covariance matrix for the CMB compressed likelihood, we obtain a $4\times4$ Fisher matrix, $S_{nm}^{\rm CMB}$, for the four parameters.
We then convert the constraints on $\bfq$ vector to those for $\boldsymbol{\theta}_{\rm p}$ vector using the following projection,
\be
F_{\alpha\beta}^{\rm CMB} = \sum_{n,m}^{4}\frac{\partial q_n}{\partial \theta_\alpha} S_{nm}^{\rm CMB} \frac{\partial q_m}{\partial \theta_\beta}\ (\alpha,\beta \le n_{\rm p}),
\label{eq:projection_fisher}
\ee
and then obtain $n_{\rm p}\times n_{\rm p}$ Fisher matrix, $\bfF^{\rm CMB}$, from the CMB prior. The combined cosmological parameter constraints from galaxy surveys and CMB prior are then given
\be
\bfF=\tilde{\bfF}^{\rm LSS}+\bfF^{\rm CMB}\ .
\ee

\subsection{Error covariance matrix}
The expected error covariance matrix, $\bfC$, for the cosmological parameters
is obtained by inverting the combined Fisher matrix $\bfF$,
\be
\bfC=C_{\alpha\beta}=[\bfF]^{-1}_{\alpha\beta}\ (\alpha,\beta\le n_{\rm p}).\label{eq:errcov}
\ee
Then, the 1D-marginalized errors on the cosmological parameters are the diagonal components of the error covariance matrix, $\sigma_{\alpha}=\sqrt{C_{\alpha\alpha}}$.

Mutual relations between two cosmological parameters $\theta_{\alpha}$ and $\theta_{\beta}$ can be examined by extracting the corresponding $2\times2$ submatrix from the error covariance matrix. It can then be visualized with the 2D confidence ellipse contours. The strength of the correlation between two parameters is quantified with the correlation coefficient defined as,
\be
\rho_{\alpha\beta} \equiv \frac{C_{\alpha\beta}}{\sigma_{\alpha}\sigma_{\beta}}.\label{eq:correlcoeffi}
\ee
To quantitatively gauge the constraining power on multiple cosmological parameters, we compute Figure-of-Merit (FoM) \cite{albrecht+06} adopting a more generalized \FoM{} definition for more than two parameters \cite{TO20, OT22}, defined as
\be
{\rm FoM} \equiv \left\{{\rm det}(\bfF)\right\}^{1/n_{\rm p}}
\label{eq:FoM}.
\ee
Using this definition, \FoM{} can be defined for an arbitrary number of parameters, yielding a measure that is inversely proportional to the mean radius of the $n_{\rm p}$-dimensional sphere with the error volume of those parameters \cite{OT22}.

\subsection{Survey setup}\label{subsec:setup}
We consider two complementary types of galaxy surveys for our forecast; a deep survey with narrow coverage and a wide survey with shallow depth. For the former and latter, we assume the Subaru PFS and Euclid covering redshift ranges, $0.6\le z<2.4$ and $0.9\le z <1.8$, respectively. Because both surveys observe emission line galaxies (ELGs) at $z\approx1-2$, we consider the IA power spectra measured with the estimator proposed in \cite{shi+21a}, effectively capturing the IA signal for the host halos of ELGs. The redshift bins, survey volume, galaxy number density, and bias for the PFS and Euclid are adopted from the references \cite{pfs+14} and \cite{euclid_prep+20}, respectively. 
Following \cite{shi+21a, OT22}, we set $A_{\rm IA}=18$ assuming it to be redshift-independent \footnote{We also tested two cases assuming $A_{\rm IA}(z)$, weakly evolving with redshift as $(1+z)^{0.3}$. For such cases, we set $A_{\rm IA}(z=1.0)=18$ and $A_{\rm IA}(z=1.5)=18$. Our conclusion is robust against such weakly varying intrinsic alignment amplitudes, regardless of the survey assumed.}. The shape noise parameters, $\sigma_{\gamma}$, are set to a slightly smaller value for the PFS since higher-quality shape information is expected from the imaging survey of the HSC \cite{hsc+18,hsc2+18}. We set $\sigma_{\gamma}=0.2$ for the PFS \cite{hikage+19} and $\sigma_{\gamma}=0.3$ for the Euclid \cite{euclid_prep+20}. We also discuss the impact of changing the $\AIA$ and $\sigma_{\gamma}$ in section~\ref{sec:discussion} and Appendix~\ref{app:setup}.

\section{Results}\label{sec:results}
\subsection{Figure-of-Merit analysis}

We first investigate how the constraining power on cosmological parameters changes when different cosmological models are considered, 
focusing on the clustering-only information.
More specifically, we illustrate \FoM{} change for two specific parameter vectors; the five common parameters in all models, $\boldsymbol{\theta}_{\rm base}=(\Omega_{\rm b}, \Omega_{\rm c}, h, A_{\rm s}, n_{\rm s})$, and the dark energy parameters, $\boldsymbol{\theta}_{\rm DE}=(w_{0}, w_{\rm a})$, as we widen the entire parameter space of a model by additionally considering extra free parameters.
We compare \FoM{} obtained from the clustering-only information with the CMB prior.

\begin{figure}
\includegraphics[width=\columnwidth]{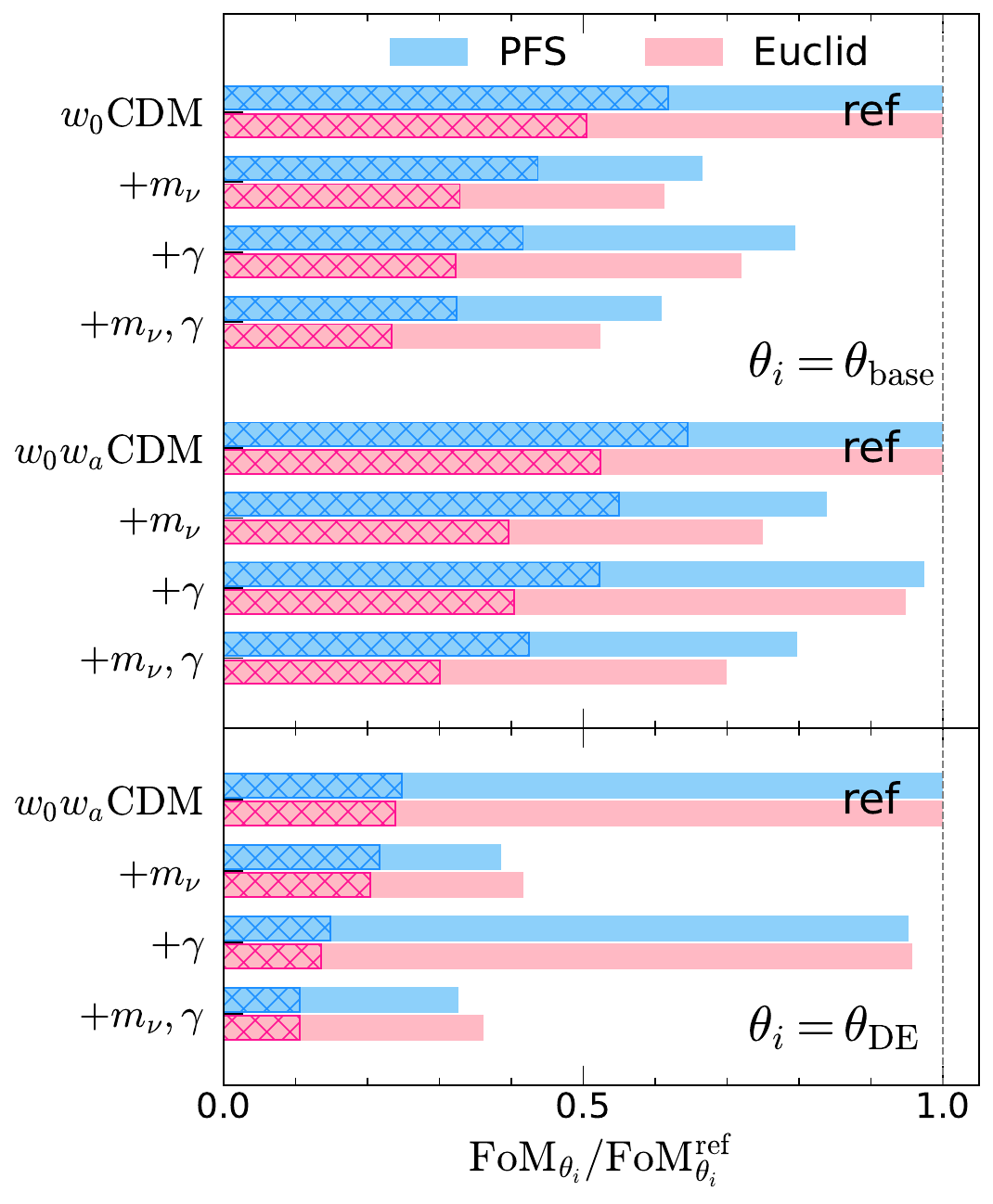}
    \caption{\FoM{} ratios for various models relative to a given reference model (denoted `ref'). Filled and hatched bars represent cases without and with curvature ($+$\Omk). Top panel: \FoM{} for the five base parameters \bfthetabase{} with the reference model \basemodel{} (\baseplusmodel) for the upper (lower) eight bars. Bottom panel: \FoM{} for the dark energy parameters \bfthetaDE{} with the reference model, \baseplusmodel. Clustering-only information from PFS-like and Euclid-like surveys combined with the CMB prior are used. }\label{fig:FoMdec}
\end{figure}

In the top panel of Fig.~\ref{fig:FoMdec}, we show \FoM{} for the five base parameters, \FoMbase, for various models normalized by that of a reference model. Here, we consider two reference models -- flat-\basemodel{} and flat-\baseplusmodel{} -- where the former and latter assume constant and time-varying dark energy EOS, respectively. The upper and lower eight bars in the top panel correspond to the normalized \FoMbase{}, or \FoMbase{} ratios, for the extensions of \basemodel{} and \baseplusmodel{} models, respectively. As \FoM{} is inversely proportional to the mean radius of the volume of the parameter uncertainties (see eqn.~\ref{eq:FoM}), FoM ratio smaller than unity indicates the overall constraint on \bfthetabase{} is degraded as extra free parameters are considered. Expectedly, constraints on \bfthetabase{} become weaker in all extended models investigated, i.e., ${\rm FoM}_{\rm base}/{\rm FoM}_{\rm base}^{\rm ref}<1$, as we additionally consider extra free parameters; \mnu, \gam, and \Omk. This is because extra degeneracies between the cosmological parameters are added when new parameters are included.

Let us first describe the behavior of one-parameter extensions of a given reference model.
Compared to the flat-\basemodel{} reference model, allowing non-standard gravity decreases \FoMbase{} by $20-28\%$, while the inclusion of the massive neutrinos degrades by $33-39\%$.
When adding \wa{}, we find that \FoMbase{} shrinks by $35-38\%$, although this is not directly inferable from this figure.
Such a \FoM{} decrease is more evident when assuming nonflat geometry. By allowing nonzero curvature, the \FoMbase{} becomes $38-50\%$ smaller, showing the most significant reduction than those due to any single extra parameter.
Considering the change in the \FoMbase{} relative to the flat-\basemodel, the impact of introducing new parameters is strongest for \Omk{}, followed by \mnu{} and \wa{}, and is weakest for \gam.

The \FoMbase{} reduction tends to increase with the number of extra free parameters.
For a two-parameter extension to the \basemodel{}, for example, simultaneously considering the modified gravity and massive neutrinos decreases \FoMbase{} by $39-48\%$. Freeing both \gam{} (or \mnu) and \Omk{} shrinks \FoMbase{} more than twofold compared to the flat-\basemodel{} model. The maximum reduction becomes a factor of $4.3$ when three extra parameters are included, i.e., nonflat and non-GR \basemodel{} with massive neutrinos.

A qualitatively similar trend of decreasing \FoMbase{} is reproduced in the extensions of the simplest dynamical dark energy model, i.e., flat-\baseplusmodel{}. However, the \FoMbase{} reductions in these models are less significant than those in the \basemodel{} models. 
In the most extended model, i.e., \baseplusmodel$+($\Omk, \mnu, \gam$)$, \FoMbase{} decreases by a factor of $\approx5.3$ compared to the flat-\basemodel. Finally, we note that \FoMbase{} in a PFS-like deep and narrow survey tends to be less degraded than a Euclid-like wide and shallow survey when adding extra parameters into consideration.

In the bottom panel of Fig.~\ref{fig:FoMdec}, we repeat the same analysis for \FoMDE{} for the dark energy parameters with the flat-\baseplusmodel{} model as the reference. We again observe the consistent behavior in \FoM{} when extra components are included; the \FoMDE{} reduction is the largest with \Omk, intermediate with \mnu, and least with \gam. For example, allowing the curvature alone, \FoMDE{} decreases approximately fourfold.
The impact of including massive neutrinos on \FoMDE{} is larger than that on \FoMbase, showing $\sim60\%$ \FoMDE{} reduction.
Given such impact due to \Omk{} and \mnu{} on the dark energy constraints, the recent $2-3\sigma$ preference for the time-evolving dark energy EOS based on the DESI Y1-data \cite{desi_BAO+24, desi_FS+24} could have been relaxed if both were taken into account.
On the other hand, the \FoMDE{} degradation is only around $5\%$ level with the modification of gravity in flat models. Accounting for multiple extra parameters significantly shrinks \FoMDE, even yielding an order of magnitude smaller \FoMDE{}, for example, in the nonflat-\baseplusmodel{}+(\mnu{} ,\gam).

\begin{figure}
\includegraphics[width=\columnwidth]{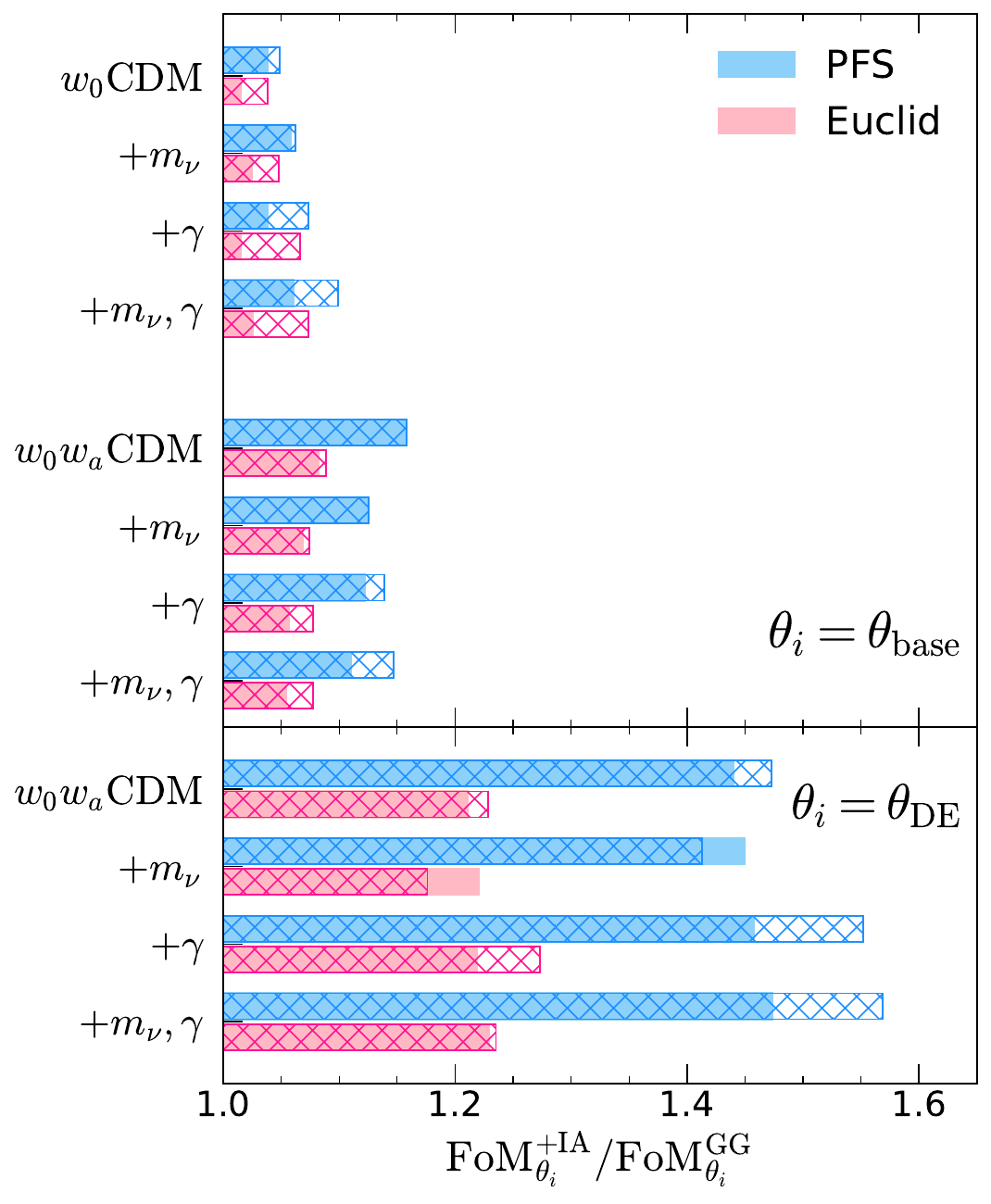}
    \caption{FoM improvement with IA in each model relative to its clustering-only \FoM{}. Filled and hatched bars represent cases without and with curvature (+\Omk). \FoM{} improvement for the five base parameters \bfthetabase{} and dark energy parameters \bfthetaDE{} are shown in the top and bottom panels, respectively. CMB prior is included.}\label{fig:FoMimp}
\end{figure}

Let us now examine the impact of combining galaxy IA with clustering. We compute the ratio of \FoM{} from the joint (clustering+IA) analysis over that from the clustering-only analysis. The \FoM{} ratio greater than unity then indicates a gain in \FoM{}, or an improvement in the constraining power. In the top panel of Fig.~\ref{fig:FoMimp}, we show the \FoM{} ratios for the base parameters. It is shown that combining IA with clustering increases \FoMbase{}, indicating improved constraints on \bfthetabase. The \FoMbase{} improvement in one-parameter extensions of \basemodel{} model is most significant with \wa{} ($16\%$) followed by \mnu{} ($6\%$), \Omk{} ($5\%$), and \gam{} ($4\%$). Although not substantial, the improvement becomes more noticeable in the \baseplusmodel{} extensions independent of extra parameters; around $15\%$ level for a PFS-like survey and about $7-9\%$ level for an Euclid-like survey.

On the other hand, the galaxy IA significantly improves the constraining power on the dark energy parameters as shown in the lower panel of Fig.~\ref{fig:FoMimp}. The gain in \FoMDE{} with IA is significantly larger in PFS-like surveys, achieving roughly $50\%$ improvements, whereas it is around $20\%$ improvements in Euclid-like surveys. Such a noticeable difference in the \FoMDE{} improvement is largely due to the different choices of the shape noise for the two surveys. The \FoM{} gain with IA sensitively depends on the shape noise \cite{TO20, OT22} and such dependence on survey parameters will be discussed in section~\ref{sec:discussion}. Note that \FoMDE{} improvement in Euclid-like survey becomes comparable when the shape noise is set identically to the PFS case. We note that the improvement with IA remains consistently significant even without the CMB prior\footnote{Without CMB prior, adding IA improves \FoMbase{} and \FoMDE{} by $24-33\%$ and $19-47\%$ depending on models, respectively.}, aligning with the findings of Refs.~\cite{TO20, OT22}.

\begin{figure}
\includegraphics[width=\columnwidth]{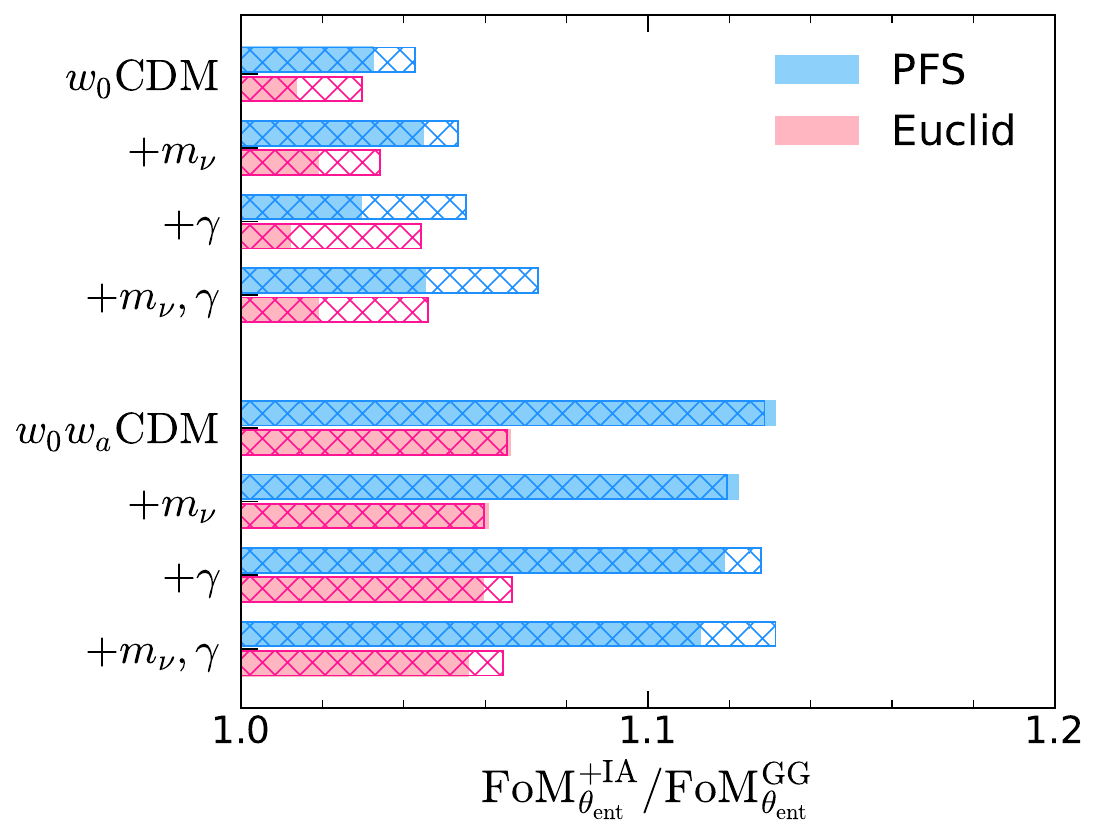}
\caption{Same as Fig.~\ref{fig:FoMimp}, but \FoM{} for the entire parameters $\boldsymbol{\theta}_{\rm ent}$. Note that the number of parameters used to calculate \FoMent{} differs in different models.}\label{fig:FoMent}
\end{figure}
In Fig.~\ref{fig:FoMent}, we examine the \FoM{} improvement with IA for the entire parameter space, $\boldsymbol{\theta}_{\rm ent}$, of the models, i.e., the base plus extra parameters. Regardless of the survey type, the overall improvement tends to enhance in nonflat models.
The improvement is again shown to be larger for PFS-like surveys. 
For PFS-like deep surveys, the \FoMent{} improvement is at maximum $\sim7\%$ for \basemodel{} extensions but roughly doubles to $\sim13\%$ for \baseplusmodel{} extensions. With Euclid-like surveys, the largest \FoMent{} gain is only about $6\%$ even when nonzero curvature is considered.

\subsection{Parameter constraints and degeneracies}
%
\begin{figure*}
\includegraphics[width=2\columnwidth]{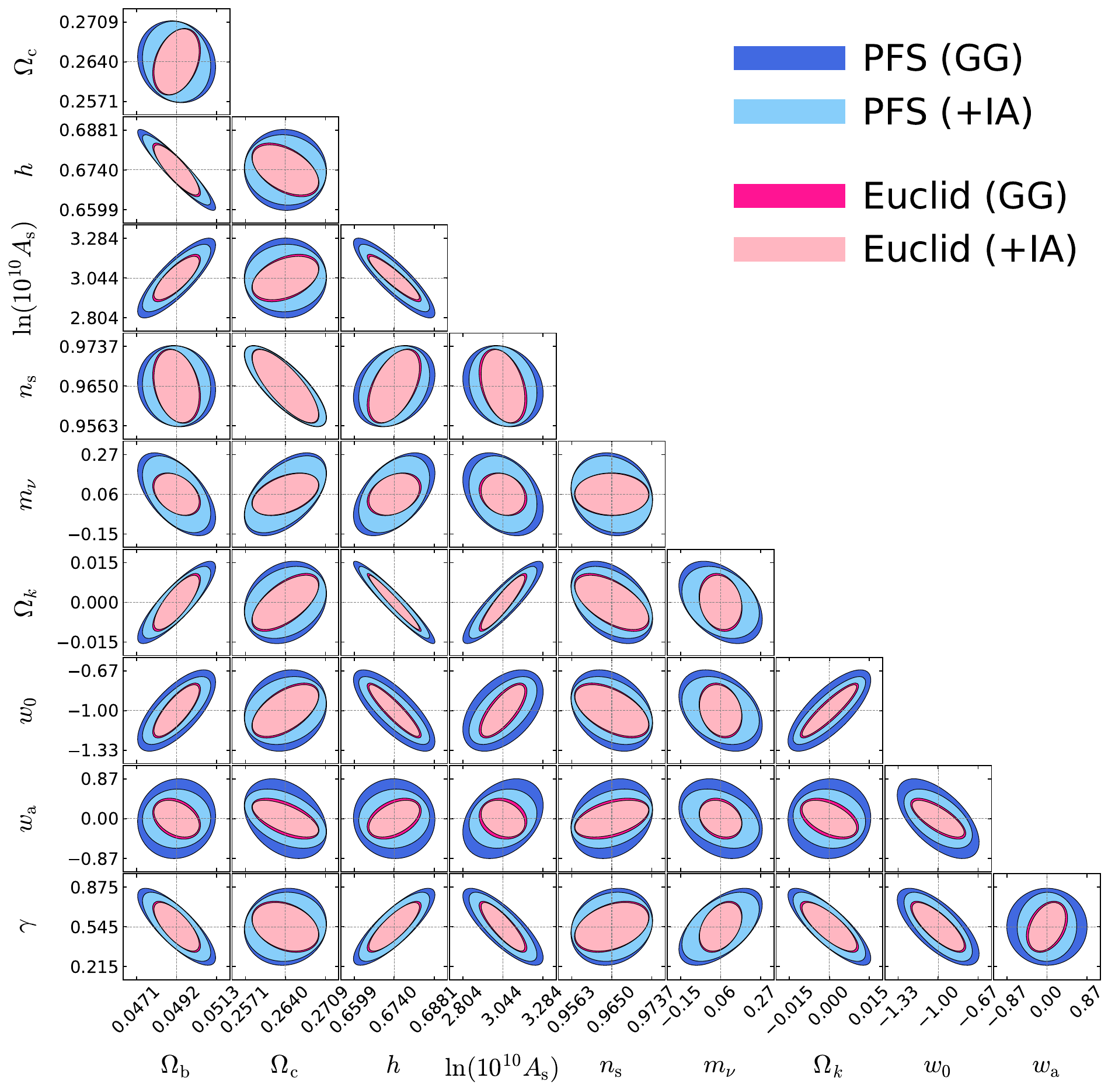}\caption{2D-confidence ellipses for 10 cosmological parameters of the most extended model, \baseplusmodel{}+(\Omk, \mnu, \gam). Contours represent 1-$\sigma$ confidence regions. Each axis spans twice the 1D-marginalized error in both directions from the corresponding parameter's fiducial value. Ticks are located at $1.5$ times the error away from the fiducial values. 1D-marginalized errors obtained from PFS-like surveys are used for this plot.}\label{fig:contour10par}
\end{figure*}
Extending the \FoM{} analysis that focused on the behavior of collective constraints on a group of parameters, 
we now focus on individual-level parameter constraints and degeneracies between parameters. 
We show confidence ellipse contours for the pairs of cosmological parameters in the most extended cosmological model, i.e., \baseplusmodel{}+(\Omk, \mnu, \gam) in Fig.~\ref{fig:contour10par}, as an illustrative example for revealing the impact of IA on 2D-marginalized parameter constraints and degeneracies between cosmological parameters.
We can visually confirm that the ellipses become smaller when adding the IA information, implying that the joint information tightens the parameter constraints. In particular, we find that ellipses for parameter pairs involving \wO{} or \wa{} contracted more noticeably, as can be expected from the substantial increase in \FoM{} for the dark energy parameters. Similar examples for one-parameter extensions of the \baseplusmodel{} model in Fig.2 of Ref.~\cite{shim+25}.

The degeneracy direction between parameters is also affected if IA information is combined; some of the ellipses from the joint analysis are rotated from their clustering-only counterpart. For example, for \wa-- \Omc{} and \wa-- \mnu{}, the directions of the inner ellipses are tilted anti-clockwise than their outer ellipses, while it is rotated in the opposite direction for \wa-- \ns. This indicates that IA information can contain different parameter degeneracies from clustering information,
even though both IA and clustering probe the same matter fluctuation.
This trend is commonly found in both PFS-like and Euclid-like surveys although the size reduction and direction change of contour ellipses are less noticeable in the Euclid-like surveys, due to a larger shape noise adopted for the IA statistics. We also find that degeneracy directions for a given parameter pair could be different depending on the survey. This is probably due to their different redshift coverages as parameter degeneracies can non-negligibly evolve at higher redshift \cite{matsubara04}.

%
\begin{figure*}
\includegraphics[width=2\columnwidth]{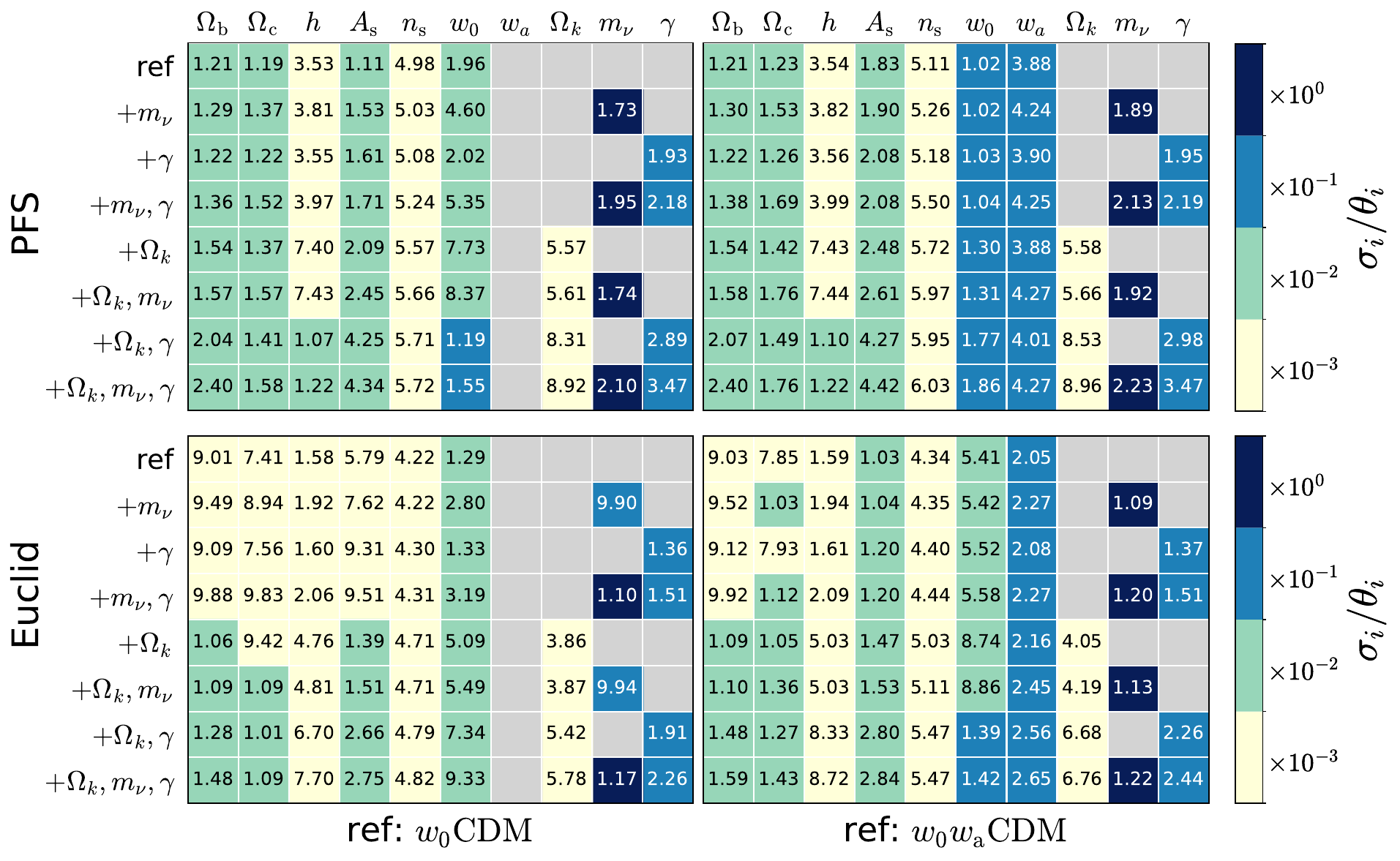}
\caption{1D-marginalized constraints on cosmological parameters from PFS-like (upper) and Euclid-like (lower) surveys, including CMB prior. Joint constraints as fractional errors, $\sigma_{i}/\theta_{i}$, are shown. For \wa{} and \Omk{}, we show their actual errors, $\sigma_{i}$, since their fiducial values, $\theta_{i}$, are zero. Extensions of \basemodel{} and \baseplusmodel{} models are in the left and right panels, respectively. Color represents the fractional error in powers of 10, aiding visual perception. Fixed cosmological parameters are left blank in grey.}\label{fig:TabVal}
\end{figure*}

We now examine individual 1D-parameter constraints in Fig.~\ref{fig:TabVal}, with the rest of the parameters marginalized over.
For each survey, we tabulate the joint marginalized constraints as fractional errors, $\sigma_{i}/\theta_{i}$, where $\sigma_{i}$ and $\theta_{i}$ are the 1D-marginalized error and fiducial value of a parameter. The parameter constraints are always tighter in the Euclid-like survey than in the PFS-like survey. For the base parameters in flat-\basemodel{} models, their constraints become sub-percent level in the Euclid-like surveys whereas some of these are slightly above one percent in the PFS-like survey. The constraints on the time-constant dark energy parameter \wO{} also improve and drop below $10\%$ level in the flat-\baseplusmodel{} models. This is likely due to the survey volume difference, implying that Euclid's volume is significantly larger than the PFS-like survey, overcoming its narrower redshift coverage.

Let us now examine how parameter constraints change in different models. We again find that increasing the number of extra free parameters always weakens constraints on cosmological parameters; e.g., \bfthetabase{} and \bfthetaDE, as expected from the trend found in the \FoM{} analysis (Fig.~\ref{fig:FoMdec}). Furthermore, we observe an interesting feature -- the degeneracy-dependent worsening of parameter constraints -- that is not captured in the \FoM{} analysis. 
For example, the average error on \bfthetabase{} is expected to be larger in \basemodel{}+\mnu{} model than in \basemodel{}+\gam{} model according to the \FoM{} analysis.
However, individual errors on \As{} and \ns{} are larger in the latter, while the constraints on the rest of the common parameters, e.g., \Omb, \Omc, \h, and \wO{}, are larger in the former. Such a behavior can be understood as the consequence of different parameter degeneracies among parameters. 
For a qualitative explanation for the phenomena, we compute correlation coefficients defined in Eq.(\ref{eq:correlcoeffi}) for one-parameter extensions of the \basemodel{} model in Table~\ref{tab:corrcoeffi}. The absolute value of the correlation coefficients for \As{} and \ns{} are larger in \basemodel{}+\gam{} model than \basemodel{}+\mnu{} model, indicating \gam{} is more degenerate with those two parameters. Therefore, constraints on those parameters become weaker in the \basemodel{}+\gam{} model. On the other hand, the remaining four common parameters show stronger degeneracies with \mnu, thus yielding larger uncertainties for those four parameters in the presence of massive neutrinos.

Analogously, a similar argument applies when comparing the 1D-constraints for \basemodel{}+\wa{} and \basemodel{}+\Omk{} models.
In the \baseplusmodel{} model, the time-varying dark energy parameter \wa{} is almost perfectly degenerate with \wO{}, whereas the rest of the correlations are much weaker. On the contrary, in the $w_{0}\Omega_{k}{\rm CDM}$ model, the curvature parameter is strongly degenerate with all parameters but with slightly weaker degeneracy with \wO{}. Consequently, while allowing nonzero curvature most significantly degrades constraints on all base parameters, the error on \wO{} becomes much larger when adding \wa{}, leaving other parameter constraints relatively much less impacted compared to those in the $w_{0}\Omega_{k}{\rm CDM}$ case.

\begin{table}
\caption{Correlation coefficients between extra (first column) and common parameters (first row) in one-parameter extensions of \basemodel{} model, using joint information from PFS-like survey.}
\begin{center}
\begin{tabular}{|c|c|c|c|c|c|c|}
\hline 
& \Omb &  \Omc & \h & \As & \ns & \wO \\
\hline
\gam & $-0.15$ & $0.23$ & $0.10$ & $-0.72$ & $-0.20$ & $-0.24$\\
\mnu & $-0.34$ & $0.50$ & $0.37$ & $0.69$ & $-0.13$ & $-0.91$\\
\wa &  $0.01$ & $-0.28$ & $0.08$ & $0.79$ & $0.22$ & $-0.98$\\
\Omk & $0.62$ & $0.50$ & $-0.88$ & $0.85$ & $-0.45$ & $0.97$\\
\hline
\end{tabular}
\label{tab:corrcoeffi}
\end{center}
\end{table}
\begin{table}
\caption{Same as Table~\ref{tab:corrcoeffi}, but for one-parameter extensions of $w_{0}\Omega_{k}{\rm CDM}$ model.}
\begin{center}
\begin{tabular}{|c|c|c|c|c|c|c|c|}
\hline 
& \Omb &  \Omc & \h & \As & \ns & \wO & \Omk \\
\hline
\gam & $-0.65$ & $-0.24$ & $0.72$ & $-0.87$ & $0.22$ & $-0.76$ & $-0.74$\\
\mnu & $-0.20$ & $0.50$ & $0.08$ & $0.53$ & $-0.18$ & $-0.39$ & $0.13$ \\
\wa &  $-0.03$ & $-0.27$ & $0.09$ & $0.54$ & $0.23$ & $-0.80$ & $-0.06$\\
\hline
\end{tabular}
\label{tab:corrcoeffi2}
\end{center}
\end{table}

In two-parameter extensions, the impact on parameter constraints by the addition of a particular extra parameter can also change in the presence of another extra parameter.
For example, when comparing the 1D-marginalized constraints for the $w_{0}\Omega_{k}{\rm CDM}$ extensions, the constraints tend to degrade the most if \gam{} is additionally considered. This is in contrast to the trend found in the one-parameter extensions of \basemodel{} model, where adding \gam{} parameter showed minimal impact on parameter constraints.
Again, this can be explained almost entirely based on the parameter degeneracy argument.
In the presence of curvature, the absolute values of correlation coefficients for parameter pairs involving \gam{} become much larger than those involving \mnu{} or \wa{}, as shown in Tab.~\ref{tab:corrcoeffi2}. This is obviously because \gam{} shows the strongest degeneracy with \Omk{} than \mnu{} and \wa{}. As the curvature parameter is strongly degenerate with all common parameters (Tab.~\ref{tab:corrcoeffi}), adding another extra parameter that is strongly degenerate with \Omk{} effectively builds mutual degeneracy among three parameters.
Consequently, the parameter constraints become the most uncertain with \gam{} among those three nonflat-\basemodel{} models.
The change in 1D-marginalized constraints is dominantly determined by mutual parameter degeneracies, and hence, investigating the degrees and directions of such correlations is important in constraining more generalized cosmological models.

\begin{figure*}
\includegraphics[width=2\columnwidth]{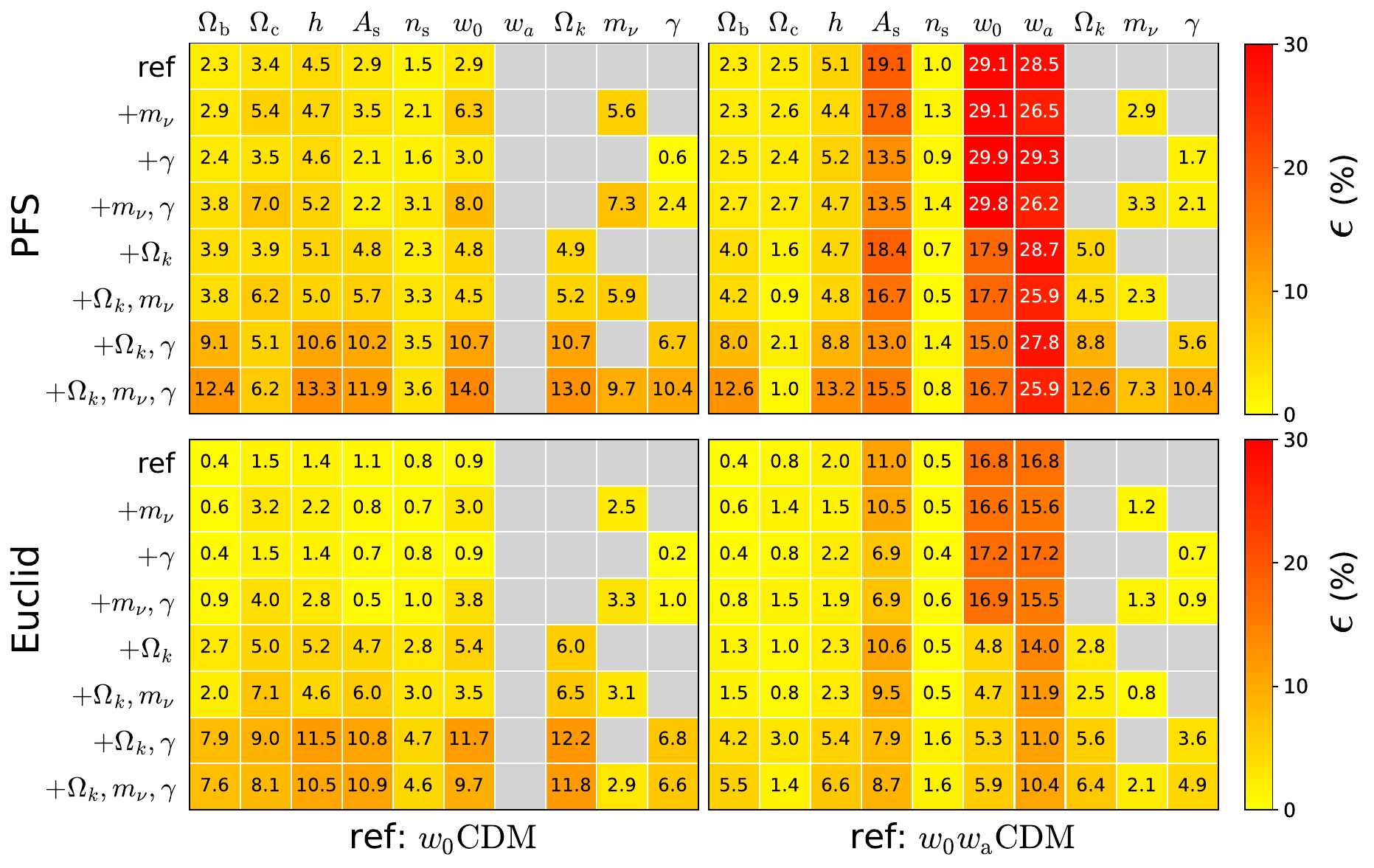}
\caption{Improvement in 1D-marginalized constraints on cosmological parameters with IA, relative to clustering-only constraints. Again, the extensions of \basemodel{} and \baseplusmodel{} models are in the left
and right panels, respectively. Color represents the level of improvement, aiding visual perception. Fixed cosmological parameters are left blank in grey.}\label{fig:TabImp}
\end{figure*}

\subsection{Improvement in parameter constraints with IA}
Let us now examine how significantly the IA information improves the 1D-marginalized constraints. In Fig.~\ref{fig:TabImp}, we display the improvement in the joint analysis relative to the clustering-only constraints.
We observe some particular models exhibiting pronounced improvement. For example in the PFS-like case, the most significant improvement is detected in \baseplusmodel{} models. In these models, adding the IA statistics particularly contributes to tightening the constraints on dark energy and power spectrum amplitude. For the time-constant part of the dark energy EOS, the marginalized error shrinks more than $15\%$ in nonflat cases but becomes even tighter by $\sim30\%$ in flat cases. The gain in \wa{} constraints exceeds at least $25\%$ in all extensions of the \baseplusmodel{} model. Such a relatively smaller improvement in \wO{} constraints in nonflat models can be attributed to the tight correlation between \wO{} and \Omk{} (e.g., see Tab.~\ref{tab:corrcoeffi}).
On the other hand, \wa{} is only negligibly degenerate with \Omk{} (e.g., Table~\ref{tab:corrcoeffi2}), allowing its improvement to be almost independent of the curvature. We note that the correlation coefficient for \wO{}-\wa{} pair negligibly reduces with the addition of IA (e.g., see Fig.~\ref{fig:ellipse_w0wa}).

In addition to dark energy parameters, the marginalized error on \As{} also noticeably reduces in \baseplusmodel{} models, yielding $\sim19\%$ improvement at maximum. In these cases, it should be worth noting that \wO{}--\As{} and \wa{}--\As{} degeneracies are broken noticeably with IA (see correlation coefficients in Fig.~\ref{fig:ellipse_w0wa}).
When adding \gam{} to \baseplusmodel{} models, improvements become less effective and decrease to $\sim13\%$. Considering the correlation coefficients for \gam{}--\As{} and \mnu{}--\As{} in \baseplusmodel{} models (Fig.~\ref{fig:ellipse_w0wa}), such weaker improvement in models with \gam{} happens due to the stronger \gam{}--\As{} degeneracy.

The benefit of IA also becomes significant in models including both \Omk{} and \gam{} parameters, i.e. nonflat-MG models. In such cases, improvements for these extra parameters and some of the base parameters -- e.g., \Omb{} and \h{} -- become at least about twice larger than in models where \Omk{} and \gam{} are individually considered. This trend consistently appears in both \basemodel{} and \baseplusmodel{} models. 
On the other hand, the gain from IA is minimal for the spectral index, \ns{}, in most of the models investigated.
Improvement for \ns{} is on average only at $\sim1\%$ level in the extensions of the \baseplusmodel{} models.
The error improvement in Euclid-like surveys also shows a similar trend. However, the improvement is less significant due to the larger shape-noise parameter for the Euclid-like survey.
The improvements in \wO, \wa, and \As{} in \baseplusmodel{} models are on average $\sim40\%$ smaller than those in the PFS-like case. In nonflat-\baseplusmodel{} models, the improvement becomes weaker at least more than by $\sim65\%$ for \wO.

\subsection{Comparison to other parameter constraints without IA}
We briefly compare our joint full-shape forecast with actual cosmological constraints obtained without IA information. Since different analyses utilize various surveys and modeling approaches, this comparison serves as a rough gauge of the effectiveness of the full-shape IA.
In the \basemodel{} model, the PFS-like survey provides constraints that are at least $30\%$ tighter for \Omc{}, \h{}, and \wO{}, though weaker for \As{} and \ns{} compared to the full-shape clustering constraints from eBOSS QSOs combined with luminous red galaxies, external BAO, and supernovae data along with Planck CMB \cite{simon+23}. However, the constraints from an Euclid-like survey are tighter for all parameters than Ref.~\cite{simon+23}. When focusing on dark energy parameters in both the \basemodel{} and \baseplusmodel{} models, our joint analysis assuming a PFS-like survey provides dark energy constraints comparable to those from analyses that do not employ full-shape information, but include supernova data that directly probe the expansion \cite{planck+18, alam+21}. Our joint constraints on \wO{} and \wa{} are weaker only by $12-35\%$ and $5-40\%$, in \baseplusmodel{} or \baseplusmodel{}+\mnu{} models when compared to the DESI's full-shape clustering constraints \cite{desi_FS+24}, which adopted nonlinear modeling and combined with DESI BAO \cite{desi_BAO+24}, CMB, and supernovae results.

For the simplest MG model, the \basemodel{}+\gam{} model, our joint constraints on \Omc{} and \h{} are tighter than the constraints combining full-shape clustering information from BOSS DR12 with nonlinear modeling, Planck CMB, and big bang nucleosynthesis \cite{aviles24}.
However, our joint constraint on \As{} is weaker, likely due to a strong degeneracy between the \gam{} parameter and \As{}. In contrast, Ref.~\cite{aviles24} directly calculates the scale-dependent linear growth rate without relying on the \gam{}-parameterization, which may account for their tighter constraints on \As{}.
In the context of massive neutrinos within MG models, our forecast yields tighter constraints on \h{} and \ns{}, but weaker constraints on \As{} and \gam{}, compared to the full-shape clustering constraints that use nonlinear modeling, assuming DESI \cite{moretti+23}. Given that Ref.~\cite{moretti+23} considers a simpler $\Lambda$CDM+(\mnu, \gam{}) model and applies CMB priors to \As{} and \ns{} constraints, unlike our forecast, the weaker constraints on \As{} and \gam{} in our results may be mainly attributed to the absence of such priors on \As{}. Our constraints on massive neutrinos in the extended \baseplusmodel{}+(\mnu, \Omk, \gam) model are consistent with those derived from a full-shape analysis that employs linear modeling for clustering, while adopting similar CMB priors along with additional parameter-specific priors \cite{boyle&komatsu+18}. These comparisons demonstrate the effectiveness of IA for tightening cosmological constraints when combined with clustering information.

\section{Discussion}\label{sec:discussion}
\subsection{Full-shape vs. Geometric \& Dynamic constraints}
In this subsection, we describe how different the full-shape constraints are from the geometric/dynamical constraints. 
More specifically, we compare our results with forecasts from the geometric/dynamical constraints in Ref.~\cite{OT22} with the same survey setups. 
In geometric/dynamical analysis of Ref.~\cite{OT22}, constraints are first placed on geometric and dynamical parameters-- e.g., $H(z), D_{\rm A}(z)$, and $f(z)$-- by leveraging RSD and AP effect. These constraints are then projected onto cosmological parameters of interest. In contrast, our full-shape analysis directly constrains cosmological parameters without the projection. Most importantly, our analysis further utilizes the broadband shape information of the power spectrum in addition to the information from the RSD and AP effect.
However, it is not straightforward to perform a quantitative comparison due to the differences in the parameter space, their fiducial values, and the treatment of CMB prior.
Thus, we focus only on the common parameters in the two approaches and perform the crude order-of-magnitude comparison between the full-shape and geometric/dynamical constraints.

We first compare the 1D-marginalized errors on \Omc, \h, \wO, \wa, \Omk, and \gam{} in the extended \baseplusmodel{} models with curvature and/or gravity parameters (see Table V. in Ref.~\cite{OT22}). Note that we compare the error for the dark matter density parameter \Omc{} in the current full-shape analysis with that for the total matter (baryon+dark matter) density parameter in \cite{OT22}.
In the full-shape analysis, \Omc, \wO, \wa, and \h{} are more tightly constrained than in the geometric/dynamical constraints.
The difference in constraints is most pronounced for the matter density and Hubble constant, yielding an order-of-magnitude improvement at maximum in the full-shape analysis. For example, in a PFS-like survey, \Omc{} constraints are around $1\%$-level in our full-shape forecast, whereas they are $\sim10\%$-level in the geometric/dynamical constraints. For \baseplusmodel{}+\gam{} model, the full-shape \Omc{}-constraint can be an order of magnitude tighter in a Euclid-like survey.
The Hubble constant constraints are typically a factor of $5-10$ tighter in the full-shape analysis regardless of the survey types. For the dark energy parameters, the discrepancy is relatively smaller. The full-shape marginalized errors for dark energy parameters are $2-4$ times smaller.

For the constraints on \Omk{} and \gam{} parameters, the full-shape forecast yields similar or slightly weaker constraints compared to the geometric/dynamical constraints. Considering that CMB places strong constraints on the curvature parameter, such slightly weaker full-shape constraints on the curvature may be due to the different CMB prior and its treatment.
One possibility for a similar constraint on \gam{} in the full-shape forecast may be related to the different growth factor $D(z)$ treatment for modified gravity models.
Our analysis further considers the effect of modified gravity on the growth factor to modulate the linear matter power spectrum at a given redshift, while this is not considered in Ref.~\cite{OT22}. 

A potentially valuable comparison that has not yet been explored is the improvements with IA in the full-shape and geometric/dynamical forecasts. 
In the full-shape approach, the improvements for \Omc{} and \h{} are $2-3$ times smaller than in Ref~\cite{OT22} in both surveys, except for MG models. For MG models, such discrepancies between the two approaches are much larger because IA significantly improved their constraints in the geometric/dynamical analysis.
Such a smaller improvement in the full-shape forecast implies that the full-shape clustering analysis already provides tight constraints on these parameters. Hence, the relative contribution of the full-shape IA analysis becomes less significant than that in the geometric/dynamical information-based forecast.
The improvement in \gam{} constraint is also much weaker in the current analysis, smaller by a factor of $5-10$. This may be due to the different linear growth rate modeling in MG models, where we consider scale-dependent linear growth rate, while scale-independent parametrization was adopted in Ref.\cite{OT22}. 
On the other hand, the full-shape improvement for the curvature is similar to the geometric/dynamical cases. The dark energy parameter improvements in the full-shape forecast are larger by $2-4$ times than those in the geometric/dynamical cases, unless MG is simultaneously considered. This implies that IA can still significantly contribute to constraining dark energy models in the full-shape analysis.

\subsection{Model-dependent paramater degeneracies}
\begin{figure*}
\includegraphics[width=\columnwidth,valign=t]{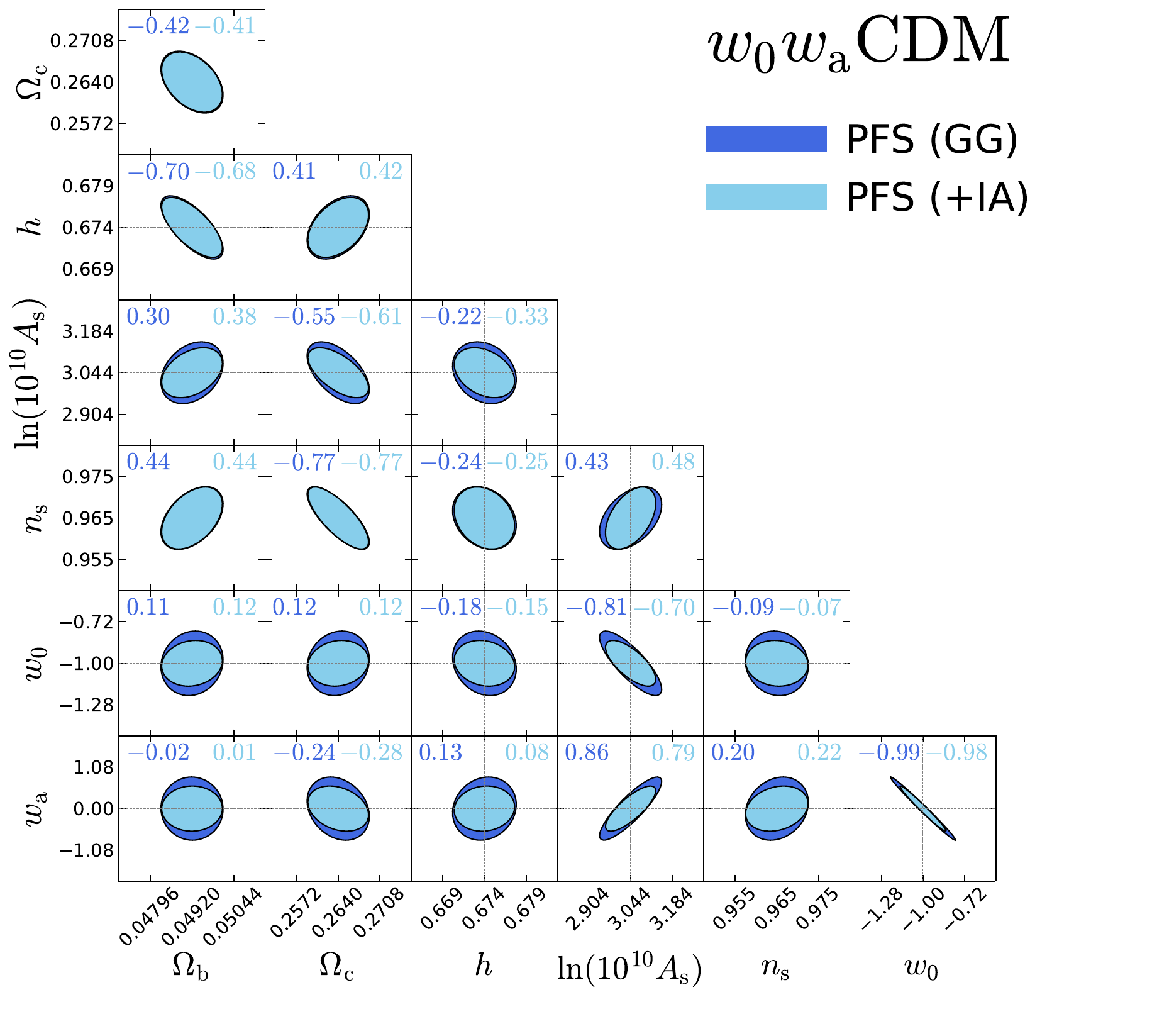}
\includegraphics[width=\columnwidth,valign=t]{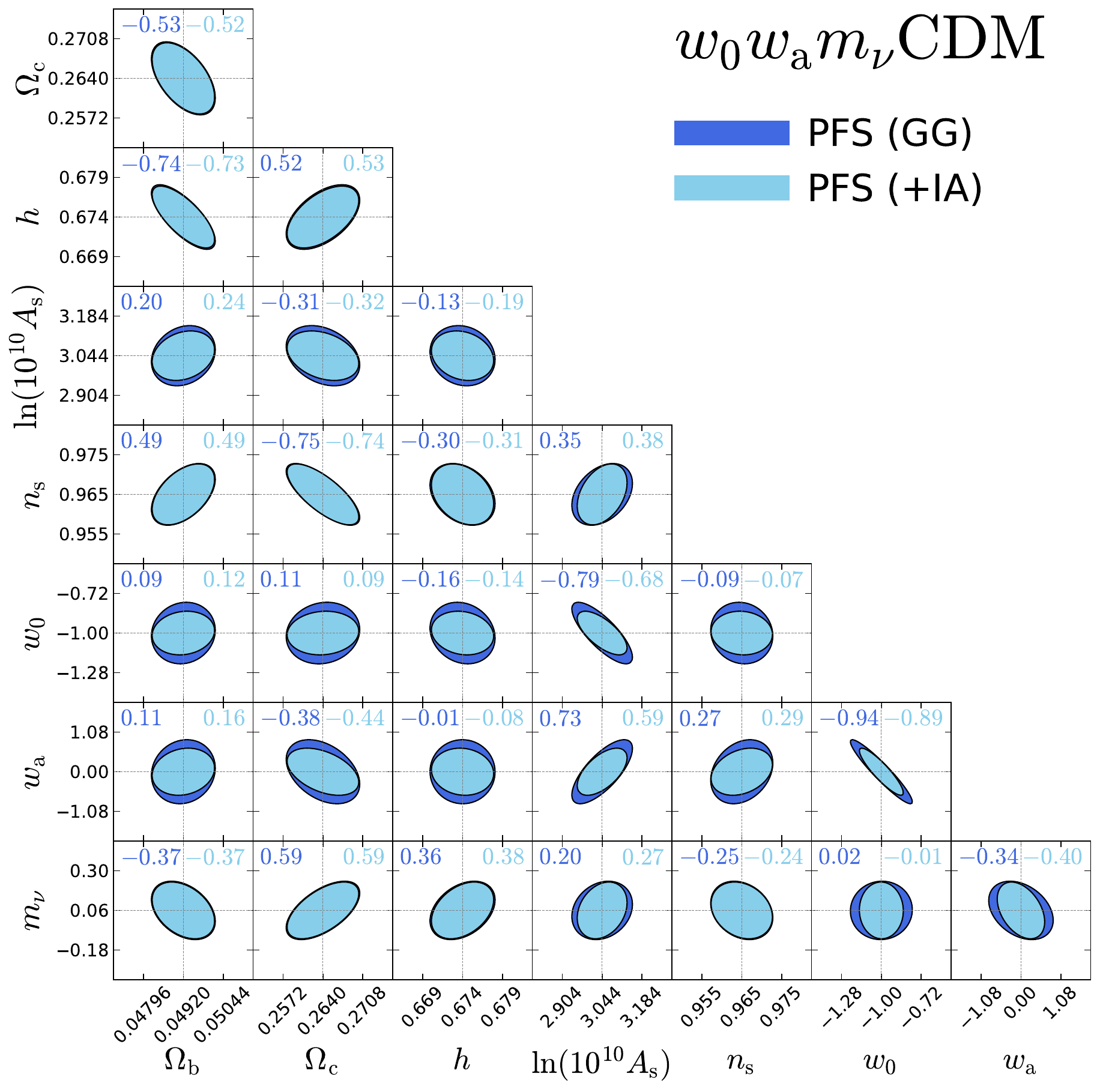}
\includegraphics[clip,width=\columnwidth]{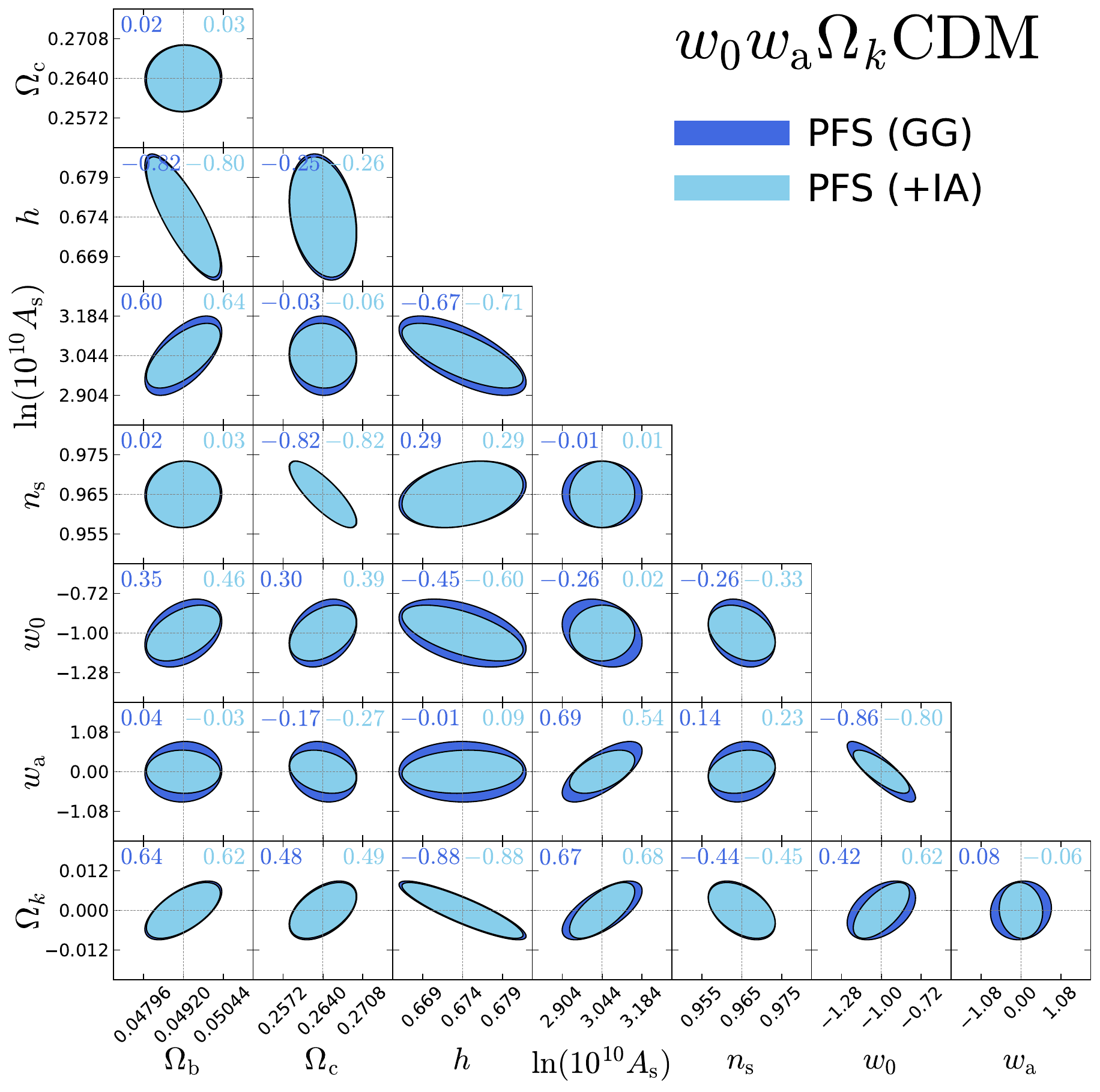}
\includegraphics[clip,width=\columnwidth]{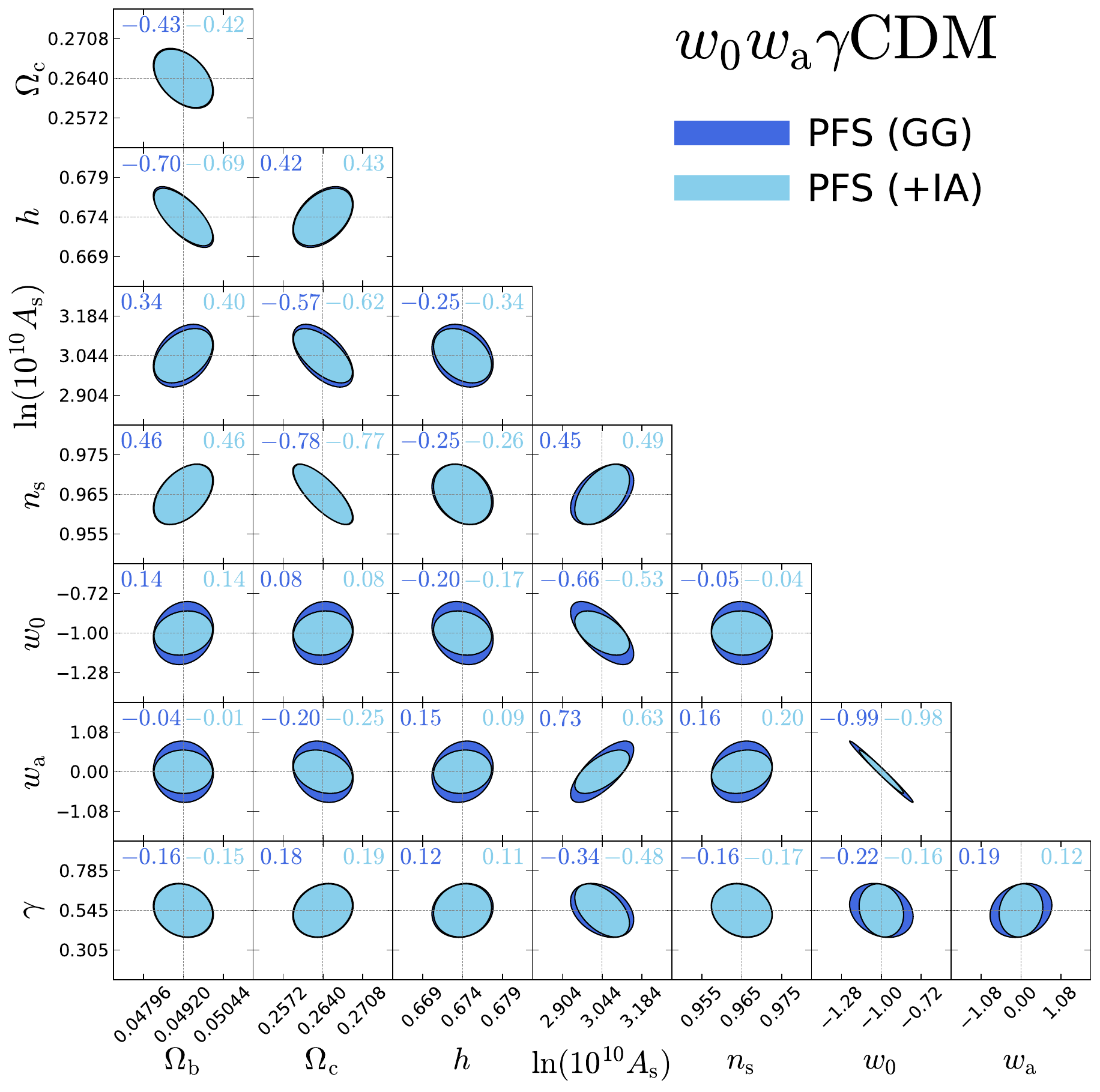}
\caption{2D-confidence ellipse contours for \baseplusmodel{} and its one-parameter extension, including CMB prior. 1$\sigma$-confidence contours are shown. Correlation coefficients between two parameters without (top-left corner) and with (top-right corner) IA information are shown in each panel, respectively. All axes are scaled with the 1D-marginalized errors of the corresponding parameters for the \baseplusmodel{} model so that ticks mark $1.5\sigma$ deviations from their fiducial values.}\label{fig:ellipse_w0wa}
\end{figure*}
As parameter constraints sensitively depend on the degree and direction of degeneracy between parameters as discussed in section~\ref{sec:results}, let us further investigate the behavior of parameter degeneracies in various models and how they are impacted by IA.
Fig.~\ref{fig:ellipse_w0wa} visually exemplifies the impact of IA on parameter degeneracies and their variations in \baseplusmodel{} model extensions, with the correlation coefficients. As can be expected from Figs.~\ref{fig:TabVal} and ~\ref{fig:TabImp}, while the addition of IA always shrinks ellipse contours, the degree of such change differs depending on parameters. When focusing on ellipses with relatively more noticeable contraction, e.g., those for parameter pairs involving \As{} or dark energy parameters, we find a more discernible degree of rotation of the inner ellipses, likely accompanying a larger correlation coefficient change. This implies that IA has different parameter degeneracy from galaxy clustering and plays an effective role in breaking those parameter degeneracies.

On the other hand, it is also shown that the shape, size, and orientation of the ellipses for a given parameter pair change in different models, indicating a model-dependent parameter degeneracy. Such a model-dependent variation can be similarly seen in the correlation coefficients (Eq.(\ref{eq:correlcoeffi})). For instance, the shape of \Omc{}-\Omb{} contour in the \baseplusmodel{}+\mnu{} models is more elongated than that in the \baseplusmodel{}, yielding a larger absolute value of correlation coefficient, indicating a stronger linear degeneracy. On the contrary, the contour becomes much rounder in the \baseplusmodel{}+\Omk{} model with an almost vanishing correlation coefficient, indicating a weak linear degeneracy.

\begin{figure}
\includegraphics[width=\columnwidth]{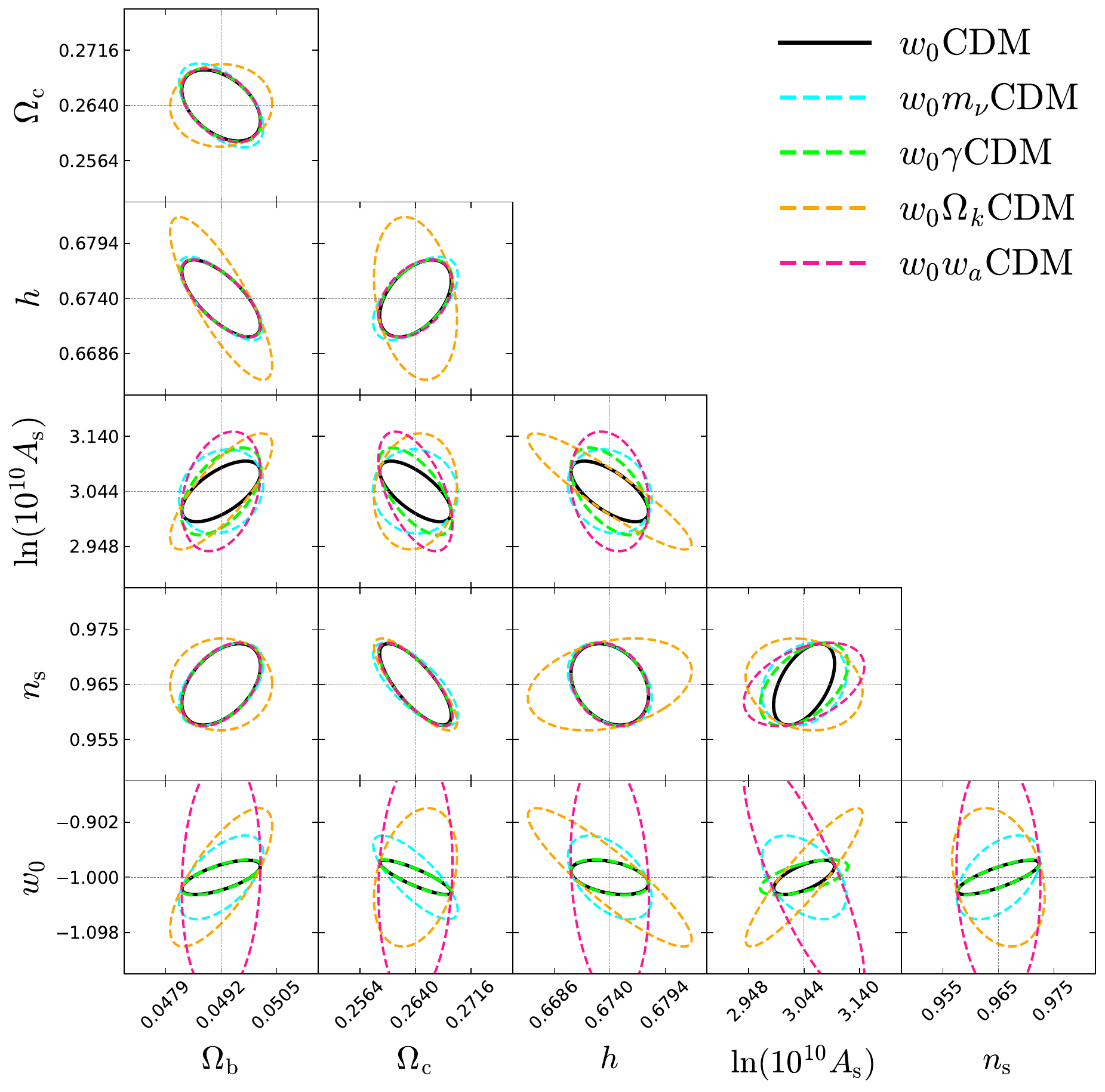}
\caption{2D-confidence ellipse contours for the base parameters and \wO{} in one-parameter extensions of \basemodel{} model. $1\sigma$ -confidence contours for clustering-only constraints from the PFS-like survey including CMB prior.}\label{fig:ellipse_w0}
\end{figure}
In addition to the strength of linear degeneracy, the direction of degeneracy can significantly vary model-dependently. In Fig.~\ref{fig:ellipse_w0}, we display the confidence ellipse contours for the six parameters, $\boldsymbol{\theta}=(\boldsymbol{\theta}_{\rm base},w_{0})$, in \basemodel{}-variant models.
It can be seen that the direction and degree of parameter degeneracy do not remain identical when a new extra parameter is taken into consideration.
In extreme cases, the correlation direction changes from positive to negative, or vice versa. This is clearly shown in the contours for \As-\wO{} parameter pair. In \basemodel{} models with either gravity or curvature parameter, \wO{} and \As{} are positively correlated as in the \basemodel{} model, which can be physically understood since increasing the amplitude of the matter power spectrum accompanies a larger \wO{} parameter so that the structure growth remains intact. However, we observe anti-correlation between \wO{} and \As{} in \baseplusmodel{} and \basemodel{}+\mnu{} models.
This demonstrates that the degeneracies between parameters can non-negligibly change depending on the cosmological models.

\subsection{Impact of IA vs. Fiducial setup}

\begin{figure*}
\includegraphics[width=2\columnwidth]{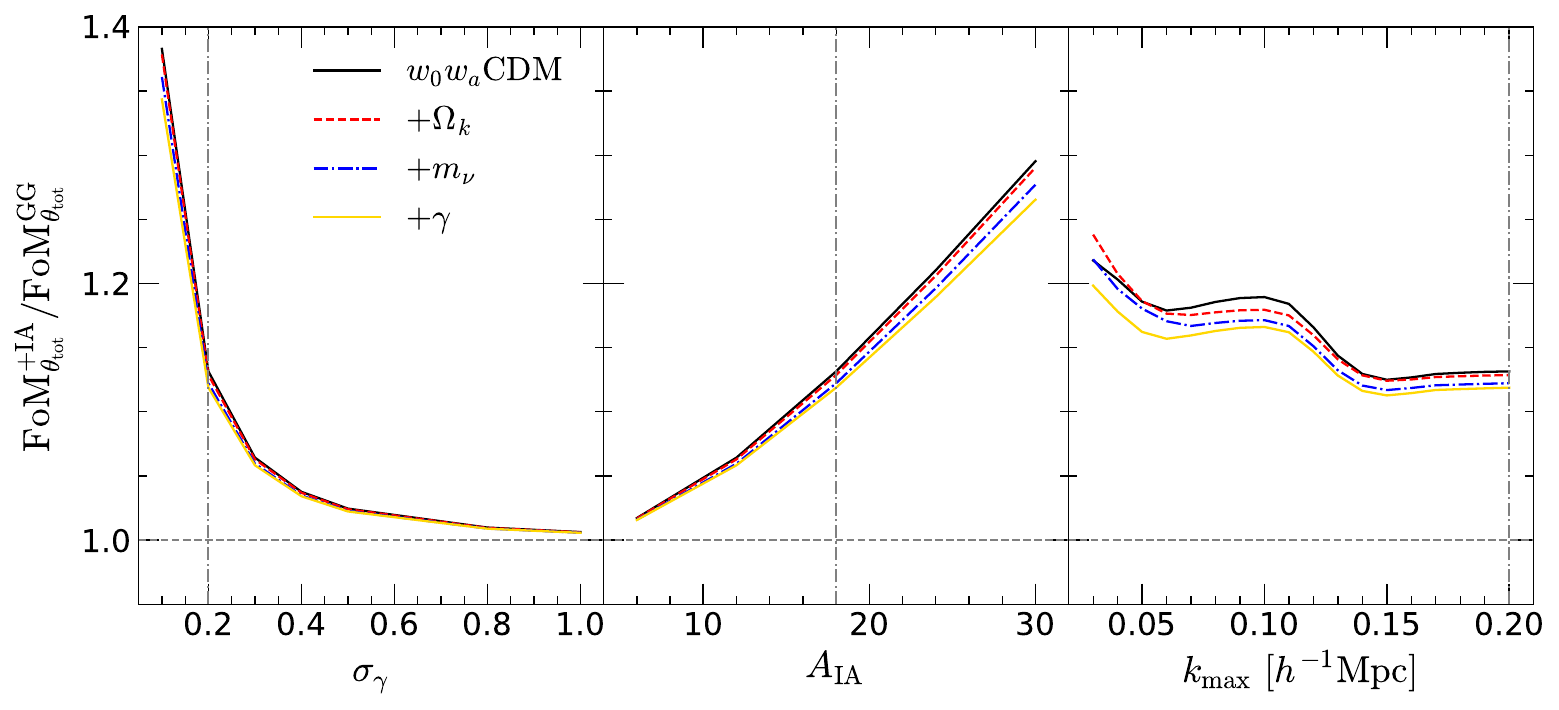}
\caption{\FoM{} gain by combining with IA relative to the clustering-only analysis as a function of three different survey parameters. The results for the PFS-like survey are shown. Cases for the shape-noise, $\sigma_{\gamma}$, IA amplitude, $A_{\rm IA}$, and the maximum wavenumber, $k_{\rm max}$, are shown in panels from left to right. Vertical lines mark the fiducial values adopted in the main analysis. The other two parameters are kept identical to the fiducial values. }\label{fig:survey_par_dep}
\end{figure*}

Finally, we examine how the relative impact of galaxy IA depends on the fiducial setup of our analysis, i.e., the shape-noise, IA amplitude, and the maximum wavenumber of the full-shape analysis. In Fig.~\ref{fig:survey_par_dep}, we show the ratios of \FoM{} for the entire parameters in one-parameter extensions of the \baseplusmodel{} model for the PFS-like survey. We observe that the benefit of IA strongly depends on the shape-noise and IA amplitude in all models considered. The gain increases when the shape-noise decreases or the IA amplitude increases. In particular, the gain can be drastically improved if one can reduce $\sigma_{\gamma}$ to below $0.2$. These tendencies are consistent with the results found in the compressed analyses \cite{TO20, OT22}.
With increasing $k_{\rm max}$, the impact of IA tends to decrease, showing oscillatory features. Such oscillatory features were also detected in the \FoM{} ratios between the full-2D power spectrum and the sum of its lower multipoles \cite{taruya+11}.
This indicates that the relative contribution of IA compared to the galaxy clustering is not constant and is increasing more rapidly than the clustering toward smaller $k_{\rm max}$.
Finally, we note that the impact of IA remains consistently significant even in a conservative scenario, where both 
$A_{\rm IA}$ and  $k_{\rm max}$ are simultaneously lowered. Appendix \ref{app:setup} provides a further discussion on this setup, including how the 1D-marginalized constraints and the improvements with IA are affected under these conditions.

\section{Conclusions \label{sec:conclusion}}

With the advent of upcoming galaxy redshift surveys aided with high-quality imaging and recent methods for efficiently detecting IA signals from high-redshift galaxies \cite{shi+21a,lamman+24}, the cosmological significance of IA is set to be escalated.
In light of this, we performed a Fisher forecast on cosmological parameters by utilizing galaxy clustering and IA information within the full-shape framework.
This is the first study to explore the cosmological impact of IA using the full-shape approach.
Focusing on the Figure-of-Merit and marginalized parameter constraints as metrics, we assessed the cosmological gain due to the IA, relative to clustering-only constraints.
We explored parameter constraints for various cosmological models, with the most extended one simultaneously incorporating dynamical dark energy, massive neutrinos, curvature, and modified gravity on top of the standard $\Lambda$CDM model. 
Our forecast assumed two different surveys; a PFS-like deep survey and an Euclid-like wide survey to study the impact of the survey design and setup.

The key findings of this paper can be summarized as follows:
\begin{enumerate}[(i)]
    \item IA significantly tightens constraints on dynamical dark energy models, particularly for \As{}, \wO, and \wa{} parameters, yielding $13-19\%$, $15-30\%$, and $26-29\%$ improvements in PFS-like surveys (or $8-11\%$, $5-17\%$, and $10-17\%$ in Euclid-like surveys) (see Figs.~\ref{fig:FoMimp} and \ref{fig:TabImp}).
    \item IA substantially improves all cosmological parameter constraints for nonflat-MG models by $10-14\%$ ($6-10\%$) on average, except for \Omc{} and \ns{}, in PFS-like (Euclid-like) surveys (see Fig.~\ref{fig:TabImp}).
    \item Joint full-shape constraints achieve subpercent-level precision for curvature, percent (subpercent)-level for the Hubble constant, ten percent-level for dark energy and gravity parameters, and the same order precision for massive neutrinos in the most extended \baseplusmodel{}+(\Omk, \mnu, \gam{}) model assuming PFS-like (Euclid-like) surveys (see Fig.~\ref{fig:TabVal}).
    
    \item Improvement with IA in full-shape constraints on dark energy and curvature parameters is as significant as that in geometric/dynamical constraints. Joint full-shape constraints on the Hubble constant and dark energy parameters are about $10$ and $3$ times tighter than the joint geometric/dynamical constraints. 
    
    \item Including an extra cosmological parameter degrades all parameter constraints for a given model, with more severe weakening for parameters that exhibit stronger degeneracy with the extra parameter (see Fig.~\ref{fig:ellipse_w0wa}).
    \item Parameter degeneracy for a given parameter pair can significantly vary in different cosmological models, particularly in \baseplusmodel{} and $w_{0}\Omega_{k}{\rm CDM}$ (see Fig.~\ref{fig:ellipse_w0}).
    \item Cosmological gain with IA increases for a larger IA amplitude or smaller shape noise and maximum wavenumber (see Fig.~\ref{fig:survey_par_dep}).
\end{enumerate}

We have demonstrated that combining galaxy IA with clustering is indeed beneficial in the full-shape analysis, as in compressed analyses focusing on geometric/dynamical measurements \cite{TO20, OT22, OT23}.
While the current analysis is based on the linear theory description of the clustering and alignment statistics, it would be interesting to investigate whether we expect further gain with nonlinear modeling, given the recent development of the beyond-linear descriptions of the galaxy shape statistics \cite{blazek+19,vlah+20,bakx+23,okumura+24,taruya+24}. 
Since scale-dependent imprints of massive neutrinos and modified gravity models on the matter power spectrum would be more evident on those nonlinear scales, their constraints may be more impacted than other cosmological parameter constraints. Another relevant avenue to be pursued would be examining whether the benefit of IA can be further enhanced with the higher-order shape statistics \cite{pyne+22,linke+24}. As non-gaussianity develops in the matter distribution due to the nonlinear gravitational evolution, it is better captured with higher-order statistics, and hence, we may expect some extra gain with IA.

This forecast assumes scale-independent MG models, using the simple, scale-independent \gam{} parameter to describe the growth of structure and the matter power spectrum. The \gam{} parameter only modifies the overall amplitude of the matter power spectrum, without affecting its scale dependence. As a result,
\gam{} is degenerate with other parameters that also influence the amplitude, such as \As{}, $\Omega_{\rm m}$, and \ns{} \cite{moretti+23, aviles24}. As noted in Ref.~\cite{aviles24}, MG models with scale-dependent features may benefit more from full-shape analysis, as their scale dependence helps disentangle the effects of amplitude-modulating parameters from those of modified gravity, such as the $f_{R0}$ parameter in $f(R)$ gravity \cite{hu&sawicki07}. Therefore, models with scale-dependent MG features are expected to yield tighter parameter constraints and more significant improvements. We leave such investigations as future work.

\begin{acknowledgments}
We thank an anonymous referee for useful comments that helped improve the original manuscript.
JS acknowledges the support by
Academia Sinica Institute of Astronomy and Astrophysics. TO acknowledges the support of the Taiwan National Science and Technology Council under Grants No. NSTC 112-2112-M-001-034- and NSTC 113-2112-M-001-011-, and the Academia Sinica Investigator Project Grant (AS-IV-114-M03) for the period of 2025-2029. 
This work was supported by MEXT/JSPS KAKENHI Grant Numbers JP20H05861 and JP21H01081 (AT).
\end{acknowledgments}

\appendix
\section{Forecasts with conservative setup}\label{app:setup}
We comment on how IA's ability to constrain cosmological parameters changes in a more conservative setup with different $A_{\rm IA}$ and $k_{\rm max}$. Specifically, we assume a much weaker IA amplitude, setting $A_{\rm IA}=10$. This accounts for the fact that the IA amplitude, determined using the shape estimator developed for blue galaxies \citep{shi+21a}, has not been measured in observations and may be lower than the fiducial value used in our main analysis.
We also study the case with the maximum wavenumber set to a lower value, $k_{\rm max}=0.1 h{\rm Mpc}^{-1}$, restricting the analysis to a more conservative linear regime. Finally, we consider the combined case with $A_{\rm IA}=10$ and $k_{\rm max}=0.1h{\rm Mpc}^{-1}$ as the conservative setup.

\begin{figure*}
\includegraphics[width=2\columnwidth]{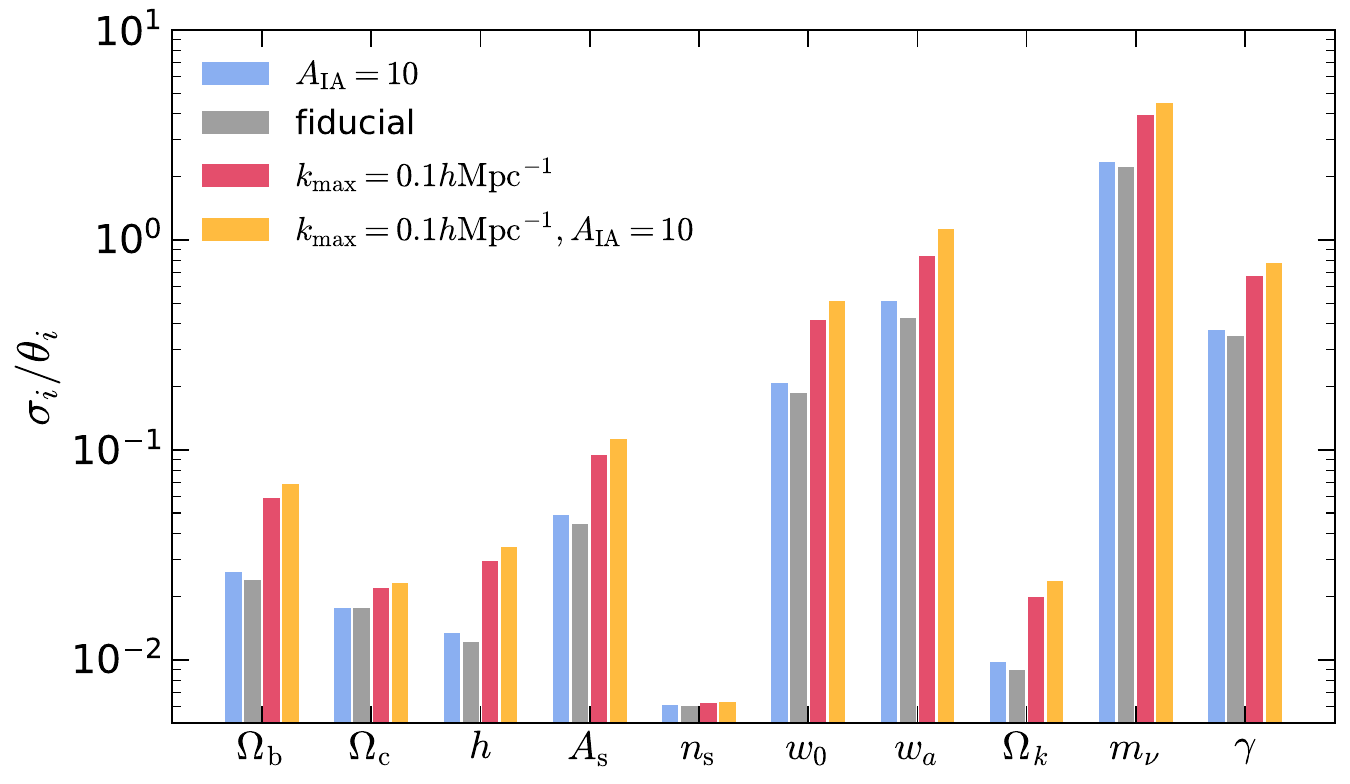}
\includegraphics[width=2\columnwidth]{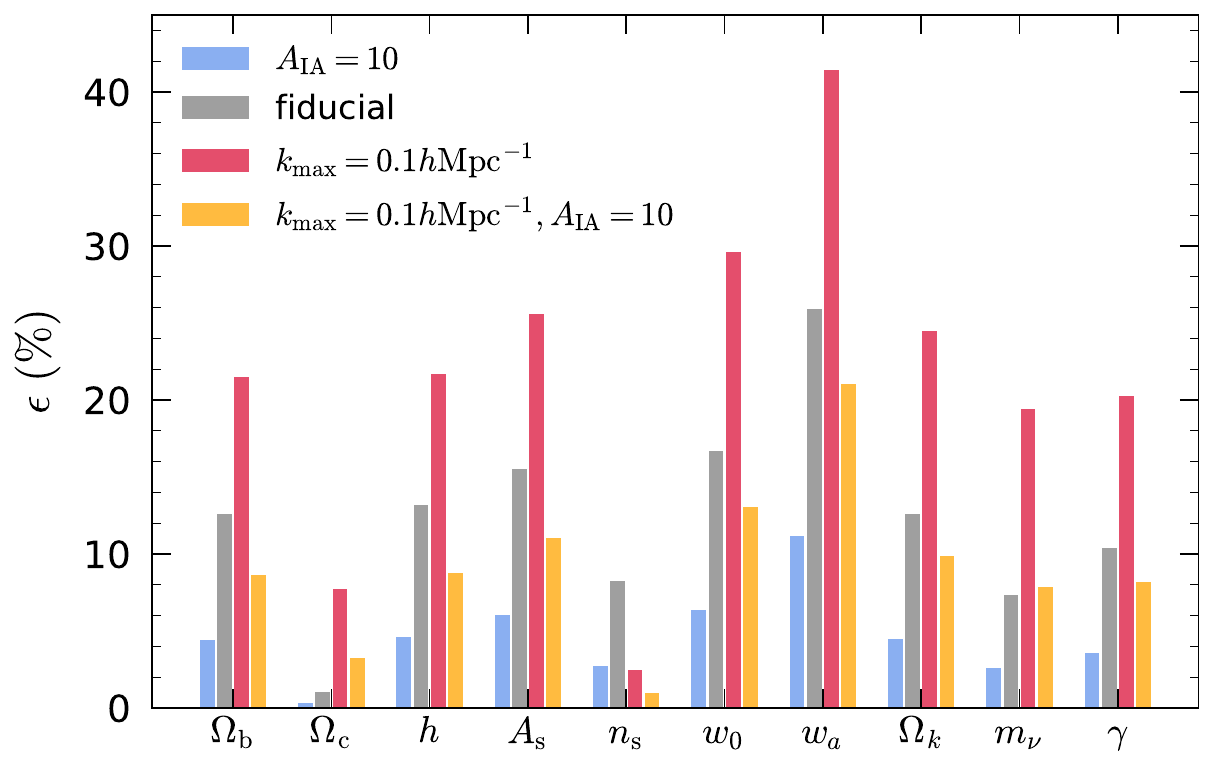}
\caption{Joint 1D-marginalized parameter constraints (upper) and improvements with IA relative to clustering-only constraints (lower) and fractional errors (lower) with conservative choices for $A_{\rm IA}$ and $k_{\rm max}$. The most extended model, \baseplusmodel{}+(\mnu, \Omk, \gam), is considered assuming PFS-like surveys. Note again that for \wa{} and \Omk, we show their actual errors, $\sigma_{i}$, since their fiducial values, $\theta_{i}$, are zero. The improvement for \ns{} has been multiplied by 10 for better visibility. We note that consistent trends are found for an Euclid-like survey but with tighter marginalized constraints and weaker improvements than the PFS-like case.}\label{fig:app1}
\end{figure*}

For illustration, we show in Fig.~\ref{fig:app1} the joint 1D-marginalized parameter constraints and improvements relative to clustering-only constraints in the most extended model under these analysis setups in a PFS-like survey. For a comparison, we include the fiducial case with $A_{\rm IA}=18$ and $k_{\rm max}=0.2h{\rm Mpc}^{-1}$. As expected, in the upper panel, limiting $k_{\rm max}$ and $A_{\rm IA}$ to lower values worsens the joint constraints for all parameters, on average by $94\%$ and $9\%$ compared to the fiducial case, respectively. Particularly, for the dark energy parameters, their fractional errors increase by $110\%$ and $16\%$ in the cases with $k_{\rm max}=0.1h{\rm Mpc}^{-1}$ and $A_{\rm IA}=10$, respectively. Lowering both $A_{\rm IA}$ and $k_{\rm max}$ expectedly yields the weakest constraints.

On the other hand, the change in the improvement follows a different trend when decreasing $A_{\rm IA}$ and $k_{\rm max}$. As shown in the lower panel of Fig.~\ref{fig:app1}, for all parameters except \ns, the improvement becomes weaker than the fiducial case when lowering $A_{\rm IA}$, whereas it becomes more significant for a lower $k_{\rm max}$. More specifically, when focusing on the dark energy parameters, their average improvement with $A_{\rm IA}=10$, relative to the clustering-only case, is approximately $9\%$. This is lower than the $21\%$ improvement observed in the fiducial case for the same model. On the other hand, the average improvement for dark energy parameters in the analysis with $k_{\rm max}=0.1h{\rm Mpc}^{-1}$ increases to $35\%$. Simultaneously adopting lower $A_{\rm IA}$ and $k_{\rm max}$ yields $21\%$ improvement.
These results suggest that lowering $A_{\rm IA}$ to 10 while keeping  $k_{\rm max}=0.2h{\rm Mpc}^{-1}$ reduces the impact of IA since the clustering-only constraints already contain substantial information. However, even with this pessimistic assumption on $A_{\rm IA}$, the improvement remains comparable to the fiducial case when the maximum wavenumber is restricted to $k_{\rm max}=0.1h{\rm Mpc}^{-1}$. All of these trends are also consistently found in the Euclid-like survey. This demonstrates that the relative cosmological impact of full-shape IA compared to full-shape clustering remains robust even under conservative assumptions, such as lower $A_{\rm IA}$ and $k_{\rm max}$. We finally note that while we adopt a common $k_{\rm max}$ in this work, separate $k_{\rm max}$ values tailored to the valid ranges of the clustering and IA modelings could be used in practice.

\bibliography{main.bib}

\begin{thebibliography}{137}
\expandafter\ifx\csname natexlab\endcsname\relax\def\natexlab#1{#1}\fi
\expandafter\ifx\csname bibnamefont\endcsname\relax
  \def\bibnamefont#1{#1}\fi
\expandafter\ifx\csname bibfnamefont\endcsname\relax
  \def\bibfnamefont#1{#1}\fi
\expandafter\ifx\csname citenamefont\endcsname\relax
  \def\citenamefont#1{#1}\fi
\expandafter\ifx\csname url\endcsname\relax
  \def\url#1{\texttt{#1}}\fi
\expandafter\ifx\csname urlprefix\endcsname\relax\def\urlprefix{URL }\fi
\providecommand{\bibinfo}[2]{#2}
\providecommand{\eprint}[2][]{\url{#2}}

\bibitem[{\citenamefont{{Peebles} and {Yu}}(1970)}]{peebles&yu70}
\bibinfo{author}{\bibfnamefont{P.~J.~E.} \bibnamefont{{Peebles}}} \bibnamefont{and} \bibinfo{author}{\bibfnamefont{J.~T.} \bibnamefont{{Yu}}}, \bibinfo{journal}{\apj} \textbf{\bibinfo{volume}{162}}, \bibinfo{pages}{815} (\bibinfo{year}{1970}).

\bibitem[{\citenamefont{{Eisenstein} and {Hu}}(1998)}]{eisenstein&hu98}
\bibinfo{author}{\bibfnamefont{D.~J.} \bibnamefont{{Eisenstein}}} \bibnamefont{and} \bibinfo{author}{\bibfnamefont{W.}~\bibnamefont{{Hu}}}, \bibinfo{journal}{\apj} \textbf{\bibinfo{volume}{496}}, \bibinfo{pages}{605} (\bibinfo{year}{1998}), \eprint{astro-ph/9709112}.

\bibitem[{\citenamefont{{Cole} et~al.}(2005)\citenamefont{{Cole}, {Percival}, {Peacock}, {Norberg}, {Baugh}, {Frenk}, {Baldry}, {Bland-Hawthorn}, {Bridges}, {Cannon} et~al.}}]{cole+05}
\bibinfo{author}{\bibfnamefont{S.}~\bibnamefont{{Cole}}}, \bibinfo{author}{\bibfnamefont{W.~J.} \bibnamefont{{Percival}}}, \bibinfo{author}{\bibfnamefont{J.~A.} \bibnamefont{{Peacock}}}, \bibinfo{author}{\bibfnamefont{P.}~\bibnamefont{{Norberg}}}, \bibinfo{author}{\bibfnamefont{C.~M.} \bibnamefont{{Baugh}}}, \bibinfo{author}{\bibfnamefont{C.~S.} \bibnamefont{{Frenk}}}, \bibinfo{author}{\bibfnamefont{I.}~\bibnamefont{{Baldry}}}, \bibinfo{author}{\bibfnamefont{J.}~\bibnamefont{{Bland-Hawthorn}}}, \bibinfo{author}{\bibfnamefont{T.}~\bibnamefont{{Bridges}}}, \bibinfo{author}{\bibfnamefont{R.}~\bibnamefont{{Cannon}}}, \bibnamefont{et~al.}, \bibinfo{journal}{\mnras} \textbf{\bibinfo{volume}{362}}, \bibinfo{pages}{505} (\bibinfo{year}{2005}), \eprint{astro-ph/0501174}.

\bibitem[{\citenamefont{{Eisenstein} et~al.}(2005)\citenamefont{{Eisenstein}, {Zehavi}, {Hogg}, {Scoccimarro}, {Blanton}, {Nichol}, {Scranton}, {Seo}, {Tegmark}, {Zheng} et~al.}}]{eisenstein+05}
\bibinfo{author}{\bibfnamefont{D.~J.} \bibnamefont{{Eisenstein}}}, \bibinfo{author}{\bibfnamefont{I.}~\bibnamefont{{Zehavi}}}, \bibinfo{author}{\bibfnamefont{D.~W.} \bibnamefont{{Hogg}}}, \bibinfo{author}{\bibfnamefont{R.}~\bibnamefont{{Scoccimarro}}}, \bibinfo{author}{\bibfnamefont{M.~R.} \bibnamefont{{Blanton}}}, \bibinfo{author}{\bibfnamefont{R.~C.} \bibnamefont{{Nichol}}}, \bibinfo{author}{\bibfnamefont{R.}~\bibnamefont{{Scranton}}}, \bibinfo{author}{\bibfnamefont{H.-J.} \bibnamefont{{Seo}}}, \bibinfo{author}{\bibfnamefont{M.}~\bibnamefont{{Tegmark}}}, \bibinfo{author}{\bibfnamefont{Z.}~\bibnamefont{{Zheng}}}, \bibnamefont{et~al.}, \bibinfo{journal}{\apj} \textbf{\bibinfo{volume}{633}}, \bibinfo{pages}{560} (\bibinfo{year}{2005}), \eprint{astro-ph/0501171}.

\bibitem[{\citenamefont{{Jackson}}(1972)}]{jackson72}
\bibinfo{author}{\bibfnamefont{J.~C.} \bibnamefont{{Jackson}}}, \bibinfo{journal}{\mnras} \textbf{\bibinfo{volume}{156}}, \bibinfo{pages}{1P} (\bibinfo{year}{1972}), \eprint{0810.3908}.

\bibitem[{\citenamefont{{Sargent} and {Turner}}(1977)}]{sargent&turner77}
\bibinfo{author}{\bibfnamefont{W.~L.~W.} \bibnamefont{{Sargent}}} \bibnamefont{and} \bibinfo{author}{\bibfnamefont{E.~L.} \bibnamefont{{Turner}}}, \bibinfo{journal}{\apjl} \textbf{\bibinfo{volume}{212}}, \bibinfo{pages}{L3} (\bibinfo{year}{1977}).

\bibitem[{\citenamefont{{Kaiser}}(1987)}]{kaiser87}
\bibinfo{author}{\bibfnamefont{N.}~\bibnamefont{{Kaiser}}}, \bibinfo{journal}{\mnras} \textbf{\bibinfo{volume}{227}}, \bibinfo{pages}{1} (\bibinfo{year}{1987}).

\bibitem[{\citenamefont{{Hamilton}}(1992)}]{hamilton92}
\bibinfo{author}{\bibfnamefont{A.~J.~S.} \bibnamefont{{Hamilton}}}, \bibinfo{journal}{\apjl} \textbf{\bibinfo{volume}{385}}, \bibinfo{pages}{L5} (\bibinfo{year}{1992}).

\bibitem[{\citenamefont{{Peacock} et~al.}(2001)\citenamefont{{Peacock}, {Cole}, {Norberg}, {Baugh}, {Bland-Hawthorn}, {Bridges}, {Cannon}, {Colless}, {Collins}, {Couch} et~al.}}]{peacock01}
\bibinfo{author}{\bibfnamefont{J.~A.} \bibnamefont{{Peacock}}}, \bibinfo{author}{\bibfnamefont{S.}~\bibnamefont{{Cole}}}, \bibinfo{author}{\bibfnamefont{P.}~\bibnamefont{{Norberg}}}, \bibinfo{author}{\bibfnamefont{C.~M.} \bibnamefont{{Baugh}}}, \bibinfo{author}{\bibfnamefont{J.}~\bibnamefont{{Bland-Hawthorn}}}, \bibinfo{author}{\bibfnamefont{T.}~\bibnamefont{{Bridges}}}, \bibinfo{author}{\bibfnamefont{R.~D.} \bibnamefont{{Cannon}}}, \bibinfo{author}{\bibfnamefont{M.}~\bibnamefont{{Colless}}}, \bibinfo{author}{\bibfnamefont{C.}~\bibnamefont{{Collins}}}, \bibinfo{author}{\bibfnamefont{W.}~\bibnamefont{{Couch}}}, \bibnamefont{et~al.}, \bibinfo{journal}{\nat} \textbf{\bibinfo{volume}{410}}, \bibinfo{pages}{169} (\bibinfo{year}{2001}), \eprint{astro-ph/0103143}.

\bibitem[{\citenamefont{{Seo} and {Eisenstein}}(2003)}]{seo&eisenstein03}
\bibinfo{author}{\bibfnamefont{H.-J.} \bibnamefont{{Seo}}} \bibnamefont{and} \bibinfo{author}{\bibfnamefont{D.~J.} \bibnamefont{{Eisenstein}}}, \bibinfo{journal}{\apj} \textbf{\bibinfo{volume}{598}}, \bibinfo{pages}{720} (\bibinfo{year}{2003}), \eprint{astro-ph/0307460}.

\bibitem[{\citenamefont{{Tegmark} et~al.}(2004)\citenamefont{{Tegmark}, {Strauss}, {Blanton}, {Abazajian}, {Dodelson}, {Sandvik}, {Wang}, {Weinberg}, {Zehavi}, {Bahcall} et~al.}}]{tegmark+04}
\bibinfo{author}{\bibfnamefont{M.}~\bibnamefont{{Tegmark}}}, \bibinfo{author}{\bibfnamefont{M.~A.} \bibnamefont{{Strauss}}}, \bibinfo{author}{\bibfnamefont{M.~R.} \bibnamefont{{Blanton}}}, \bibinfo{author}{\bibfnamefont{K.}~\bibnamefont{{Abazajian}}}, \bibinfo{author}{\bibfnamefont{S.}~\bibnamefont{{Dodelson}}}, \bibinfo{author}{\bibfnamefont{H.}~\bibnamefont{{Sandvik}}}, \bibinfo{author}{\bibfnamefont{X.}~\bibnamefont{{Wang}}}, \bibinfo{author}{\bibfnamefont{D.~H.} \bibnamefont{{Weinberg}}}, \bibinfo{author}{\bibfnamefont{I.}~\bibnamefont{{Zehavi}}}, \bibinfo{author}{\bibfnamefont{N.~A.} \bibnamefont{{Bahcall}}}, \bibnamefont{et~al.}, \bibinfo{journal}{\prd} \textbf{\bibinfo{volume}{69}}, \bibinfo{eid}{103501} (\bibinfo{year}{2004}), \eprint{astro-ph/0310723}.

\bibitem[{\citenamefont{{Okumura} et~al.}(2008)\citenamefont{{Okumura}, {Matsubara}, {Eisenstein}, {Kayo}, {Hikage}, {Szalay}, and {Schneider}}}]{okumura+08}
\bibinfo{author}{\bibfnamefont{T.}~\bibnamefont{{Okumura}}}, \bibinfo{author}{\bibfnamefont{T.}~\bibnamefont{{Matsubara}}}, \bibinfo{author}{\bibfnamefont{D.~J.} \bibnamefont{{Eisenstein}}}, \bibinfo{author}{\bibfnamefont{I.}~\bibnamefont{{Kayo}}}, \bibinfo{author}{\bibfnamefont{C.}~\bibnamefont{{Hikage}}}, \bibinfo{author}{\bibfnamefont{A.~S.} \bibnamefont{{Szalay}}}, \bibnamefont{and} \bibinfo{author}{\bibfnamefont{D.~P.} \bibnamefont{{Schneider}}}, \bibinfo{journal}{\apj} \textbf{\bibinfo{volume}{676}}, \bibinfo{pages}{889} (\bibinfo{year}{2008}), \eprint{0711.3640}.

\bibitem[{\citenamefont{{Guzzo} et~al.}(2008)\citenamefont{{Guzzo}, {Pierleoni}, {Meneux}, {Branchini}, {Le F{\`e}vre}, {Marinoni}, {Garilli}, {Blaizot}, {De Lucia}, {Pollo} et~al.}}]{guzzo+08}
\bibinfo{author}{\bibfnamefont{L.}~\bibnamefont{{Guzzo}}}, \bibinfo{author}{\bibfnamefont{M.}~\bibnamefont{{Pierleoni}}}, \bibinfo{author}{\bibfnamefont{B.}~\bibnamefont{{Meneux}}}, \bibinfo{author}{\bibfnamefont{E.}~\bibnamefont{{Branchini}}}, \bibinfo{author}{\bibfnamefont{O.}~\bibnamefont{{Le F{\`e}vre}}}, \bibinfo{author}{\bibfnamefont{C.}~\bibnamefont{{Marinoni}}}, \bibinfo{author}{\bibfnamefont{B.}~\bibnamefont{{Garilli}}}, \bibinfo{author}{\bibfnamefont{J.}~\bibnamefont{{Blaizot}}}, \bibinfo{author}{\bibfnamefont{G.}~\bibnamefont{{De Lucia}}}, \bibinfo{author}{\bibfnamefont{A.}~\bibnamefont{{Pollo}}}, \bibnamefont{et~al.}, \bibinfo{journal}{\nat} \textbf{\bibinfo{volume}{451}}, \bibinfo{pages}{541} (\bibinfo{year}{2008}), \eprint{0802.1944}.

\bibitem[{\citenamefont{{Beutler} et~al.}(2012)\citenamefont{{Beutler}, {Blake}, {Colless}, {Jones}, {Staveley-Smith}, {Poole}, {Campbell}, {Parker}, {Saunders}, and {Watson}}}]{beutler+12}
\bibinfo{author}{\bibfnamefont{F.}~\bibnamefont{{Beutler}}}, \bibinfo{author}{\bibfnamefont{C.}~\bibnamefont{{Blake}}}, \bibinfo{author}{\bibfnamefont{M.}~\bibnamefont{{Colless}}}, \bibinfo{author}{\bibfnamefont{D.~H.} \bibnamefont{{Jones}}}, \bibinfo{author}{\bibfnamefont{L.}~\bibnamefont{{Staveley-Smith}}}, \bibinfo{author}{\bibfnamefont{G.~B.} \bibnamefont{{Poole}}}, \bibinfo{author}{\bibfnamefont{L.}~\bibnamefont{{Campbell}}}, \bibinfo{author}{\bibfnamefont{Q.}~\bibnamefont{{Parker}}}, \bibinfo{author}{\bibfnamefont{W.}~\bibnamefont{{Saunders}}}, \bibnamefont{and} \bibinfo{author}{\bibfnamefont{F.}~\bibnamefont{{Watson}}}, \bibinfo{journal}{\mnras} \textbf{\bibinfo{volume}{423}}, \bibinfo{pages}{3430} (\bibinfo{year}{2012}), \eprint{1204.4725}.

\bibitem[{\citenamefont{{Blake} et~al.}(2011)\citenamefont{{Blake}, {Brough}, {Colless}, {Contreras}, {Couch}, {Croom}, {Davis}, {Drinkwater}, {Forster}, {Gilbank} et~al.}}]{blake+12}
\bibinfo{author}{\bibfnamefont{C.}~\bibnamefont{{Blake}}}, \bibinfo{author}{\bibfnamefont{S.}~\bibnamefont{{Brough}}}, \bibinfo{author}{\bibfnamefont{M.}~\bibnamefont{{Colless}}}, \bibinfo{author}{\bibfnamefont{C.}~\bibnamefont{{Contreras}}}, \bibinfo{author}{\bibfnamefont{W.}~\bibnamefont{{Couch}}}, \bibinfo{author}{\bibfnamefont{S.}~\bibnamefont{{Croom}}}, \bibinfo{author}{\bibfnamefont{T.}~\bibnamefont{{Davis}}}, \bibinfo{author}{\bibfnamefont{M.~J.} \bibnamefont{{Drinkwater}}}, \bibinfo{author}{\bibfnamefont{K.}~\bibnamefont{{Forster}}}, \bibinfo{author}{\bibfnamefont{D.}~\bibnamefont{{Gilbank}}}, \bibnamefont{et~al.}, \bibinfo{journal}{\mnras} \textbf{\bibinfo{volume}{415}}, \bibinfo{pages}{2876} (\bibinfo{year}{2011}), \eprint{1104.2948}.

\bibitem[{\citenamefont{{Reid} et~al.}(2012)\citenamefont{{Reid}, {Samushia}, {White}, {Percival}, {Manera}, {Padmanabhan}, {Ross}, {S{\'a}nchez}, {Bailey}, {Bizyaev} et~al.}}]{reid+12}
\bibinfo{author}{\bibfnamefont{B.~A.} \bibnamefont{{Reid}}}, \bibinfo{author}{\bibfnamefont{L.}~\bibnamefont{{Samushia}}}, \bibinfo{author}{\bibfnamefont{M.}~\bibnamefont{{White}}}, \bibinfo{author}{\bibfnamefont{W.~J.} \bibnamefont{{Percival}}}, \bibinfo{author}{\bibfnamefont{M.}~\bibnamefont{{Manera}}}, \bibinfo{author}{\bibfnamefont{N.}~\bibnamefont{{Padmanabhan}}}, \bibinfo{author}{\bibfnamefont{A.~J.} \bibnamefont{{Ross}}}, \bibinfo{author}{\bibfnamefont{A.~G.} \bibnamefont{{S{\'a}nchez}}}, \bibinfo{author}{\bibfnamefont{S.}~\bibnamefont{{Bailey}}}, \bibinfo{author}{\bibfnamefont{D.}~\bibnamefont{{Bizyaev}}}, \bibnamefont{et~al.}, \bibinfo{journal}{\mnras} \textbf{\bibinfo{volume}{426}}, \bibinfo{pages}{2719} (\bibinfo{year}{2012}), \eprint{1203.6641}.

\bibitem[{\citenamefont{{Samushia} et~al.}(2013)\citenamefont{{Samushia}, {Reid}, {White}, {Percival}, {Cuesta}, {Lombriser}, {Manera}, {Nichol}, {Schneider}, {Bizyaev} et~al.}}]{samushia+13}
\bibinfo{author}{\bibfnamefont{L.}~\bibnamefont{{Samushia}}}, \bibinfo{author}{\bibfnamefont{B.~A.} \bibnamefont{{Reid}}}, \bibinfo{author}{\bibfnamefont{M.}~\bibnamefont{{White}}}, \bibinfo{author}{\bibfnamefont{W.~J.} \bibnamefont{{Percival}}}, \bibinfo{author}{\bibfnamefont{A.~J.} \bibnamefont{{Cuesta}}}, \bibinfo{author}{\bibfnamefont{L.}~\bibnamefont{{Lombriser}}}, \bibinfo{author}{\bibfnamefont{M.}~\bibnamefont{{Manera}}}, \bibinfo{author}{\bibfnamefont{R.~C.} \bibnamefont{{Nichol}}}, \bibinfo{author}{\bibfnamefont{D.~P.} \bibnamefont{{Schneider}}}, \bibinfo{author}{\bibfnamefont{D.}~\bibnamefont{{Bizyaev}}}, \bibnamefont{et~al.}, \bibinfo{journal}{\mnras} \textbf{\bibinfo{volume}{429}}, \bibinfo{pages}{1514} (\bibinfo{year}{2013}), \eprint{1206.5309}.

\bibitem[{\citenamefont{{Beutler} et~al.}(2014)\citenamefont{{Beutler}, {Saito}, {Seo}, {Brinkmann}, {Dawson}, {Eisenstein}, {Font-Ribera}, {Ho}, {McBride}, {Montesano} et~al.}}]{beutler+14}
\bibinfo{author}{\bibfnamefont{F.}~\bibnamefont{{Beutler}}}, \bibinfo{author}{\bibfnamefont{S.}~\bibnamefont{{Saito}}}, \bibinfo{author}{\bibfnamefont{H.-J.} \bibnamefont{{Seo}}}, \bibinfo{author}{\bibfnamefont{J.}~\bibnamefont{{Brinkmann}}}, \bibinfo{author}{\bibfnamefont{K.~S.} \bibnamefont{{Dawson}}}, \bibinfo{author}{\bibfnamefont{D.~J.} \bibnamefont{{Eisenstein}}}, \bibinfo{author}{\bibfnamefont{A.}~\bibnamefont{{Font-Ribera}}}, \bibinfo{author}{\bibfnamefont{S.}~\bibnamefont{{Ho}}}, \bibinfo{author}{\bibfnamefont{C.~K.} \bibnamefont{{McBride}}}, \bibinfo{author}{\bibfnamefont{F.}~\bibnamefont{{Montesano}}}, \bibnamefont{et~al.}, \bibinfo{journal}{\mnras} \textbf{\bibinfo{volume}{443}}, \bibinfo{pages}{1065} (\bibinfo{year}{2014}), \eprint{1312.4611}.

\bibitem[{\citenamefont{{Aubourg} et~al.}(2015)\citenamefont{{Aubourg}, {Bailey}, {Bautista}, {Beutler}, {Bhardwaj}, {Bizyaev}, {Blanton}, {Blomqvist}, {Bolton}, {Bovy} et~al.}}]{aubourg+15}
\bibinfo{author}{\bibfnamefont{{\'E}.}~\bibnamefont{{Aubourg}}}, \bibinfo{author}{\bibfnamefont{S.}~\bibnamefont{{Bailey}}}, \bibinfo{author}{\bibfnamefont{J.~E.} \bibnamefont{{Bautista}}}, \bibinfo{author}{\bibfnamefont{F.}~\bibnamefont{{Beutler}}}, \bibinfo{author}{\bibfnamefont{V.}~\bibnamefont{{Bhardwaj}}}, \bibinfo{author}{\bibfnamefont{D.}~\bibnamefont{{Bizyaev}}}, \bibinfo{author}{\bibfnamefont{M.}~\bibnamefont{{Blanton}}}, \bibinfo{author}{\bibfnamefont{M.}~\bibnamefont{{Blomqvist}}}, \bibinfo{author}{\bibfnamefont{A.~S.} \bibnamefont{{Bolton}}}, \bibinfo{author}{\bibfnamefont{J.}~\bibnamefont{{Bovy}}}, \bibnamefont{et~al.}, \bibinfo{journal}{\prd} \textbf{\bibinfo{volume}{92}}, \bibinfo{eid}{123516} (\bibinfo{year}{2015}), \eprint{1411.1074}.

\bibitem[{\citenamefont{{Okumura} et~al.}(2016)\citenamefont{{Okumura}, {Hikage}, {Totani}, {Tonegawa}, {Okada}, {Glazebrook}, {Blake}, {Ferreira}, {More}, {Taruya} et~al.}}]{okumura+16}
\bibinfo{author}{\bibfnamefont{T.}~\bibnamefont{{Okumura}}}, \bibinfo{author}{\bibfnamefont{C.}~\bibnamefont{{Hikage}}}, \bibinfo{author}{\bibfnamefont{T.}~\bibnamefont{{Totani}}}, \bibinfo{author}{\bibfnamefont{M.}~\bibnamefont{{Tonegawa}}}, \bibinfo{author}{\bibfnamefont{H.}~\bibnamefont{{Okada}}}, \bibinfo{author}{\bibfnamefont{K.}~\bibnamefont{{Glazebrook}}}, \bibinfo{author}{\bibfnamefont{C.}~\bibnamefont{{Blake}}}, \bibinfo{author}{\bibfnamefont{P.~G.} \bibnamefont{{Ferreira}}}, \bibinfo{author}{\bibfnamefont{S.}~\bibnamefont{{More}}}, \bibinfo{author}{\bibfnamefont{A.}~\bibnamefont{{Taruya}}}, \bibnamefont{et~al.}, \bibinfo{journal}{\pasj} \textbf{\bibinfo{volume}{68}}, \bibinfo{eid}{38} (\bibinfo{year}{2016}), \eprint{1511.08083}.

\bibitem[{\citenamefont{{Alam} et~al.}(2017)\citenamefont{{Alam}, {Ata}, {Bailey}, {Beutler}, {Bizyaev}, {Blazek}, {Bolton}, {Brownstein}, {Burden}, {Chuang} et~al.}}]{alam+17}
\bibinfo{author}{\bibfnamefont{S.}~\bibnamefont{{Alam}}}, \bibinfo{author}{\bibfnamefont{M.}~\bibnamefont{{Ata}}}, \bibinfo{author}{\bibfnamefont{S.}~\bibnamefont{{Bailey}}}, \bibinfo{author}{\bibfnamefont{F.}~\bibnamefont{{Beutler}}}, \bibinfo{author}{\bibfnamefont{D.}~\bibnamefont{{Bizyaev}}}, \bibinfo{author}{\bibfnamefont{J.~A.} \bibnamefont{{Blazek}}}, \bibinfo{author}{\bibfnamefont{A.~S.} \bibnamefont{{Bolton}}}, \bibinfo{author}{\bibfnamefont{J.~R.} \bibnamefont{{Brownstein}}}, \bibinfo{author}{\bibfnamefont{A.}~\bibnamefont{{Burden}}}, \bibinfo{author}{\bibfnamefont{C.-H.} \bibnamefont{{Chuang}}}, \bibnamefont{et~al.}, \bibinfo{journal}{\mnras} \textbf{\bibinfo{volume}{470}}, \bibinfo{pages}{2617} (\bibinfo{year}{2017}), \eprint{1607.03155}.

\bibitem[{\citenamefont{{Beutler} et~al.}(2017)\citenamefont{{Beutler}, {Seo}, {Ross}, {McDonald}, {Saito}, {Bolton}, {Brownstein}, {Chuang}, {Cuesta}, {Eisenstein} et~al.}}]{beutler+17}
\bibinfo{author}{\bibfnamefont{F.}~\bibnamefont{{Beutler}}}, \bibinfo{author}{\bibfnamefont{H.-J.} \bibnamefont{{Seo}}}, \bibinfo{author}{\bibfnamefont{A.~J.} \bibnamefont{{Ross}}}, \bibinfo{author}{\bibfnamefont{P.}~\bibnamefont{{McDonald}}}, \bibinfo{author}{\bibfnamefont{S.}~\bibnamefont{{Saito}}}, \bibinfo{author}{\bibfnamefont{A.~S.} \bibnamefont{{Bolton}}}, \bibinfo{author}{\bibfnamefont{J.~R.} \bibnamefont{{Brownstein}}}, \bibinfo{author}{\bibfnamefont{C.-H.} \bibnamefont{{Chuang}}}, \bibinfo{author}{\bibfnamefont{A.~J.} \bibnamefont{{Cuesta}}}, \bibinfo{author}{\bibfnamefont{D.~J.} \bibnamefont{{Eisenstein}}}, \bibnamefont{et~al.}, \bibinfo{journal}{\mnras} \textbf{\bibinfo{volume}{464}}, \bibinfo{pages}{3409} (\bibinfo{year}{2017}), \eprint{1607.03149}.

\bibitem[{\citenamefont{{Gil-Mar{\'\i}n} et~al.}(2017)\citenamefont{{Gil-Mar{\'\i}n}, {Percival}, {Verde}, {Brownstein}, {Chuang}, {Kitaura}, {Rodr{\'\i}guez-Torres}, and {Olmstead}}}]{gilmarin+17}
\bibinfo{author}{\bibfnamefont{H.}~\bibnamefont{{Gil-Mar{\'\i}n}}}, \bibinfo{author}{\bibfnamefont{W.~J.} \bibnamefont{{Percival}}}, \bibinfo{author}{\bibfnamefont{L.}~\bibnamefont{{Verde}}}, \bibinfo{author}{\bibfnamefont{J.~R.} \bibnamefont{{Brownstein}}}, \bibinfo{author}{\bibfnamefont{C.-H.} \bibnamefont{{Chuang}}}, \bibinfo{author}{\bibfnamefont{F.-S.} \bibnamefont{{Kitaura}}}, \bibinfo{author}{\bibfnamefont{S.~A.} \bibnamefont{{Rodr{\'\i}guez-Torres}}}, \bibnamefont{and} \bibinfo{author}{\bibfnamefont{M.~D.} \bibnamefont{{Olmstead}}}, \bibinfo{journal}{\mnras} \textbf{\bibinfo{volume}{465}}, \bibinfo{pages}{1757} (\bibinfo{year}{2017}), \eprint{1606.00439}.

\bibitem[{\citenamefont{{Hawken} et~al.}(2017)\citenamefont{{Hawken}, {Granett}, {Iovino}, {Guzzo}, {Peacock}, {de la Torre}, {Garilli}, {Bolzonella}, {Scodeggio}, {Abbas} et~al.}}]{hawken+17}
\bibinfo{author}{\bibfnamefont{A.~J.} \bibnamefont{{Hawken}}}, \bibinfo{author}{\bibfnamefont{B.~R.} \bibnamefont{{Granett}}}, \bibinfo{author}{\bibfnamefont{A.}~\bibnamefont{{Iovino}}}, \bibinfo{author}{\bibfnamefont{L.}~\bibnamefont{{Guzzo}}}, \bibinfo{author}{\bibfnamefont{J.~A.} \bibnamefont{{Peacock}}}, \bibinfo{author}{\bibfnamefont{S.}~\bibnamefont{{de la Torre}}}, \bibinfo{author}{\bibfnamefont{B.}~\bibnamefont{{Garilli}}}, \bibinfo{author}{\bibfnamefont{M.}~\bibnamefont{{Bolzonella}}}, \bibinfo{author}{\bibfnamefont{M.}~\bibnamefont{{Scodeggio}}}, \bibinfo{author}{\bibfnamefont{U.}~\bibnamefont{{Abbas}}}, \bibnamefont{et~al.}, \bibinfo{journal}{\aap} \textbf{\bibinfo{volume}{607}}, \bibinfo{eid}{A54} (\bibinfo{year}{2017}), \eprint{1611.07046}.

\bibitem[{\citenamefont{{Hou} et~al.}(2021)\citenamefont{{Hou}, {S{\'a}nchez}, {Ross}, {Smith}, {Neveux}, {Bautista}, {Burtin}, {Zhao}, {Scoccimarro}, {Dawson} et~al.}}]{hou+21}
\bibinfo{author}{\bibfnamefont{J.}~\bibnamefont{{Hou}}}, \bibinfo{author}{\bibfnamefont{A.~G.} \bibnamefont{{S{\'a}nchez}}}, \bibinfo{author}{\bibfnamefont{A.~J.} \bibnamefont{{Ross}}}, \bibinfo{author}{\bibfnamefont{A.}~\bibnamefont{{Smith}}}, \bibinfo{author}{\bibfnamefont{R.}~\bibnamefont{{Neveux}}}, \bibinfo{author}{\bibfnamefont{J.}~\bibnamefont{{Bautista}}}, \bibinfo{author}{\bibfnamefont{E.}~\bibnamefont{{Burtin}}}, \bibinfo{author}{\bibfnamefont{C.}~\bibnamefont{{Zhao}}}, \bibinfo{author}{\bibfnamefont{R.}~\bibnamefont{{Scoccimarro}}}, \bibinfo{author}{\bibfnamefont{K.~S.} \bibnamefont{{Dawson}}}, \bibnamefont{et~al.}, \bibinfo{journal}{\mnras} \textbf{\bibinfo{volume}{500}}, \bibinfo{pages}{1201} (\bibinfo{year}{2021}), \eprint{2007.08998}.

\bibitem[{\citenamefont{{Aubert} et~al.}(2022)\citenamefont{{Aubert}, {Cousinou}, {Escoffier}, {Hawken}, {Nadathur}, {Alam}, {Bautista}, {Burtin}, {Chuang}, {de la Macorra} et~al.}}]{aubert+22}
\bibinfo{author}{\bibfnamefont{M.}~\bibnamefont{{Aubert}}}, \bibinfo{author}{\bibfnamefont{M.-C.} \bibnamefont{{Cousinou}}}, \bibinfo{author}{\bibfnamefont{S.}~\bibnamefont{{Escoffier}}}, \bibinfo{author}{\bibfnamefont{A.~J.} \bibnamefont{{Hawken}}}, \bibinfo{author}{\bibfnamefont{S.}~\bibnamefont{{Nadathur}}}, \bibinfo{author}{\bibfnamefont{S.}~\bibnamefont{{Alam}}}, \bibinfo{author}{\bibfnamefont{J.}~\bibnamefont{{Bautista}}}, \bibinfo{author}{\bibfnamefont{E.}~\bibnamefont{{Burtin}}}, \bibinfo{author}{\bibfnamefont{C.-H.} \bibnamefont{{Chuang}}}, \bibinfo{author}{\bibfnamefont{A.}~\bibnamefont{{de la Macorra}}}, \bibnamefont{et~al.}, \bibinfo{journal}{\mnras} \textbf{\bibinfo{volume}{513}}, \bibinfo{pages}{186} (\bibinfo{year}{2022}), \eprint{2007.09013}.

\bibitem[{\citenamefont{{Anderson} et~al.}(2014)\citenamefont{{Anderson}, {Aubourg}, {Bailey}, {Beutler}, {Bhardwaj}, {Blanton}, {Bolton}, {Brinkmann}, {Brownstein}, {Burden} et~al.}}]{anderson+14}
\bibinfo{author}{\bibfnamefont{L.}~\bibnamefont{{Anderson}}}, \bibinfo{author}{\bibfnamefont{{\'E}.}~\bibnamefont{{Aubourg}}}, \bibinfo{author}{\bibfnamefont{S.}~\bibnamefont{{Bailey}}}, \bibinfo{author}{\bibfnamefont{F.}~\bibnamefont{{Beutler}}}, \bibinfo{author}{\bibfnamefont{V.}~\bibnamefont{{Bhardwaj}}}, \bibinfo{author}{\bibfnamefont{M.}~\bibnamefont{{Blanton}}}, \bibinfo{author}{\bibfnamefont{A.~S.} \bibnamefont{{Bolton}}}, \bibinfo{author}{\bibfnamefont{J.}~\bibnamefont{{Brinkmann}}}, \bibinfo{author}{\bibfnamefont{J.~R.} \bibnamefont{{Brownstein}}}, \bibinfo{author}{\bibfnamefont{A.}~\bibnamefont{{Burden}}}, \bibnamefont{et~al.}, \bibinfo{journal}{\mnras} \textbf{\bibinfo{volume}{441}}, \bibinfo{pages}{24} (\bibinfo{year}{2014}), \eprint{1312.4877}.

\bibitem[{\citenamefont{{Nadathur} et~al.}(2019)\citenamefont{{Nadathur}, {Carter}, {Percival}, {Winther}, and {Bautista}}}]{nadathur+19}
\bibinfo{author}{\bibfnamefont{S.}~\bibnamefont{{Nadathur}}}, \bibinfo{author}{\bibfnamefont{P.~M.} \bibnamefont{{Carter}}}, \bibinfo{author}{\bibfnamefont{W.~J.} \bibnamefont{{Percival}}}, \bibinfo{author}{\bibfnamefont{H.~A.} \bibnamefont{{Winther}}}, \bibnamefont{and} \bibinfo{author}{\bibfnamefont{J.~E.} \bibnamefont{{Bautista}}}, \bibinfo{journal}{\prd} \textbf{\bibinfo{volume}{100}}, \bibinfo{eid}{023504} (\bibinfo{year}{2019}), \eprint{1904.01030}.

\bibitem[{\citenamefont{{Zhao} et~al.}(2022)\citenamefont{{Zhao}, {Variu}, {He}, {Forero-S{\'a}nchez}, {Tamone}, {Chuang}, {Kitaura}, {Tao}, {Yu}, {Kneib} et~al.}}]{zhao+22}
\bibinfo{author}{\bibfnamefont{C.}~\bibnamefont{{Zhao}}}, \bibinfo{author}{\bibfnamefont{A.}~\bibnamefont{{Variu}}}, \bibinfo{author}{\bibfnamefont{M.}~\bibnamefont{{He}}}, \bibinfo{author}{\bibfnamefont{D.}~\bibnamefont{{Forero-S{\'a}nchez}}}, \bibinfo{author}{\bibfnamefont{A.}~\bibnamefont{{Tamone}}}, \bibinfo{author}{\bibfnamefont{C.-H.} \bibnamefont{{Chuang}}}, \bibinfo{author}{\bibfnamefont{F.-S.} \bibnamefont{{Kitaura}}}, \bibinfo{author}{\bibfnamefont{C.}~\bibnamefont{{Tao}}}, \bibinfo{author}{\bibfnamefont{J.}~\bibnamefont{{Yu}}}, \bibinfo{author}{\bibfnamefont{J.-P.} \bibnamefont{{Kneib}}}, \bibnamefont{et~al.}, \bibinfo{journal}{\mnras} \textbf{\bibinfo{volume}{511}}, \bibinfo{pages}{5492} (\bibinfo{year}{2022}), \eprint{2110.03824}.

\bibitem[{\citenamefont{{DESI Collaboration} et~al.}(2024{\natexlab{a}})\citenamefont{{DESI Collaboration}, {Adame}, {Aguilar}, {Ahlen}, {Alam}, {Alexander}, {Alvarez}, {Alves}, {Anand}, {Andrade} et~al.}}]{desi_BAO+24}
\bibinfo{author}{\bibnamefont{{DESI Collaboration}}}, \bibinfo{author}{\bibfnamefont{A.~G.} \bibnamefont{{Adame}}}, \bibinfo{author}{\bibfnamefont{J.}~\bibnamefont{{Aguilar}}}, \bibinfo{author}{\bibfnamefont{S.}~\bibnamefont{{Ahlen}}}, \bibinfo{author}{\bibfnamefont{S.}~\bibnamefont{{Alam}}}, \bibinfo{author}{\bibfnamefont{D.~M.} \bibnamefont{{Alexander}}}, \bibinfo{author}{\bibfnamefont{M.}~\bibnamefont{{Alvarez}}}, \bibinfo{author}{\bibfnamefont{O.}~\bibnamefont{{Alves}}}, \bibinfo{author}{\bibfnamefont{A.}~\bibnamefont{{Anand}}}, \bibinfo{author}{\bibfnamefont{U.}~\bibnamefont{{Andrade}}}, \bibnamefont{et~al.}, \bibinfo{journal}{arXiv e-prints} \bibinfo{eid}{arXiv:2404.03002} (\bibinfo{year}{2024}{\natexlab{a}}), \eprint{2404.03002}.

\bibitem[{\citenamefont{{S{\'a}nchez} et~al.}(2009)\citenamefont{{S{\'a}nchez}, {Crocce}, {Cabr{\'e}}, {Baugh}, and {Gazta{\~n}aga}}}]{sanchez+09}
\bibinfo{author}{\bibfnamefont{A.~G.} \bibnamefont{{S{\'a}nchez}}}, \bibinfo{author}{\bibfnamefont{M.}~\bibnamefont{{Crocce}}}, \bibinfo{author}{\bibfnamefont{A.}~\bibnamefont{{Cabr{\'e}}}}, \bibinfo{author}{\bibfnamefont{C.~M.} \bibnamefont{{Baugh}}}, \bibnamefont{and} \bibinfo{author}{\bibfnamefont{E.}~\bibnamefont{{Gazta{\~n}aga}}}, \bibinfo{journal}{\mnras} \textbf{\bibinfo{volume}{400}}, \bibinfo{pages}{1643} (\bibinfo{year}{2009}), \eprint{0901.2570}.

\bibitem[{\citenamefont{{Montesano} et~al.}(2010)\citenamefont{{Montesano}, {S{\'a}nchez}, and {Phleps}}}]{montesano+10}
\bibinfo{author}{\bibfnamefont{F.}~\bibnamefont{{Montesano}}}, \bibinfo{author}{\bibfnamefont{A.~G.} \bibnamefont{{S{\'a}nchez}}}, \bibnamefont{and} \bibinfo{author}{\bibfnamefont{S.}~\bibnamefont{{Phleps}}}, \bibinfo{journal}{\mnras} \textbf{\bibinfo{volume}{408}}, \bibinfo{pages}{2397} (\bibinfo{year}{2010}), \eprint{1007.0755}.

\bibitem[{\citenamefont{{Montesano} et~al.}(2012)\citenamefont{{Montesano}, {S{\'a}nchez}, and {Phleps}}}]{montesano+12}
\bibinfo{author}{\bibfnamefont{F.}~\bibnamefont{{Montesano}}}, \bibinfo{author}{\bibfnamefont{A.~G.} \bibnamefont{{S{\'a}nchez}}}, \bibnamefont{and} \bibinfo{author}{\bibfnamefont{S.}~\bibnamefont{{Phleps}}}, \bibinfo{journal}{\mnras} \textbf{\bibinfo{volume}{421}}, \bibinfo{pages}{2656} (\bibinfo{year}{2012}), \eprint{1107.4097}.

\bibitem[{\citenamefont{{S{\'a}nchez} et~al.}(2013)\citenamefont{{S{\'a}nchez}, {Kazin}, {Beutler}, {Chuang}, {Cuesta}, {Eisenstein}, {Manera}, {Montesano}, {Nichol}, {Padmanabhan} et~al.}}]{sanchez+13}
\bibinfo{author}{\bibfnamefont{A.~G.} \bibnamefont{{S{\'a}nchez}}}, \bibinfo{author}{\bibfnamefont{E.~A.} \bibnamefont{{Kazin}}}, \bibinfo{author}{\bibfnamefont{F.}~\bibnamefont{{Beutler}}}, \bibinfo{author}{\bibfnamefont{C.-H.} \bibnamefont{{Chuang}}}, \bibinfo{author}{\bibfnamefont{A.~J.} \bibnamefont{{Cuesta}}}, \bibinfo{author}{\bibfnamefont{D.~J.} \bibnamefont{{Eisenstein}}}, \bibinfo{author}{\bibfnamefont{M.}~\bibnamefont{{Manera}}}, \bibinfo{author}{\bibfnamefont{F.}~\bibnamefont{{Montesano}}}, \bibinfo{author}{\bibfnamefont{R.~C.} \bibnamefont{{Nichol}}}, \bibinfo{author}{\bibfnamefont{N.}~\bibnamefont{{Padmanabhan}}}, \bibnamefont{et~al.}, \bibinfo{journal}{\mnras} \textbf{\bibinfo{volume}{433}}, \bibinfo{pages}{1202} (\bibinfo{year}{2013}), \eprint{1303.4396}.

\bibitem[{\citenamefont{{Ivanov} et~al.}(2020)\citenamefont{{Ivanov}, {Simonovi{\'c}}, and {Zaldarriaga}}}]{ivanov+20}
\bibinfo{author}{\bibfnamefont{M.~M.} \bibnamefont{{Ivanov}}}, \bibinfo{author}{\bibfnamefont{M.}~\bibnamefont{{Simonovi{\'c}}}}, \bibnamefont{and} \bibinfo{author}{\bibfnamefont{M.}~\bibnamefont{{Zaldarriaga}}}, \bibinfo{journal}{\jcap} \textbf{\bibinfo{volume}{2020}}, \bibinfo{eid}{042} (\bibinfo{year}{2020}), \eprint{1909.05277}.

\bibitem[{\citenamefont{{Nunes} et~al.}(2022)\citenamefont{{Nunes}, {Vagnozzi}, {Kumar}, {Di Valentino}, and {Mena}}}]{nunes+22}
\bibinfo{author}{\bibfnamefont{R.~C.} \bibnamefont{{Nunes}}}, \bibinfo{author}{\bibfnamefont{S.}~\bibnamefont{{Vagnozzi}}}, \bibinfo{author}{\bibfnamefont{S.}~\bibnamefont{{Kumar}}}, \bibinfo{author}{\bibfnamefont{E.}~\bibnamefont{{Di Valentino}}}, \bibnamefont{and} \bibinfo{author}{\bibfnamefont{O.}~\bibnamefont{{Mena}}}, \bibinfo{journal}{\prd} \textbf{\bibinfo{volume}{105}}, \bibinfo{eid}{123506} (\bibinfo{year}{2022}), \eprint{2203.08093}.

\bibitem[{\citenamefont{{Philcox} and {Ivanov}}(2022)}]{philcox&ivanov22}
\bibinfo{author}{\bibfnamefont{O.~H.~E.} \bibnamefont{{Philcox}}} \bibnamefont{and} \bibinfo{author}{\bibfnamefont{M.~M.} \bibnamefont{{Ivanov}}}, \bibinfo{journal}{\prd} \textbf{\bibinfo{volume}{105}}, \bibinfo{eid}{043517} (\bibinfo{year}{2022}), \eprint{2112.04515}.

\bibitem[{\citenamefont{{Simon} et~al.}(2023)\citenamefont{{Simon}, {Zhang}, and {Poulin}}}]{simon+23}
\bibinfo{author}{\bibfnamefont{T.}~\bibnamefont{{Simon}}}, \bibinfo{author}{\bibfnamefont{P.}~\bibnamefont{{Zhang}}}, \bibnamefont{and} \bibinfo{author}{\bibfnamefont{V.}~\bibnamefont{{Poulin}}}, \bibinfo{journal}{\jcap} \textbf{\bibinfo{volume}{2023}}, \bibinfo{eid}{041} (\bibinfo{year}{2023}), \eprint{2210.14931}.

\bibitem[{\citenamefont{{Gsponer} et~al.}(2024)\citenamefont{{Gsponer}, {Zhao}, {Donald-McCann}, {Bacon}, {Koyama}, {Crittenden}, {Simon}, and {Mueller}}}]{Gsponer+24}
\bibinfo{author}{\bibfnamefont{R.}~\bibnamefont{{Gsponer}}}, \bibinfo{author}{\bibfnamefont{R.}~\bibnamefont{{Zhao}}}, \bibinfo{author}{\bibfnamefont{J.}~\bibnamefont{{Donald-McCann}}}, \bibinfo{author}{\bibfnamefont{D.}~\bibnamefont{{Bacon}}}, \bibinfo{author}{\bibfnamefont{K.}~\bibnamefont{{Koyama}}}, \bibinfo{author}{\bibfnamefont{R.}~\bibnamefont{{Crittenden}}}, \bibinfo{author}{\bibfnamefont{T.}~\bibnamefont{{Simon}}}, \bibnamefont{and} \bibinfo{author}{\bibfnamefont{E.-M.} \bibnamefont{{Mueller}}}, \bibinfo{journal}{\mnras} \textbf{\bibinfo{volume}{530}}, \bibinfo{pages}{3075} (\bibinfo{year}{2024}), \eprint{2312.01977}.

\bibitem[{\citenamefont{{Ramirez} et~al.}(2024)\citenamefont{{Ramirez}, {Icaza-Lizaola}, {Fromenteau}, {Vargas-Maga{\~n}a}, and {Aviles}}}]{ramirez+24}
\bibinfo{author}{\bibfnamefont{S.}~\bibnamefont{{Ramirez}}}, \bibinfo{author}{\bibfnamefont{M.}~\bibnamefont{{Icaza-Lizaola}}}, \bibinfo{author}{\bibfnamefont{S.}~\bibnamefont{{Fromenteau}}}, \bibinfo{author}{\bibfnamefont{M.}~\bibnamefont{{Vargas-Maga{\~n}a}}}, \bibnamefont{and} \bibinfo{author}{\bibfnamefont{A.}~\bibnamefont{{Aviles}}}, \bibinfo{journal}{\jcap} \textbf{\bibinfo{volume}{2024}}, \bibinfo{eid}{049} (\bibinfo{year}{2024}), \eprint{2310.17834}.

\bibitem[{\citenamefont{{DESI Collaboration} et~al.}(2024{\natexlab{b}})\citenamefont{{DESI Collaboration}, {Adame}, {Aguilar}, {Ahlen}, {Alam}, {Alexander}, {Allende Prieto}, {Alvarez}, {Alves}, {Anand} et~al.}}]{desi_FS+24}
\bibinfo{author}{\bibnamefont{{DESI Collaboration}}}, \bibinfo{author}{\bibfnamefont{A.~G.} \bibnamefont{{Adame}}}, \bibinfo{author}{\bibfnamefont{J.}~\bibnamefont{{Aguilar}}}, \bibinfo{author}{\bibfnamefont{S.}~\bibnamefont{{Ahlen}}}, \bibinfo{author}{\bibfnamefont{S.}~\bibnamefont{{Alam}}}, \bibinfo{author}{\bibfnamefont{D.~M.} \bibnamefont{{Alexander}}}, \bibinfo{author}{\bibfnamefont{C.}~\bibnamefont{{Allende Prieto}}}, \bibinfo{author}{\bibfnamefont{M.}~\bibnamefont{{Alvarez}}}, \bibinfo{author}{\bibfnamefont{O.}~\bibnamefont{{Alves}}}, \bibinfo{author}{\bibfnamefont{A.}~\bibnamefont{{Anand}}}, \bibnamefont{et~al.}, \bibinfo{journal}{arXiv e-prints} \bibinfo{eid}{arXiv:2411.12022} (\bibinfo{year}{2024}{\natexlab{b}}), \eprint{2411.12022}.

\bibitem[{\citenamefont{{Philcox} et~al.}(2020)\citenamefont{{Philcox}, {Ivanov}, {Simonovi{\'c}}, and {Zaldarriaga}}}]{philcox+20}
\bibinfo{author}{\bibfnamefont{O.~H.~E.} \bibnamefont{{Philcox}}}, \bibinfo{author}{\bibfnamefont{M.~M.} \bibnamefont{{Ivanov}}}, \bibinfo{author}{\bibfnamefont{M.}~\bibnamefont{{Simonovi{\'c}}}}, \bibnamefont{and} \bibinfo{author}{\bibfnamefont{M.}~\bibnamefont{{Zaldarriaga}}}, \bibinfo{journal}{\jcap} \textbf{\bibinfo{volume}{2020}}, \bibinfo{eid}{032} (\bibinfo{year}{2020}), \eprint{2002.04035}.

\bibitem[{\citenamefont{{Brieden} et~al.}(2021{\natexlab{a}})\citenamefont{{Brieden}, {Gil-Mar{\'\i}n}, and {Verde}}}]{brieden+21a}
\bibinfo{author}{\bibfnamefont{S.}~\bibnamefont{{Brieden}}}, \bibinfo{author}{\bibfnamefont{H.}~\bibnamefont{{Gil-Mar{\'\i}n}}}, \bibnamefont{and} \bibinfo{author}{\bibfnamefont{L.}~\bibnamefont{{Verde}}}, \bibinfo{journal}{\jcap} \textbf{\bibinfo{volume}{2021}}, \bibinfo{eid}{054} (\bibinfo{year}{2021}{\natexlab{a}}), \eprint{2106.07641}.

\bibitem[{\citenamefont{{Brieden} et~al.}(2021{\natexlab{b}})\citenamefont{{Brieden}, {Gil-Mar{\'\i}n}, and {Verde}}}]{brieden+21b}
\bibinfo{author}{\bibfnamefont{S.}~\bibnamefont{{Brieden}}}, \bibinfo{author}{\bibfnamefont{H.}~\bibnamefont{{Gil-Mar{\'\i}n}}}, \bibnamefont{and} \bibinfo{author}{\bibfnamefont{L.}~\bibnamefont{{Verde}}}, \bibinfo{journal}{\prd} \textbf{\bibinfo{volume}{104}}, \bibinfo{eid}{L121301} (\bibinfo{year}{2021}{\natexlab{b}}), \eprint{2106.11931}.

\bibitem[{\citenamefont{{Ishak} et~al.}(2024)\citenamefont{{Ishak}, {Pan}, {Calderon}, {Lodha}, {Valogiannis}, {Aviles}, {Niz}, {Yi}, {Zheng}, {Garcia-Quintero} et~al.}}]{DESI_FS_MG+24}
\bibinfo{author}{\bibfnamefont{M.}~\bibnamefont{{Ishak}}}, \bibinfo{author}{\bibfnamefont{J.}~\bibnamefont{{Pan}}}, \bibinfo{author}{\bibfnamefont{R.}~\bibnamefont{{Calderon}}}, \bibinfo{author}{\bibfnamefont{K.}~\bibnamefont{{Lodha}}}, \bibinfo{author}{\bibfnamefont{G.}~\bibnamefont{{Valogiannis}}}, \bibinfo{author}{\bibfnamefont{A.}~\bibnamefont{{Aviles}}}, \bibinfo{author}{\bibfnamefont{G.}~\bibnamefont{{Niz}}}, \bibinfo{author}{\bibfnamefont{L.}~\bibnamefont{{Yi}}}, \bibinfo{author}{\bibfnamefont{C.}~\bibnamefont{{Zheng}}}, \bibinfo{author}{\bibfnamefont{C.}~\bibnamefont{{Garcia-Quintero}}}, \bibnamefont{et~al.}, \bibinfo{journal}{arXiv e-prints} \bibinfo{eid}{arXiv:2411.12026} (\bibinfo{year}{2024}), \eprint{2411.12026}.

\bibitem[{\citenamefont{{Boyle} and {Komatsu}}(2018)}]{boyle&komatsu+18}
\bibinfo{author}{\bibfnamefont{A.}~\bibnamefont{{Boyle}}} \bibnamefont{and} \bibinfo{author}{\bibfnamefont{E.}~\bibnamefont{{Komatsu}}}, \bibinfo{journal}{\jcap} \textbf{\bibinfo{volume}{2018}}, \bibinfo{eid}{035} (\bibinfo{year}{2018}), \eprint{1712.01857}.

\bibitem[{\citenamefont{{Kumar} et~al.}(2022)\citenamefont{{Kumar}, {Nunes}, and {Yadav}}}]{kumar+22}
\bibinfo{author}{\bibfnamefont{S.}~\bibnamefont{{Kumar}}}, \bibinfo{author}{\bibfnamefont{R.~C.} \bibnamefont{{Nunes}}}, \bibnamefont{and} \bibinfo{author}{\bibfnamefont{P.}~\bibnamefont{{Yadav}}}, \bibinfo{journal}{\jcap} \textbf{\bibinfo{volume}{2022}}, \bibinfo{eid}{060} (\bibinfo{year}{2022}), \eprint{2205.04292}.

\bibitem[{\citenamefont{{Moretti} et~al.}(2023)\citenamefont{{Moretti}, {Tsedrik}, {Carrilho}, and {Pourtsidou}}}]{moretti+23}
\bibinfo{author}{\bibfnamefont{C.}~\bibnamefont{{Moretti}}}, \bibinfo{author}{\bibfnamefont{M.}~\bibnamefont{{Tsedrik}}}, \bibinfo{author}{\bibfnamefont{P.}~\bibnamefont{{Carrilho}}}, \bibnamefont{and} \bibinfo{author}{\bibfnamefont{A.}~\bibnamefont{{Pourtsidou}}}, \bibinfo{journal}{\jcap} \textbf{\bibinfo{volume}{2023}}, \bibinfo{eid}{025} (\bibinfo{year}{2023}), \eprint{2306.09275}.

\bibitem[{\citenamefont{{Rodriguez-Meza} et~al.}(2024)\citenamefont{{Rodriguez-Meza}, {Aviles}, {Noriega}, {Ruan}, {Li}, {Vargas-Maga{\~n}a}, and {Cervantes-Cota}}}]{rodriguez-meza+24}
\bibinfo{author}{\bibfnamefont{M.~A.} \bibnamefont{{Rodriguez-Meza}}}, \bibinfo{author}{\bibfnamefont{A.}~\bibnamefont{{Aviles}}}, \bibinfo{author}{\bibfnamefont{H.~E.} \bibnamefont{{Noriega}}}, \bibinfo{author}{\bibfnamefont{C.-Z.} \bibnamefont{{Ruan}}}, \bibinfo{author}{\bibfnamefont{B.}~\bibnamefont{{Li}}}, \bibinfo{author}{\bibfnamefont{M.}~\bibnamefont{{Vargas-Maga{\~n}a}}}, \bibnamefont{and} \bibinfo{author}{\bibfnamefont{J.~L.} \bibnamefont{{Cervantes-Cota}}}, \bibinfo{journal}{\jcap} \textbf{\bibinfo{volume}{2024}}, \bibinfo{eid}{049} (\bibinfo{year}{2024}), \eprint{2312.10510}.

\bibitem[{\citenamefont{{Carroll} et~al.}(2004)\citenamefont{{Carroll}, {Duvvuri}, {Trodden}, and {Turner}}}]{carroll+04}
\bibinfo{author}{\bibfnamefont{S.~M.} \bibnamefont{{Carroll}}}, \bibinfo{author}{\bibfnamefont{V.}~\bibnamefont{{Duvvuri}}}, \bibinfo{author}{\bibfnamefont{M.}~\bibnamefont{{Trodden}}}, \bibnamefont{and} \bibinfo{author}{\bibfnamefont{M.~S.} \bibnamefont{{Turner}}}, \bibinfo{journal}{\prd} \textbf{\bibinfo{volume}{70}}, \bibinfo{eid}{043528} (\bibinfo{year}{2004}), \eprint{astro-ph/0306438}.

\bibitem[{\citenamefont{{Lesgourgues} and {Pastor}}(2006)}]{lesgourgues&pastor06}
\bibinfo{author}{\bibfnamefont{J.}~\bibnamefont{{Lesgourgues}}} \bibnamefont{and} \bibinfo{author}{\bibfnamefont{S.}~\bibnamefont{{Pastor}}}, \bibinfo{journal}{\physrep} \textbf{\bibinfo{volume}{429}}, \bibinfo{pages}{307} (\bibinfo{year}{2006}), \eprint{astro-ph/0603494}.

\bibitem[{\citenamefont{{Hu} and {Sawicki}}(2007)}]{hu&sawicki07}
\bibinfo{author}{\bibfnamefont{W.}~\bibnamefont{{Hu}}} \bibnamefont{and} \bibinfo{author}{\bibfnamefont{I.}~\bibnamefont{{Sawicki}}}, \bibinfo{journal}{\prd} \textbf{\bibinfo{volume}{76}}, \bibinfo{eid}{064004} (\bibinfo{year}{2007}), \eprint{0705.1158}.

\bibitem[{\citenamefont{{Kaiser}}(1992)}]{kaiser92}
\bibinfo{author}{\bibfnamefont{N.}~\bibnamefont{{Kaiser}}}, \bibinfo{journal}{\apj} \textbf{\bibinfo{volume}{388}}, \bibinfo{pages}{272} (\bibinfo{year}{1992}).

\bibitem[{\citenamefont{{Bartelmann} and {Schneider}}(2001)}]{bartelmann&schneider01}
\bibinfo{author}{\bibfnamefont{M.}~\bibnamefont{{Bartelmann}}} \bibnamefont{and} \bibinfo{author}{\bibfnamefont{P.}~\bibnamefont{{Schneider}}}, \bibinfo{journal}{\physrep} \textbf{\bibinfo{volume}{340}}, \bibinfo{pages}{291} (\bibinfo{year}{2001}), \eprint{astro-ph/9912508}.

\bibitem[{\citenamefont{{Catelan} et~al.}(2001)\citenamefont{{Catelan}, {Kamionkowski}, and {Blandford}}}]{catelan+01}
\bibinfo{author}{\bibfnamefont{P.}~\bibnamefont{{Catelan}}}, \bibinfo{author}{\bibfnamefont{M.}~\bibnamefont{{Kamionkowski}}}, \bibnamefont{and} \bibinfo{author}{\bibfnamefont{R.~D.} \bibnamefont{{Blandford}}}, \bibinfo{journal}{\mnras} \textbf{\bibinfo{volume}{320}}, \bibinfo{pages}{L7} (\bibinfo{year}{2001}), \eprint{astro-ph/0005470}.

\bibitem[{\citenamefont{{Hirata} and {Seljak}}(2004)}]{hirata&seljak04}
\bibinfo{author}{\bibfnamefont{C.~M.} \bibnamefont{{Hirata}}} \bibnamefont{and} \bibinfo{author}{\bibfnamefont{U.}~\bibnamefont{{Seljak}}}, \bibinfo{journal}{\prd} \textbf{\bibinfo{volume}{70}}, \bibinfo{eid}{063526} (\bibinfo{year}{2004}), \eprint{astro-ph/0406275}.

\bibitem[{\citenamefont{{Brown} et~al.}(2002)\citenamefont{{Brown}, {Taylor}, {Hambly}, and {Dye}}}]{brown+02}
\bibinfo{author}{\bibfnamefont{M.~L.} \bibnamefont{{Brown}}}, \bibinfo{author}{\bibfnamefont{A.~N.} \bibnamefont{{Taylor}}}, \bibinfo{author}{\bibfnamefont{N.~C.} \bibnamefont{{Hambly}}}, \bibnamefont{and} \bibinfo{author}{\bibfnamefont{S.}~\bibnamefont{{Dye}}}, \bibinfo{journal}{\mnras} \textbf{\bibinfo{volume}{333}}, \bibinfo{pages}{501} (\bibinfo{year}{2002}), \eprint{astro-ph/0009499}.

\bibitem[{\citenamefont{{Mandelbaum} et~al.}(2006)\citenamefont{{Mandelbaum}, {Hirata}, {Ishak}, {Seljak}, and {Brinkmann}}}]{mandelbaum+06}
\bibinfo{author}{\bibfnamefont{R.}~\bibnamefont{{Mandelbaum}}}, \bibinfo{author}{\bibfnamefont{C.~M.} \bibnamefont{{Hirata}}}, \bibinfo{author}{\bibfnamefont{M.}~\bibnamefont{{Ishak}}}, \bibinfo{author}{\bibfnamefont{U.}~\bibnamefont{{Seljak}}}, \bibnamefont{and} \bibinfo{author}{\bibfnamefont{J.}~\bibnamefont{{Brinkmann}}}, \bibinfo{journal}{\mnras} \textbf{\bibinfo{volume}{367}}, \bibinfo{pages}{611} (\bibinfo{year}{2006}), \eprint{astro-ph/0509026}.

\bibitem[{\citenamefont{{Hirata} et~al.}(2007)\citenamefont{{Hirata}, {Mandelbaum}, {Ishak}, {Seljak}, {Nichol}, {Pimbblet}, {Ross}, and {Wake}}}]{hirata+07}
\bibinfo{author}{\bibfnamefont{C.~M.} \bibnamefont{{Hirata}}}, \bibinfo{author}{\bibfnamefont{R.}~\bibnamefont{{Mandelbaum}}}, \bibinfo{author}{\bibfnamefont{M.}~\bibnamefont{{Ishak}}}, \bibinfo{author}{\bibfnamefont{U.}~\bibnamefont{{Seljak}}}, \bibinfo{author}{\bibfnamefont{R.}~\bibnamefont{{Nichol}}}, \bibinfo{author}{\bibfnamefont{K.~A.} \bibnamefont{{Pimbblet}}}, \bibinfo{author}{\bibfnamefont{N.~P.} \bibnamefont{{Ross}}}, \bibnamefont{and} \bibinfo{author}{\bibfnamefont{D.}~\bibnamefont{{Wake}}}, \bibinfo{journal}{\mnras} \textbf{\bibinfo{volume}{381}}, \bibinfo{pages}{1197} (\bibinfo{year}{2007}), \eprint{astro-ph/0701671}.

\bibitem[{\citenamefont{{Okumura} et~al.}(2009)\citenamefont{{Okumura}, {Jing}, and {Li}}}]{okumura+09}
\bibinfo{author}{\bibfnamefont{T.}~\bibnamefont{{Okumura}}}, \bibinfo{author}{\bibfnamefont{Y.~P.} \bibnamefont{{Jing}}}, \bibnamefont{and} \bibinfo{author}{\bibfnamefont{C.}~\bibnamefont{{Li}}}, \bibinfo{journal}{\apj} \textbf{\bibinfo{volume}{694}}, \bibinfo{pages}{214} (\bibinfo{year}{2009}), \eprint{0809.3790}.

\bibitem[{\citenamefont{{Tonegawa} and {Okumura}}(2022)}]{tonegawa+22}
\bibinfo{author}{\bibfnamefont{M.}~\bibnamefont{{Tonegawa}}} \bibnamefont{and} \bibinfo{author}{\bibfnamefont{T.}~\bibnamefont{{Okumura}}}, \bibinfo{journal}{\apjl} \textbf{\bibinfo{volume}{924}}, \bibinfo{eid}{L3} (\bibinfo{year}{2022}), \eprint{2109.14297}.

\bibitem[{\citenamefont{{Tsaprazi} et~al.}(2022)\citenamefont{{Tsaprazi}, {Nguyen}, {Jasche}, {Schmidt}, and {Lavaux}}}]{tsaprazi+22}
\bibinfo{author}{\bibfnamefont{E.}~\bibnamefont{{Tsaprazi}}}, \bibinfo{author}{\bibfnamefont{N.-M.} \bibnamefont{{Nguyen}}}, \bibinfo{author}{\bibfnamefont{J.}~\bibnamefont{{Jasche}}}, \bibinfo{author}{\bibfnamefont{F.}~\bibnamefont{{Schmidt}}}, \bibnamefont{and} \bibinfo{author}{\bibfnamefont{G.}~\bibnamefont{{Lavaux}}}, \bibinfo{journal}{\jcap} \textbf{\bibinfo{volume}{2022}}, \bibinfo{eid}{003} (\bibinfo{year}{2022}), \eprint{2112.04484}.

\bibitem[{\citenamefont{{Okumura} and {Taruya}}(2023)}]{OT23}
\bibinfo{author}{\bibfnamefont{T.}~\bibnamefont{{Okumura}}} \bibnamefont{and} \bibinfo{author}{\bibfnamefont{A.}~\bibnamefont{{Taruya}}}, \bibinfo{journal}{\apjl} \textbf{\bibinfo{volume}{945}}, \bibinfo{eid}{L30} (\bibinfo{year}{2023}), \eprint{2301.06273}.

\bibitem[{\citenamefont{{Zhou} et~al.}(2023)\citenamefont{{Zhou}, {Tong}, {Troxel}, {Blazek}, {Lin}, {Bacon}, {Bleem}, {Chang}, {Costanzi}, {DeRose} et~al.}}]{zhou+23}
\bibinfo{author}{\bibfnamefont{C.}~\bibnamefont{{Zhou}}}, \bibinfo{author}{\bibfnamefont{A.}~\bibnamefont{{Tong}}}, \bibinfo{author}{\bibfnamefont{M.~A.} \bibnamefont{{Troxel}}}, \bibinfo{author}{\bibfnamefont{J.}~\bibnamefont{{Blazek}}}, \bibinfo{author}{\bibfnamefont{C.}~\bibnamefont{{Lin}}}, \bibinfo{author}{\bibfnamefont{D.}~\bibnamefont{{Bacon}}}, \bibinfo{author}{\bibfnamefont{L.}~\bibnamefont{{Bleem}}}, \bibinfo{author}{\bibfnamefont{C.}~\bibnamefont{{Chang}}}, \bibinfo{author}{\bibfnamefont{M.}~\bibnamefont{{Costanzi}}}, \bibinfo{author}{\bibfnamefont{J.}~\bibnamefont{{DeRose}}}, \bibnamefont{et~al.}, \bibinfo{journal}{\mnras} \textbf{\bibinfo{volume}{526}}, \bibinfo{pages}{323} (\bibinfo{year}{2023}), \eprint{2302.12325}.

\bibitem[{\citenamefont{{Schmidt} and {Jeong}}(2012)}]{schmidt&jeong12}
\bibinfo{author}{\bibfnamefont{F.}~\bibnamefont{{Schmidt}}} \bibnamefont{and} \bibinfo{author}{\bibfnamefont{D.}~\bibnamefont{{Jeong}}}, \bibinfo{journal}{\prd} \textbf{\bibinfo{volume}{86}}, \bibinfo{eid}{083513} (\bibinfo{year}{2012}), \eprint{1205.1514}.

\bibitem[{\citenamefont{{Faltenbacher} et~al.}(2012)\citenamefont{{Faltenbacher}, {Li}, and {Wang}}}]{faltenbacher+12}
\bibinfo{author}{\bibfnamefont{A.}~\bibnamefont{{Faltenbacher}}}, \bibinfo{author}{\bibfnamefont{C.}~\bibnamefont{{Li}}}, \bibnamefont{and} \bibinfo{author}{\bibfnamefont{J.}~\bibnamefont{{Wang}}}, \bibinfo{journal}{\apjl} \textbf{\bibinfo{volume}{751}}, \bibinfo{eid}{L2} (\bibinfo{year}{2012}), \eprint{1112.0503}.

\bibitem[{\citenamefont{{Chisari} and {Dvorkin}}(2013)}]{chisari&dvorkin13}
\bibinfo{author}{\bibfnamefont{N.~E.} \bibnamefont{{Chisari}}} \bibnamefont{and} \bibinfo{author}{\bibfnamefont{C.}~\bibnamefont{{Dvorkin}}}, \bibinfo{journal}{\jcap} \textbf{\bibinfo{volume}{2013}}, \bibinfo{eid}{029} (\bibinfo{year}{2013}), \eprint{1308.5972}.

\bibitem[{\citenamefont{{Chisari} et~al.}(2016)\citenamefont{{Chisari}, {Dvorkin}, {Schmidt}, and {Spergel}}}]{chisari+16}
\bibinfo{author}{\bibfnamefont{N.~E.} \bibnamefont{{Chisari}}}, \bibinfo{author}{\bibfnamefont{C.}~\bibnamefont{{Dvorkin}}}, \bibinfo{author}{\bibfnamefont{F.}~\bibnamefont{{Schmidt}}}, \bibnamefont{and} \bibinfo{author}{\bibfnamefont{D.~N.} \bibnamefont{{Spergel}}}, \bibinfo{journal}{\prd} \textbf{\bibinfo{volume}{94}}, \bibinfo{eid}{123507} (\bibinfo{year}{2016}), \eprint{1607.05232}.

\bibitem[{\citenamefont{{Kogai} et~al.}(2018)\citenamefont{{Kogai}, {Matsubara}, {Nishizawa}, and {Urakawa}}}]{kogai+18}
\bibinfo{author}{\bibfnamefont{K.}~\bibnamefont{{Kogai}}}, \bibinfo{author}{\bibfnamefont{T.}~\bibnamefont{{Matsubara}}}, \bibinfo{author}{\bibfnamefont{A.~J.} \bibnamefont{{Nishizawa}}}, \bibnamefont{and} \bibinfo{author}{\bibfnamefont{Y.}~\bibnamefont{{Urakawa}}}, \bibinfo{journal}{\jcap} \textbf{\bibinfo{volume}{2018}}, \bibinfo{eid}{014} (\bibinfo{year}{2018}), \eprint{1804.06284}.

\bibitem[{\citenamefont{{Biagetti} and {Orlando}}(2020)}]{biagetti&orland020}
\bibinfo{author}{\bibfnamefont{M.}~\bibnamefont{{Biagetti}}} \bibnamefont{and} \bibinfo{author}{\bibfnamefont{G.}~\bibnamefont{{Orlando}}}, \bibinfo{journal}{\jcap} \textbf{\bibinfo{volume}{2020}}, \bibinfo{eid}{005} (\bibinfo{year}{2020}), \eprint{2001.05930}.

\bibitem[{\citenamefont{{Okumura} and {Taruya}}(2020)}]{OT20b}
\bibinfo{author}{\bibfnamefont{T.}~\bibnamefont{{Okumura}}} \bibnamefont{and} \bibinfo{author}{\bibfnamefont{A.}~\bibnamefont{{Taruya}}}, \bibinfo{journal}{\mnras} \textbf{\bibinfo{volume}{493}}, \bibinfo{pages}{L124} (\bibinfo{year}{2020}), \eprint{1912.04118}.

\bibitem[{\citenamefont{{Okumura} et~al.}(2020)\citenamefont{{Okumura}, {Taruya}, and {Nishimichi}}}]{okumura+20}
\bibinfo{author}{\bibfnamefont{T.}~\bibnamefont{{Okumura}}}, \bibinfo{author}{\bibfnamefont{A.}~\bibnamefont{{Taruya}}}, \bibnamefont{and} \bibinfo{author}{\bibfnamefont{T.}~\bibnamefont{{Nishimichi}}}, \bibinfo{journal}{\mnras} \textbf{\bibinfo{volume}{494}}, \bibinfo{pages}{694} (\bibinfo{year}{2020}), \eprint{2001.05302}.

\bibitem[{\citenamefont{{Chuang} et~al.}(2022)\citenamefont{{Chuang}, {Okumura}, and {Shirasaki}}}]{chuang+22}
\bibinfo{author}{\bibfnamefont{Y.-T.} \bibnamefont{{Chuang}}}, \bibinfo{author}{\bibfnamefont{T.}~\bibnamefont{{Okumura}}}, \bibnamefont{and} \bibinfo{author}{\bibfnamefont{M.}~\bibnamefont{{Shirasaki}}}, \bibinfo{journal}{\mnras} \textbf{\bibinfo{volume}{515}}, \bibinfo{pages}{4464} (\bibinfo{year}{2022}), \eprint{2111.01417}.

\bibitem[{\citenamefont{{Akitsu} et~al.}(2023)\citenamefont{{Akitsu}, {Li}, and {Okumura}}}]{akitsu+23}
\bibinfo{author}{\bibfnamefont{K.}~\bibnamefont{{Akitsu}}}, \bibinfo{author}{\bibfnamefont{Y.}~\bibnamefont{{Li}}}, \bibnamefont{and} \bibinfo{author}{\bibfnamefont{T.}~\bibnamefont{{Okumura}}}, \bibinfo{journal}{\prd} \textbf{\bibinfo{volume}{107}}, \bibinfo{eid}{063531} (\bibinfo{year}{2023}), \eprint{2209.06226}.

\bibitem[{\citenamefont{{Shiraishi} et~al.}(2023)\citenamefont{{Shiraishi}, {Okumura}, and {Akitsu}}}]{shiraishi+23}
\bibinfo{author}{\bibfnamefont{M.}~\bibnamefont{{Shiraishi}}}, \bibinfo{author}{\bibfnamefont{T.}~\bibnamefont{{Okumura}}}, \bibnamefont{and} \bibinfo{author}{\bibfnamefont{K.}~\bibnamefont{{Akitsu}}}, \bibinfo{journal}{\jcap} \textbf{\bibinfo{volume}{2023}}, \bibinfo{eid}{013} (\bibinfo{year}{2023}), \eprint{2303.10890}.

\bibitem[{\citenamefont{{Philcox} et~al.}(2024)\citenamefont{{Philcox}, {K{\"o}nig}, {Alexander}, and {Spergel}}}]{philcox+24}
\bibinfo{author}{\bibfnamefont{O.~H.~E.} \bibnamefont{{Philcox}}}, \bibinfo{author}{\bibfnamefont{M.~J.} \bibnamefont{{K{\"o}nig}}}, \bibinfo{author}{\bibfnamefont{S.}~\bibnamefont{{Alexander}}}, \bibnamefont{and} \bibinfo{author}{\bibfnamefont{D.~N.} \bibnamefont{{Spergel}}}, \bibinfo{journal}{\prd} \textbf{\bibinfo{volume}{109}}, \bibinfo{eid}{063541} (\bibinfo{year}{2024}), \eprint{2309.08653}.

\bibitem[{\citenamefont{{Saga} et~al.}(2024)\citenamefont{{Saga}, {Shiraishi}, {Akitsu}, and {Okumura}}}]{saga+24}
\bibinfo{author}{\bibfnamefont{S.}~\bibnamefont{{Saga}}}, \bibinfo{author}{\bibfnamefont{M.}~\bibnamefont{{Shiraishi}}}, \bibinfo{author}{\bibfnamefont{K.}~\bibnamefont{{Akitsu}}}, \bibnamefont{and} \bibinfo{author}{\bibfnamefont{T.}~\bibnamefont{{Okumura}}}, \bibinfo{journal}{\prd} \textbf{\bibinfo{volume}{109}}, \bibinfo{eid}{043520} (\bibinfo{year}{2024}), \eprint{2312.16316}.

\bibitem[{\citenamefont{{Taruya} and {Okumura}}(2020)}]{TO20}
\bibinfo{author}{\bibfnamefont{A.}~\bibnamefont{{Taruya}}} \bibnamefont{and} \bibinfo{author}{\bibfnamefont{T.}~\bibnamefont{{Okumura}}}, \bibinfo{journal}{\apjl} \textbf{\bibinfo{volume}{891}}, \bibinfo{eid}{L42} (\bibinfo{year}{2020}), \eprint{2001.05962}.

\bibitem[{\citenamefont{{Okumura} and {Taruya}}(2022)}]{OT22}
\bibinfo{author}{\bibfnamefont{T.}~\bibnamefont{{Okumura}}} \bibnamefont{and} \bibinfo{author}{\bibfnamefont{A.}~\bibnamefont{{Taruya}}}, \bibinfo{journal}{\prd} \textbf{\bibinfo{volume}{106}}, \bibinfo{eid}{043523} (\bibinfo{year}{2022}), \eprint{2110.11127}.

\bibitem[{\citenamefont{{Xu} et~al.}(2023)\citenamefont{{Xu}, {Jing}, {Zhao}, and {Cuesta}}}]{xu+23}
\bibinfo{author}{\bibfnamefont{K.}~\bibnamefont{{Xu}}}, \bibinfo{author}{\bibfnamefont{Y.~P.} \bibnamefont{{Jing}}}, \bibinfo{author}{\bibfnamefont{G.-B.} \bibnamefont{{Zhao}}}, \bibnamefont{and} \bibinfo{author}{\bibfnamefont{A.~J.} \bibnamefont{{Cuesta}}}, \bibinfo{journal}{Nature Astronomy} \textbf{\bibinfo{volume}{7}}, \bibinfo{pages}{1259} (\bibinfo{year}{2023}), \eprint{2306.09407}.

\bibitem[{\citenamefont{{Shim} et~al.}(2025)\citenamefont{{Shim}, {Okumura}, and {Taruya}}}]{shim+25}
\bibinfo{author}{\bibfnamefont{J.}~\bibnamefont{{Shim}}}, \bibinfo{author}{\bibfnamefont{T.}~\bibnamefont{{Okumura}}}, \bibnamefont{and} \bibinfo{author}{\bibfnamefont{A.}~\bibnamefont{{Taruya}}}, \bibinfo{journal}{in preparation}  (\bibinfo{year}{2025}).

\bibitem[{\citenamefont{{DESI Collaboration} et~al.}(2016)\citenamefont{{DESI Collaboration}, {Aghamousa}, {Aguilar}, {Ahlen}, {Alam}, {Allen}, {Allende Prieto}, {Annis}, {Bailey}, {Balland} et~al.}}]{desi+16}
\bibinfo{author}{\bibnamefont{{DESI Collaboration}}}, \bibinfo{author}{\bibfnamefont{A.}~\bibnamefont{{Aghamousa}}}, \bibinfo{author}{\bibfnamefont{J.}~\bibnamefont{{Aguilar}}}, \bibinfo{author}{\bibfnamefont{S.}~\bibnamefont{{Ahlen}}}, \bibinfo{author}{\bibfnamefont{S.}~\bibnamefont{{Alam}}}, \bibinfo{author}{\bibfnamefont{L.~E.} \bibnamefont{{Allen}}}, \bibinfo{author}{\bibfnamefont{C.}~\bibnamefont{{Allende Prieto}}}, \bibinfo{author}{\bibfnamefont{J.}~\bibnamefont{{Annis}}}, \bibinfo{author}{\bibfnamefont{S.}~\bibnamefont{{Bailey}}}, \bibinfo{author}{\bibfnamefont{C.}~\bibnamefont{{Balland}}}, \bibnamefont{et~al.}, \bibinfo{journal}{arXiv e-prints} \bibinfo{eid}{arXiv:1611.00036} (\bibinfo{year}{2016}), \eprint{1611.00036}.

\bibitem[{\citenamefont{{Takada} et~al.}(2014{\natexlab{a}})\citenamefont{{Takada}, {Ellis}, {Chiba}, {Greene}, {Aihara}, {Arimoto}, {Bundy}, {Cohen}, {Dor{\'e}}, {Graves} et~al.}}]{takada+14}
\bibinfo{author}{\bibfnamefont{M.}~\bibnamefont{{Takada}}}, \bibinfo{author}{\bibfnamefont{R.~S.} \bibnamefont{{Ellis}}}, \bibinfo{author}{\bibfnamefont{M.}~\bibnamefont{{Chiba}}}, \bibinfo{author}{\bibfnamefont{J.~E.} \bibnamefont{{Greene}}}, \bibinfo{author}{\bibfnamefont{H.}~\bibnamefont{{Aihara}}}, \bibinfo{author}{\bibfnamefont{N.}~\bibnamefont{{Arimoto}}}, \bibinfo{author}{\bibfnamefont{K.}~\bibnamefont{{Bundy}}}, \bibinfo{author}{\bibfnamefont{J.}~\bibnamefont{{Cohen}}}, \bibinfo{author}{\bibfnamefont{O.}~\bibnamefont{{Dor{\'e}}}}, \bibinfo{author}{\bibfnamefont{G.}~\bibnamefont{{Graves}}}, \bibnamefont{et~al.}, \bibinfo{journal}{\pasj} \textbf{\bibinfo{volume}{66}}, \bibinfo{eid}{R1} (\bibinfo{year}{2014}{\natexlab{a}}), \eprint{1206.0737}.

\bibitem[{\citenamefont{{Laureijs} et~al.}(2011)\citenamefont{{Laureijs}, {Amiaux}, {Arduini}, {Augu{\`e}res}, {Brinchmann}, {Cole}, {Cropper}, {Dabin}, {Duvet}, {Ealet} et~al.}}]{euclid+11}
\bibinfo{author}{\bibfnamefont{R.}~\bibnamefont{{Laureijs}}}, \bibinfo{author}{\bibfnamefont{J.}~\bibnamefont{{Amiaux}}}, \bibinfo{author}{\bibfnamefont{S.}~\bibnamefont{{Arduini}}}, \bibinfo{author}{\bibfnamefont{J.~L.} \bibnamefont{{Augu{\`e}res}}}, \bibinfo{author}{\bibfnamefont{J.}~\bibnamefont{{Brinchmann}}}, \bibinfo{author}{\bibfnamefont{R.}~\bibnamefont{{Cole}}}, \bibinfo{author}{\bibfnamefont{M.}~\bibnamefont{{Cropper}}}, \bibinfo{author}{\bibfnamefont{C.}~\bibnamefont{{Dabin}}}, \bibinfo{author}{\bibfnamefont{L.}~\bibnamefont{{Duvet}}}, \bibinfo{author}{\bibfnamefont{A.}~\bibnamefont{{Ealet}}}, \bibnamefont{et~al.}, \bibinfo{journal}{arXiv e-prints} \bibinfo{eid}{arXiv:1110.3193} (\bibinfo{year}{2011}), \eprint{1110.3193}.

\bibitem[{\citenamefont{{Euclid Collaboration} et~al.}(2020)\citenamefont{{Euclid Collaboration}, {Blanchard}, {Camera}, {Carbone}, {Cardone}, {Casas}, {Clesse}, {Ili{\'c}}, {Kilbinger}, {Kitching} et~al.}}]{euclid_prep+20}
\bibinfo{author}{\bibnamefont{{Euclid Collaboration}}}, \bibinfo{author}{\bibfnamefont{A.}~\bibnamefont{{Blanchard}}}, \bibinfo{author}{\bibfnamefont{S.}~\bibnamefont{{Camera}}}, \bibinfo{author}{\bibfnamefont{C.}~\bibnamefont{{Carbone}}}, \bibinfo{author}{\bibfnamefont{V.~F.} \bibnamefont{{Cardone}}}, \bibinfo{author}{\bibfnamefont{S.}~\bibnamefont{{Casas}}}, \bibinfo{author}{\bibfnamefont{S.}~\bibnamefont{{Clesse}}}, \bibinfo{author}{\bibfnamefont{S.}~\bibnamefont{{Ili{\'c}}}}, \bibinfo{author}{\bibfnamefont{M.}~\bibnamefont{{Kilbinger}}}, \bibinfo{author}{\bibfnamefont{T.}~\bibnamefont{{Kitching}}}, \bibnamefont{et~al.}, \bibinfo{journal}{\aap} \textbf{\bibinfo{volume}{642}}, \bibinfo{eid}{A191} (\bibinfo{year}{2020}), \eprint{1910.09273}.

\bibitem[{\citenamefont{{Spergel} et~al.}(2013)\citenamefont{{Spergel}, {Gehrels}, {Breckinridge}, {Donahue}, {Dressler}, {Gaudi}, {Greene}, {Guyon}, {Hirata}, {Kalirai} et~al.}}]{roman+13}
\bibinfo{author}{\bibfnamefont{D.}~\bibnamefont{{Spergel}}}, \bibinfo{author}{\bibfnamefont{N.}~\bibnamefont{{Gehrels}}}, \bibinfo{author}{\bibfnamefont{J.}~\bibnamefont{{Breckinridge}}}, \bibinfo{author}{\bibfnamefont{M.}~\bibnamefont{{Donahue}}}, \bibinfo{author}{\bibfnamefont{A.}~\bibnamefont{{Dressler}}}, \bibinfo{author}{\bibfnamefont{B.~S.} \bibnamefont{{Gaudi}}}, \bibinfo{author}{\bibfnamefont{T.}~\bibnamefont{{Greene}}}, \bibinfo{author}{\bibfnamefont{O.}~\bibnamefont{{Guyon}}}, \bibinfo{author}{\bibfnamefont{C.}~\bibnamefont{{Hirata}}}, \bibinfo{author}{\bibfnamefont{J.}~\bibnamefont{{Kalirai}}}, \bibnamefont{et~al.}, \bibinfo{journal}{arXiv e-prints} \bibinfo{eid}{arXiv:1305.5422} (\bibinfo{year}{2013}), \eprint{1305.5422}.

\bibitem[{\citenamefont{{The LSST Dark Energy Science Collaboration} et~al.}(2018)\citenamefont{{The LSST Dark Energy Science Collaboration}, {Mandelbaum}, {Eifler}, {Hlo{\v{z}}ek}, {Collett}, {Gawiser}, {Scolnic}, {Alonso}, {Awan}, {Biswas} et~al.}}]{lsst+18}
\bibinfo{author}{\bibnamefont{{The LSST Dark Energy Science Collaboration}}}, \bibinfo{author}{\bibfnamefont{R.}~\bibnamefont{{Mandelbaum}}}, \bibinfo{author}{\bibfnamefont{T.}~\bibnamefont{{Eifler}}}, \bibinfo{author}{\bibfnamefont{R.}~\bibnamefont{{Hlo{\v{z}}ek}}}, \bibinfo{author}{\bibfnamefont{T.}~\bibnamefont{{Collett}}}, \bibinfo{author}{\bibfnamefont{E.}~\bibnamefont{{Gawiser}}}, \bibinfo{author}{\bibfnamefont{D.}~\bibnamefont{{Scolnic}}}, \bibinfo{author}{\bibfnamefont{D.}~\bibnamefont{{Alonso}}}, \bibinfo{author}{\bibfnamefont{H.}~\bibnamefont{{Awan}}}, \bibinfo{author}{\bibfnamefont{R.}~\bibnamefont{{Biswas}}}, \bibnamefont{et~al.}, \bibinfo{journal}{arXiv e-prints} \bibinfo{eid}{arXiv:1809.01669} (\bibinfo{year}{2018}), \eprint{1809.01669}.

\bibitem[{\citenamefont{{Miyazaki} et~al.}(2018)\citenamefont{{Miyazaki}, {Komiyama}, {Kawanomoto}, {Doi}, {Furusawa}, {Hamana}, {Hayashi}, {Ikeda}, {Kamata}, {Karoji} et~al.}}]{hsc+18}
\bibinfo{author}{\bibfnamefont{S.}~\bibnamefont{{Miyazaki}}}, \bibinfo{author}{\bibfnamefont{Y.}~\bibnamefont{{Komiyama}}}, \bibinfo{author}{\bibfnamefont{S.}~\bibnamefont{{Kawanomoto}}}, \bibinfo{author}{\bibfnamefont{Y.}~\bibnamefont{{Doi}}}, \bibinfo{author}{\bibfnamefont{H.}~\bibnamefont{{Furusawa}}}, \bibinfo{author}{\bibfnamefont{T.}~\bibnamefont{{Hamana}}}, \bibinfo{author}{\bibfnamefont{Y.}~\bibnamefont{{Hayashi}}}, \bibinfo{author}{\bibfnamefont{H.}~\bibnamefont{{Ikeda}}}, \bibinfo{author}{\bibfnamefont{Y.}~\bibnamefont{{Kamata}}}, \bibinfo{author}{\bibfnamefont{H.}~\bibnamefont{{Karoji}}}, \bibnamefont{et~al.}, \bibinfo{journal}{\pasj} \textbf{\bibinfo{volume}{70}}, \bibinfo{eid}{S1} (\bibinfo{year}{2018}).

\bibitem[{\citenamefont{{Aihara} et~al.}(2018)\citenamefont{{Aihara}, {Arimoto}, {Armstrong}, {Arnouts}, {Bahcall}, {Bickerton}, {Bosch}, {Bundy}, {Capak}, {Chan} et~al.}}]{hsc2+18}
\bibinfo{author}{\bibfnamefont{H.}~\bibnamefont{{Aihara}}}, \bibinfo{author}{\bibfnamefont{N.}~\bibnamefont{{Arimoto}}}, \bibinfo{author}{\bibfnamefont{R.}~\bibnamefont{{Armstrong}}}, \bibinfo{author}{\bibfnamefont{S.}~\bibnamefont{{Arnouts}}}, \bibinfo{author}{\bibfnamefont{N.~A.} \bibnamefont{{Bahcall}}}, \bibinfo{author}{\bibfnamefont{S.}~\bibnamefont{{Bickerton}}}, \bibinfo{author}{\bibfnamefont{J.}~\bibnamefont{{Bosch}}}, \bibinfo{author}{\bibfnamefont{K.}~\bibnamefont{{Bundy}}}, \bibinfo{author}{\bibfnamefont{P.~L.} \bibnamefont{{Capak}}}, \bibinfo{author}{\bibfnamefont{J.~H.~H.} \bibnamefont{{Chan}}}, \bibnamefont{et~al.}, \bibinfo{journal}{\pasj} \textbf{\bibinfo{volume}{70}}, \bibinfo{eid}{S4} (\bibinfo{year}{2018}), \eprint{1704.05858}.

\bibitem[{\citenamefont{{Chevallier} and {Polarski}}(2001)}]{chevallier&polarski01}
\bibinfo{author}{\bibfnamefont{M.}~\bibnamefont{{Chevallier}}} \bibnamefont{and} \bibinfo{author}{\bibfnamefont{D.}~\bibnamefont{{Polarski}}}, \bibinfo{journal}{International Journal of Modern Physics D} \textbf{\bibinfo{volume}{10}}, \bibinfo{pages}{213} (\bibinfo{year}{2001}), \eprint{gr-qc/0009008}.

\bibitem[{\citenamefont{{Linder}}(2003)}]{linder03}
\bibinfo{author}{\bibfnamefont{E.~V.} \bibnamefont{{Linder}}}, \bibinfo{journal}{\prl} \textbf{\bibinfo{volume}{90}}, \bibinfo{eid}{091301} (\bibinfo{year}{2003}), \eprint{astro-ph/0208512}.

\bibitem[{\citenamefont{{Wright}}(2006)}]{wright06}
\bibinfo{author}{\bibfnamefont{E.~L.} \bibnamefont{{Wright}}}, \bibinfo{journal}{\pasp} \textbf{\bibinfo{volume}{118}}, \bibinfo{pages}{1711} (\bibinfo{year}{2006}), \eprint{astro-ph/0609593}.

\bibitem[{\citenamefont{{Komatsu} and {others}}(2011)}]{komatsu+11}
\bibinfo{author}{\bibfnamefont{E.}~\bibnamefont{{Komatsu}}} \bibnamefont{and} \bibinfo{author}{\bibnamefont{{others}}}, \bibinfo{journal}{\apjs} \textbf{\bibinfo{volume}{192}}, \bibinfo{eid}{18} (\bibinfo{year}{2011}), \eprint{1001.4538}.

\bibitem[{\citenamefont{{Jimenez} et~al.}(2010)\citenamefont{{Jimenez}, {Kitching}, {Pe{\~n}a-Garay}, and {Verde}}}]{jimenez+10}
\bibinfo{author}{\bibfnamefont{R.}~\bibnamefont{{Jimenez}}}, \bibinfo{author}{\bibfnamefont{T.}~\bibnamefont{{Kitching}}}, \bibinfo{author}{\bibfnamefont{C.}~\bibnamefont{{Pe{\~n}a-Garay}}}, \bibnamefont{and} \bibinfo{author}{\bibfnamefont{L.}~\bibnamefont{{Verde}}}, \bibinfo{journal}{\jcap} \textbf{\bibinfo{volume}{2010}}, \bibinfo{eid}{035} (\bibinfo{year}{2010}), \eprint{1003.5918}.

\bibitem[{\citenamefont{Zhao et~al.}(2018)\citenamefont{Zhao, Zhang, and Zhang}}]{zhao+18}
\bibinfo{author}{\bibfnamefont{M.-M.} \bibnamefont{Zhao}}, \bibinfo{author}{\bibfnamefont{J.-F.} \bibnamefont{Zhang}}, \bibnamefont{and} \bibinfo{author}{\bibfnamefont{X.}~\bibnamefont{Zhang}}, \bibinfo{journal}{Physics Letters B} \textbf{\bibinfo{volume}{779}}, \bibinfo{pages}{473} (\bibinfo{year}{2018}), ISSN \bibinfo{issn}{0370-2693}, \urlprefix\url{https://www.sciencedirect.com/science/article/pii/S0370269318301527}.

\bibitem[{\citenamefont{{Yang} et~al.}(2017)\citenamefont{{Yang}, {Nunes}, {Pan}, and {Mota}}}]{yang+17}
\bibinfo{author}{\bibfnamefont{W.}~\bibnamefont{{Yang}}}, \bibinfo{author}{\bibfnamefont{R.~C.} \bibnamefont{{Nunes}}}, \bibinfo{author}{\bibfnamefont{S.}~\bibnamefont{{Pan}}}, \bibnamefont{and} \bibinfo{author}{\bibfnamefont{D.~F.} \bibnamefont{{Mota}}}, \bibinfo{journal}{\prd} \textbf{\bibinfo{volume}{95}}, \bibinfo{eid}{103522} (\bibinfo{year}{2017}), \eprint{1703.02556}.

\bibitem[{\citenamefont{{Li} et~al.}(2018)\citenamefont{{Li}, {Zhang}, {Du}, {Zhou}, and {Xu}}}]{li+18}
\bibinfo{author}{\bibfnamefont{E.-K.} \bibnamefont{{Li}}}, \bibinfo{author}{\bibfnamefont{H.}~\bibnamefont{{Zhang}}}, \bibinfo{author}{\bibfnamefont{M.}~\bibnamefont{{Du}}}, \bibinfo{author}{\bibfnamefont{Z.-H.} \bibnamefont{{Zhou}}}, \bibnamefont{and} \bibinfo{author}{\bibfnamefont{L.}~\bibnamefont{{Xu}}}, \bibinfo{journal}{\jcap} \textbf{\bibinfo{volume}{2018}}, \bibinfo{eid}{042} (\bibinfo{year}{2018}), \eprint{1703.01554}.

\bibitem[{\citenamefont{{Mangano} et~al.}(2005)\citenamefont{{Mangano}, {Miele}, {Pastor}, {Pinto}, {Pisanti}, and {Serpico}}}]{mangano+05}
\bibinfo{author}{\bibfnamefont{G.}~\bibnamefont{{Mangano}}}, \bibinfo{author}{\bibfnamefont{G.}~\bibnamefont{{Miele}}}, \bibinfo{author}{\bibfnamefont{S.}~\bibnamefont{{Pastor}}}, \bibinfo{author}{\bibfnamefont{T.}~\bibnamefont{{Pinto}}}, \bibinfo{author}{\bibfnamefont{O.}~\bibnamefont{{Pisanti}}}, \bibnamefont{and} \bibinfo{author}{\bibfnamefont{P.~D.} \bibnamefont{{Serpico}}}, \bibinfo{journal}{Nuclear Physics B} \textbf{\bibinfo{volume}{729}}, \bibinfo{pages}{221} (\bibinfo{year}{2005}), \eprint{hep-ph/0506164}.

\bibitem[{\citenamefont{Kiakotou et~al.}(2008)\citenamefont{Kiakotou, Elgar\o{}y, and Lahav}}]{kiakotou+08}
\bibinfo{author}{\bibfnamefont{A.}~\bibnamefont{Kiakotou}}, \bibinfo{author}{\bibfnamefont{O.}~\bibnamefont{Elgar\o{}y}}, \bibnamefont{and} \bibinfo{author}{\bibfnamefont{O.}~\bibnamefont{Lahav}}, \bibinfo{journal}{Phys. Rev. D} \textbf{\bibinfo{volume}{77}}, \bibinfo{pages}{063005} (\bibinfo{year}{2008}), \urlprefix\url{https://link.aps.org/doi/10.1103/PhysRevD.77.063005}.

\bibitem[{\citenamefont{{Wang} and {Steinhardt}}(1998)}]{wang&steinhardt98}
\bibinfo{author}{\bibfnamefont{L.}~\bibnamefont{{Wang}}} \bibnamefont{and} \bibinfo{author}{\bibfnamefont{P.~J.} \bibnamefont{{Steinhardt}}}, \bibinfo{journal}{\apj} \textbf{\bibinfo{volume}{508}}, \bibinfo{pages}{483} (\bibinfo{year}{1998}), \eprint{astro-ph/9804015}.

\bibitem[{\citenamefont{{Linder}}(2005)}]{linder05}
\bibinfo{author}{\bibfnamefont{E.~V.} \bibnamefont{{Linder}}}, \bibinfo{journal}{\prd} \textbf{\bibinfo{volume}{72}}, \bibinfo{eid}{043529} (\bibinfo{year}{2005}), \eprint{astro-ph/0507263}.

\bibitem[{\citenamefont{{Peebles}}(1980)}]{peebles80}
\bibinfo{author}{\bibfnamefont{P.~J.~E.} \bibnamefont{{Peebles}}}, \emph{\bibinfo{title}{{The large-scale structure of the universe}}} (\bibinfo{year}{1980}).

\bibitem[{\citenamefont{{Lahav} et~al.}(1991)\citenamefont{{Lahav}, {Lilje}, {Primack}, and {Rees}}}]{lahav+91}
\bibinfo{author}{\bibfnamefont{O.}~\bibnamefont{{Lahav}}}, \bibinfo{author}{\bibfnamefont{P.~B.} \bibnamefont{{Lilje}}}, \bibinfo{author}{\bibfnamefont{J.~R.} \bibnamefont{{Primack}}}, \bibnamefont{and} \bibinfo{author}{\bibfnamefont{M.~J.} \bibnamefont{{Rees}}}, \bibinfo{journal}{\mnras} \textbf{\bibinfo{volume}{251}}, \bibinfo{pages}{128} (\bibinfo{year}{1991}).

\bibitem[{\citenamefont{{Kaiser}}(1984)}]{kaiser84}
\bibinfo{author}{\bibfnamefont{N.}~\bibnamefont{{Kaiser}}}, \bibinfo{journal}{\apjl} \textbf{\bibinfo{volume}{284}}, \bibinfo{pages}{L9} (\bibinfo{year}{1984}).

\bibitem[{\citenamefont{{Okumura} and {Jing}}(2009)}]{okumura&jing09}
\bibinfo{author}{\bibfnamefont{T.}~\bibnamefont{{Okumura}}} \bibnamefont{and} \bibinfo{author}{\bibfnamefont{Y.~P.} \bibnamefont{{Jing}}}, \bibinfo{journal}{\apjl} \textbf{\bibinfo{volume}{694}}, \bibinfo{pages}{L83} (\bibinfo{year}{2009}), \eprint{0812.2935}.

\bibitem[{\citenamefont{{Stebbins} et~al.}(1996)\citenamefont{{Stebbins}, {McKay}, and {Frieman}}}]{stebbins+96}
\bibinfo{author}{\bibfnamefont{A.}~\bibnamefont{{Stebbins}}}, \bibinfo{author}{\bibfnamefont{T.}~\bibnamefont{{McKay}}}, \bibnamefont{and} \bibinfo{author}{\bibfnamefont{J.~A.} \bibnamefont{{Frieman}}}, in \emph{\bibinfo{booktitle}{Astrophysical Applications of Gravitational Lensing}}, edited by \bibinfo{editor}{\bibfnamefont{C.~S.} \bibnamefont{{Kochanek}}} \bibnamefont{and} \bibinfo{editor}{\bibfnamefont{J.~N.} \bibnamefont{{Hewitt}}} (\bibinfo{year}{1996}), vol. \bibinfo{volume}{173} of \emph{\bibinfo{series}{IAU Symposium}}, p.~\bibinfo{pages}{75}, \eprint{astro-ph/9510012}.

\bibitem[{\citenamefont{{Kamionkowski} et~al.}(1998)\citenamefont{{Kamionkowski}, {Babul}, {Cress}, and {Refregier}}}]{kamionkowski+98}
\bibinfo{author}{\bibfnamefont{M.}~\bibnamefont{{Kamionkowski}}}, \bibinfo{author}{\bibfnamefont{A.}~\bibnamefont{{Babul}}}, \bibinfo{author}{\bibfnamefont{C.~M.} \bibnamefont{{Cress}}}, \bibnamefont{and} \bibinfo{author}{\bibfnamefont{A.}~\bibnamefont{{Refregier}}}, \bibinfo{journal}{\mnras} \textbf{\bibinfo{volume}{301}}, \bibinfo{pages}{1064} (\bibinfo{year}{1998}), \eprint{astro-ph/9712030}.

\bibitem[{\citenamefont{{Crittenden} et~al.}(2002)\citenamefont{{Crittenden}, {Natarajan}, {Pen}, and {Theuns}}}]{crittenden+02}
\bibinfo{author}{\bibfnamefont{R.~G.} \bibnamefont{{Crittenden}}}, \bibinfo{author}{\bibfnamefont{P.}~\bibnamefont{{Natarajan}}}, \bibinfo{author}{\bibfnamefont{U.-L.} \bibnamefont{{Pen}}}, \bibnamefont{and} \bibinfo{author}{\bibfnamefont{T.}~\bibnamefont{{Theuns}}}, \bibinfo{journal}{\apj} \textbf{\bibinfo{volume}{568}}, \bibinfo{pages}{20} (\bibinfo{year}{2002}), \eprint{astro-ph/0012336}.

\bibitem[{\citenamefont{{Joachimi} et~al.}(2011)\citenamefont{{Joachimi}, {Mandelbaum}, {Abdalla}, and {Bridle}}}]{joachimi+11}
\bibinfo{author}{\bibfnamefont{B.}~\bibnamefont{{Joachimi}}}, \bibinfo{author}{\bibfnamefont{R.}~\bibnamefont{{Mandelbaum}}}, \bibinfo{author}{\bibfnamefont{F.~B.} \bibnamefont{{Abdalla}}}, \bibnamefont{and} \bibinfo{author}{\bibfnamefont{S.~L.} \bibnamefont{{Bridle}}}, \bibinfo{journal}{\aap} \textbf{\bibinfo{volume}{527}}, \bibinfo{eid}{A26} (\bibinfo{year}{2011}), \eprint{1008.3491}.

\bibitem[{\citenamefont{{Kurita} et~al.}(2021)\citenamefont{{Kurita}, {Takada}, {Nishimichi}, {Takahashi}, {Osato}, and {Kobayashi}}}]{kurita+21}
\bibinfo{author}{\bibfnamefont{T.}~\bibnamefont{{Kurita}}}, \bibinfo{author}{\bibfnamefont{M.}~\bibnamefont{{Takada}}}, \bibinfo{author}{\bibfnamefont{T.}~\bibnamefont{{Nishimichi}}}, \bibinfo{author}{\bibfnamefont{R.}~\bibnamefont{{Takahashi}}}, \bibinfo{author}{\bibfnamefont{K.}~\bibnamefont{{Osato}}}, \bibnamefont{and} \bibinfo{author}{\bibfnamefont{Y.}~\bibnamefont{{Kobayashi}}}, \bibinfo{journal}{\mnras} \textbf{\bibinfo{volume}{501}}, \bibinfo{pages}{833} (\bibinfo{year}{2021}), \eprint{2004.12579}.

\bibitem[{\citenamefont{{Shi} et~al.}(2021{\natexlab{a}})\citenamefont{{Shi}, {Osato}, {Kurita}, and {Takada}}}]{shi+21a}
\bibinfo{author}{\bibfnamefont{J.}~\bibnamefont{{Shi}}}, \bibinfo{author}{\bibfnamefont{K.}~\bibnamefont{{Osato}}}, \bibinfo{author}{\bibfnamefont{T.}~\bibnamefont{{Kurita}}}, \bibnamefont{and} \bibinfo{author}{\bibfnamefont{M.}~\bibnamefont{{Takada}}}, \bibinfo{journal}{\apj} \textbf{\bibinfo{volume}{917}}, \bibinfo{eid}{109} (\bibinfo{year}{2021}{\natexlab{a}}), \eprint{2104.12329}.

\bibitem[{\citenamefont{{Shi} et~al.}(2021{\natexlab{b}})\citenamefont{{Shi}, {Kurita}, {Takada}, {Osato}, {Kobayashi}, and {Nishimichi}}}]{shi+21b}
\bibinfo{author}{\bibfnamefont{J.}~\bibnamefont{{Shi}}}, \bibinfo{author}{\bibfnamefont{T.}~\bibnamefont{{Kurita}}}, \bibinfo{author}{\bibfnamefont{M.}~\bibnamefont{{Takada}}}, \bibinfo{author}{\bibfnamefont{K.}~\bibnamefont{{Osato}}}, \bibinfo{author}{\bibfnamefont{Y.}~\bibnamefont{{Kobayashi}}}, \bibnamefont{and} \bibinfo{author}{\bibfnamefont{T.}~\bibnamefont{{Nishimichi}}}, \bibinfo{journal}{\jcap} \textbf{\bibinfo{volume}{2021}}, \bibinfo{eid}{030} (\bibinfo{year}{2021}{\natexlab{b}}), \eprint{2009.00276}.

\bibitem[{\citenamefont{{Inoue} et~al.}(2024)\citenamefont{{Inoue}, {Okumura}, {Saga}, and {Taruya}}}]{inoue+24}
\bibinfo{author}{\bibfnamefont{T.}~\bibnamefont{{Inoue}}}, \bibinfo{author}{\bibfnamefont{T.}~\bibnamefont{{Okumura}}}, \bibinfo{author}{\bibfnamefont{S.}~\bibnamefont{{Saga}}}, \bibnamefont{and} \bibinfo{author}{\bibfnamefont{A.}~\bibnamefont{{Taruya}}}, \bibinfo{journal}{arXiv e-prints} \bibinfo{eid}{arXiv:2406.19669} (\bibinfo{year}{2024}), \eprint{2406.19669}.

\bibitem[{\citenamefont{{Alcock} and {Paczynski}}(1979)}]{alcock&paczynski79}
\bibinfo{author}{\bibfnamefont{C.}~\bibnamefont{{Alcock}}} \bibnamefont{and} \bibinfo{author}{\bibfnamefont{B.}~\bibnamefont{{Paczynski}}}, \bibinfo{journal}{\nat} \textbf{\bibinfo{volume}{281}}, \bibinfo{pages}{358} (\bibinfo{year}{1979}).

\bibitem[{\citenamefont{{Tegmark}}(1997)}]{tegmark97}
\bibinfo{author}{\bibfnamefont{M.}~\bibnamefont{{Tegmark}}}, \bibinfo{journal}{\prl} \textbf{\bibinfo{volume}{79}}, \bibinfo{pages}{3806} (\bibinfo{year}{1997}), \eprint{astro-ph/9706198}.

\bibitem[{\citenamefont{{Blas} et~al.}(2011)\citenamefont{{Blas}, {Lesgourgues}, and {Tram}}}]{blas+11}
\bibinfo{author}{\bibfnamefont{D.}~\bibnamefont{{Blas}}}, \bibinfo{author}{\bibfnamefont{J.}~\bibnamefont{{Lesgourgues}}}, \bibnamefont{and} \bibinfo{author}{\bibfnamefont{T.}~\bibnamefont{{Tram}}}, \bibinfo{journal}{\jcap} \textbf{\bibinfo{volume}{2011}}, \bibinfo{eid}{034} (\bibinfo{year}{2011}), \eprint{1104.2933}.

\bibitem[{\citenamefont{{Planck Collaboration} et~al.}(2016)\citenamefont{{Planck Collaboration}, {Ade}, {Aghanim}, {Arnaud}, {Ashdown}, {Aumont}, {Baccigalupi}, {Banday}, {Barreiro}, {Bartolo} et~al.}}]{planck+15}
\bibinfo{author}{\bibnamefont{{Planck Collaboration}}}, \bibinfo{author}{\bibfnamefont{P.~A.~R.} \bibnamefont{{Ade}}}, \bibinfo{author}{\bibfnamefont{N.}~\bibnamefont{{Aghanim}}}, \bibinfo{author}{\bibfnamefont{M.}~\bibnamefont{{Arnaud}}}, \bibinfo{author}{\bibfnamefont{M.}~\bibnamefont{{Ashdown}}}, \bibinfo{author}{\bibfnamefont{J.}~\bibnamefont{{Aumont}}}, \bibinfo{author}{\bibfnamefont{C.}~\bibnamefont{{Baccigalupi}}}, \bibinfo{author}{\bibfnamefont{A.~J.} \bibnamefont{{Banday}}}, \bibinfo{author}{\bibfnamefont{R.~B.} \bibnamefont{{Barreiro}}}, \bibinfo{author}{\bibfnamefont{N.}~\bibnamefont{{Bartolo}}}, \bibnamefont{et~al.}, \bibinfo{journal}{\aap} \textbf{\bibinfo{volume}{594}}, \bibinfo{eid}{A14} (\bibinfo{year}{2016}), \eprint{1502.01590}.

\bibitem[{\citenamefont{{Wang} and {Mukherjee}}(2007)}]{wang&mukherjee07}
\bibinfo{author}{\bibfnamefont{Y.}~\bibnamefont{{Wang}}} \bibnamefont{and} \bibinfo{author}{\bibfnamefont{P.}~\bibnamefont{{Mukherjee}}}, \bibinfo{journal}{\prd} \textbf{\bibinfo{volume}{76}}, \bibinfo{eid}{103533} (\bibinfo{year}{2007}), \eprint{astro-ph/0703780}.

\bibitem[{\citenamefont{{Mukherjee} et~al.}(2008)\citenamefont{{Mukherjee}, {Kunz}, {Parkinson}, and {Wang}}}]{mukherjee+08}
\bibinfo{author}{\bibfnamefont{P.}~\bibnamefont{{Mukherjee}}}, \bibinfo{author}{\bibfnamefont{M.}~\bibnamefont{{Kunz}}}, \bibinfo{author}{\bibfnamefont{D.}~\bibnamefont{{Parkinson}}}, \bibnamefont{and} \bibinfo{author}{\bibfnamefont{Y.}~\bibnamefont{{Wang}}}, \bibinfo{journal}{\prd} \textbf{\bibinfo{volume}{78}}, \bibinfo{eid}{083529} (\bibinfo{year}{2008}), \eprint{0803.1616}.

\bibitem[{\citenamefont{{Zhai} et~al.}(2020)\citenamefont{{Zhai}, {Park}, {Wang}, and {Ratra}}}]{zhai+20}
\bibinfo{author}{\bibfnamefont{Z.}~\bibnamefont{{Zhai}}}, \bibinfo{author}{\bibfnamefont{C.-G.} \bibnamefont{{Park}}}, \bibinfo{author}{\bibfnamefont{Y.}~\bibnamefont{{Wang}}}, \bibnamefont{and} \bibinfo{author}{\bibfnamefont{B.}~\bibnamefont{{Ratra}}}, \bibinfo{journal}{\jcap} \textbf{\bibinfo{volume}{2020}}, \bibinfo{eid}{009} (\bibinfo{year}{2020}), \eprint{1912.04921}.

\bibitem[{\citenamefont{{Efstathiou} and {Bond}}(1999)}]{efstathiou&bond99}
\bibinfo{author}{\bibfnamefont{G.}~\bibnamefont{{Efstathiou}}} \bibnamefont{and} \bibinfo{author}{\bibfnamefont{J.~R.} \bibnamefont{{Bond}}}, \bibinfo{journal}{\mnras} \textbf{\bibinfo{volume}{304}}, \bibinfo{pages}{75} (\bibinfo{year}{1999}), \eprint{astro-ph/9807103}.

\bibitem[{\citenamefont{{Albrecht} et~al.}(2006)\citenamefont{{Albrecht}, {Bernstein}, {Cahn}, {Freedman}, {Hewitt}, {Hu}, {Huth}, {Kamionkowski}, {Kolb}, {Knox} et~al.}}]{albrecht+06}
\bibinfo{author}{\bibfnamefont{A.}~\bibnamefont{{Albrecht}}}, \bibinfo{author}{\bibfnamefont{G.}~\bibnamefont{{Bernstein}}}, \bibinfo{author}{\bibfnamefont{R.}~\bibnamefont{{Cahn}}}, \bibinfo{author}{\bibfnamefont{W.~L.} \bibnamefont{{Freedman}}}, \bibinfo{author}{\bibfnamefont{J.}~\bibnamefont{{Hewitt}}}, \bibinfo{author}{\bibfnamefont{W.}~\bibnamefont{{Hu}}}, \bibinfo{author}{\bibfnamefont{J.}~\bibnamefont{{Huth}}}, \bibinfo{author}{\bibfnamefont{M.}~\bibnamefont{{Kamionkowski}}}, \bibinfo{author}{\bibfnamefont{E.~W.} \bibnamefont{{Kolb}}}, \bibinfo{author}{\bibfnamefont{L.}~\bibnamefont{{Knox}}}, \bibnamefont{et~al.}, \bibinfo{journal}{arXiv e-prints} \bibinfo{eid}{astro-ph/0609591} (\bibinfo{year}{2006}), \eprint{astro-ph/0609591}.

\bibitem[{\citenamefont{{Takada} et~al.}(2014{\natexlab{b}})\citenamefont{{Takada}, {Ellis}, {Chiba}, {Greene}, {Aihara}, {Arimoto}, {Bundy}, {Cohen}, {Dor{\'e}}, {Graves} et~al.}}]{pfs+14}
\bibinfo{author}{\bibfnamefont{M.}~\bibnamefont{{Takada}}}, \bibinfo{author}{\bibfnamefont{R.~S.} \bibnamefont{{Ellis}}}, \bibinfo{author}{\bibfnamefont{M.}~\bibnamefont{{Chiba}}}, \bibinfo{author}{\bibfnamefont{J.~E.} \bibnamefont{{Greene}}}, \bibinfo{author}{\bibfnamefont{H.}~\bibnamefont{{Aihara}}}, \bibinfo{author}{\bibfnamefont{N.}~\bibnamefont{{Arimoto}}}, \bibinfo{author}{\bibfnamefont{K.}~\bibnamefont{{Bundy}}}, \bibinfo{author}{\bibfnamefont{J.}~\bibnamefont{{Cohen}}}, \bibinfo{author}{\bibfnamefont{O.}~\bibnamefont{{Dor{\'e}}}}, \bibinfo{author}{\bibfnamefont{G.}~\bibnamefont{{Graves}}}, \bibnamefont{et~al.}, \bibinfo{journal}{\pasj} \textbf{\bibinfo{volume}{66}}, \bibinfo{eid}{R1} (\bibinfo{year}{2014}{\natexlab{b}}), \eprint{1206.0737}.

\bibitem[{\citenamefont{{Hikage} et~al.}(2019)\citenamefont{{Hikage}, {Oguri}, {Hamana}, {More}, {Mandelbaum}, {Takada}, {K{\"o}hlinger}, {Miyatake}, {Nishizawa}, {Aihara} et~al.}}]{hikage+19}
\bibinfo{author}{\bibfnamefont{C.}~\bibnamefont{{Hikage}}}, \bibinfo{author}{\bibfnamefont{M.}~\bibnamefont{{Oguri}}}, \bibinfo{author}{\bibfnamefont{T.}~\bibnamefont{{Hamana}}}, \bibinfo{author}{\bibfnamefont{S.}~\bibnamefont{{More}}}, \bibinfo{author}{\bibfnamefont{R.}~\bibnamefont{{Mandelbaum}}}, \bibinfo{author}{\bibfnamefont{M.}~\bibnamefont{{Takada}}}, \bibinfo{author}{\bibfnamefont{F.}~\bibnamefont{{K{\"o}hlinger}}}, \bibinfo{author}{\bibfnamefont{H.}~\bibnamefont{{Miyatake}}}, \bibinfo{author}{\bibfnamefont{A.~J.} \bibnamefont{{Nishizawa}}}, \bibinfo{author}{\bibfnamefont{H.}~\bibnamefont{{Aihara}}}, \bibnamefont{et~al.}, \bibinfo{journal}{\pasj} \textbf{\bibinfo{volume}{71}}, \bibinfo{eid}{43} (\bibinfo{year}{2019}), \eprint{1809.09148}.

\bibitem[{\citenamefont{{Matsubara}}(2004)}]{matsubara04}
\bibinfo{author}{\bibfnamefont{T.}~\bibnamefont{{Matsubara}}}, \bibinfo{journal}{\apj} \textbf{\bibinfo{volume}{615}}, \bibinfo{pages}{573} (\bibinfo{year}{2004}), \eprint{astro-ph/0408349}.

\bibitem[{\citenamefont{{Planck Collaboration} et~al.}(2020)\citenamefont{{Planck Collaboration}, {Aghanim}, {Akrami}, {Ashdown}, {Aumont}, {Baccigalupi}, {Ballardini}, {Banday}, {Barreiro}, {Bartolo} et~al.}}]{planck+18}
\bibinfo{author}{\bibnamefont{{Planck Collaboration}}}, \bibinfo{author}{\bibfnamefont{N.}~\bibnamefont{{Aghanim}}}, \bibinfo{author}{\bibfnamefont{Y.}~\bibnamefont{{Akrami}}}, \bibinfo{author}{\bibfnamefont{M.}~\bibnamefont{{Ashdown}}}, \bibinfo{author}{\bibfnamefont{J.}~\bibnamefont{{Aumont}}}, \bibinfo{author}{\bibfnamefont{C.}~\bibnamefont{{Baccigalupi}}}, \bibinfo{author}{\bibfnamefont{M.}~\bibnamefont{{Ballardini}}}, \bibinfo{author}{\bibfnamefont{A.~J.} \bibnamefont{{Banday}}}, \bibinfo{author}{\bibfnamefont{R.~B.} \bibnamefont{{Barreiro}}}, \bibinfo{author}{\bibfnamefont{N.}~\bibnamefont{{Bartolo}}}, \bibnamefont{et~al.}, \bibinfo{journal}{\aap} \textbf{\bibinfo{volume}{641}}, \bibinfo{eid}{A6} (\bibinfo{year}{2020}), \eprint{1807.06209}.

\bibitem[{\citenamefont{{Alam} et~al.}(2021)\citenamefont{{Alam}, {Aubert}, {Avila}, {Balland}, {Bautista}, {Bershady}, {Bizyaev}, {Blanton}, {Bolton}, {Bovy} et~al.}}]{alam+21}
\bibinfo{author}{\bibfnamefont{S.}~\bibnamefont{{Alam}}}, \bibinfo{author}{\bibfnamefont{M.}~\bibnamefont{{Aubert}}}, \bibinfo{author}{\bibfnamefont{S.}~\bibnamefont{{Avila}}}, \bibinfo{author}{\bibfnamefont{C.}~\bibnamefont{{Balland}}}, \bibinfo{author}{\bibfnamefont{J.~E.} \bibnamefont{{Bautista}}}, \bibinfo{author}{\bibfnamefont{M.~A.} \bibnamefont{{Bershady}}}, \bibinfo{author}{\bibfnamefont{D.}~\bibnamefont{{Bizyaev}}}, \bibinfo{author}{\bibfnamefont{M.~R.} \bibnamefont{{Blanton}}}, \bibinfo{author}{\bibfnamefont{A.~S.} \bibnamefont{{Bolton}}}, \bibinfo{author}{\bibfnamefont{J.}~\bibnamefont{{Bovy}}}, \bibnamefont{et~al.}, \bibinfo{journal}{\prd} \textbf{\bibinfo{volume}{103}}, \bibinfo{eid}{083533} (\bibinfo{year}{2021}), \eprint{2007.08991}.

\bibitem[{\citenamefont{{Aviles}}(2024)}]{aviles24}
\bibinfo{author}{\bibfnamefont{A.}~\bibnamefont{{Aviles}}}, \bibinfo{journal}{arXiv e-prints} \bibinfo{eid}{arXiv:2409.15640} (\bibinfo{year}{2024}), \eprint{2409.15640}.

\bibitem[{\citenamefont{{Taruya} et~al.}(2011)\citenamefont{{Taruya}, {Saito}, and {Nishimichi}}}]{taruya+11}
\bibinfo{author}{\bibfnamefont{A.}~\bibnamefont{{Taruya}}}, \bibinfo{author}{\bibfnamefont{S.}~\bibnamefont{{Saito}}}, \bibnamefont{and} \bibinfo{author}{\bibfnamefont{T.}~\bibnamefont{{Nishimichi}}}, \bibinfo{journal}{\prd} \textbf{\bibinfo{volume}{83}}, \bibinfo{eid}{103527} (\bibinfo{year}{2011}), \eprint{1101.4723}.

\bibitem[{\citenamefont{{Lamman} et~al.}(2024)\citenamefont{{Lamman}, {Eisenstein}, {Forero-Romero}, {Aguilar}, {Ahlen}, {Bailey}, {Bianchi}, {Brooks}, {Claybaugh}, {de la Macorra} et~al.}}]{lamman+24}
\bibinfo{author}{\bibfnamefont{C.}~\bibnamefont{{Lamman}}}, \bibinfo{author}{\bibfnamefont{D.}~\bibnamefont{{Eisenstein}}}, \bibinfo{author}{\bibfnamefont{J.~E.} \bibnamefont{{Forero-Romero}}}, \bibinfo{author}{\bibfnamefont{J.~N.} \bibnamefont{{Aguilar}}}, \bibinfo{author}{\bibfnamefont{S.}~\bibnamefont{{Ahlen}}}, \bibinfo{author}{\bibfnamefont{S.}~\bibnamefont{{Bailey}}}, \bibinfo{author}{\bibfnamefont{D.}~\bibnamefont{{Bianchi}}}, \bibinfo{author}{\bibfnamefont{D.}~\bibnamefont{{Brooks}}}, \bibinfo{author}{\bibfnamefont{T.}~\bibnamefont{{Claybaugh}}}, \bibinfo{author}{\bibfnamefont{A.}~\bibnamefont{{de la Macorra}}}, \bibnamefont{et~al.}, \bibinfo{journal}{arXiv e-prints} \bibinfo{eid}{arXiv:2408.11056} (\bibinfo{year}{2024}), \eprint{2408.11056}.

\bibitem[{\citenamefont{{Blazek} et~al.}(2019)\citenamefont{{Blazek}, {MacCrann}, {Troxel}, and {Fang}}}]{blazek+19}
\bibinfo{author}{\bibfnamefont{J.~A.} \bibnamefont{{Blazek}}}, \bibinfo{author}{\bibfnamefont{N.}~\bibnamefont{{MacCrann}}}, \bibinfo{author}{\bibfnamefont{M.~A.} \bibnamefont{{Troxel}}}, \bibnamefont{and} \bibinfo{author}{\bibfnamefont{X.}~\bibnamefont{{Fang}}}, \bibinfo{journal}{\prd} \textbf{\bibinfo{volume}{100}}, \bibinfo{eid}{103506} (\bibinfo{year}{2019}), \eprint{1708.09247}.

\bibitem[{\citenamefont{{Vlah} et~al.}(2020)\citenamefont{{Vlah}, {Chisari}, and {Schmidt}}}]{vlah+20}
\bibinfo{author}{\bibfnamefont{Z.}~\bibnamefont{{Vlah}}}, \bibinfo{author}{\bibfnamefont{N.~E.} \bibnamefont{{Chisari}}}, \bibnamefont{and} \bibinfo{author}{\bibfnamefont{F.}~\bibnamefont{{Schmidt}}}, \bibinfo{journal}{\jcap} \textbf{\bibinfo{volume}{2020}}, \bibinfo{eid}{025} (\bibinfo{year}{2020}), \eprint{1910.08085}.

\bibitem[{\citenamefont{{Bakx} et~al.}(2023)\citenamefont{{Bakx}, {Kurita}, {Elisa Chisari}, {Vlah}, and {Schmidt}}}]{bakx+23}
\bibinfo{author}{\bibfnamefont{T.}~\bibnamefont{{Bakx}}}, \bibinfo{author}{\bibfnamefont{T.}~\bibnamefont{{Kurita}}}, \bibinfo{author}{\bibfnamefont{N.}~\bibnamefont{{Elisa Chisari}}}, \bibinfo{author}{\bibfnamefont{Z.}~\bibnamefont{{Vlah}}}, \bibnamefont{and} \bibinfo{author}{\bibfnamefont{F.}~\bibnamefont{{Schmidt}}}, \bibinfo{journal}{\jcap} \textbf{\bibinfo{volume}{2023}}, \bibinfo{eid}{005} (\bibinfo{year}{2023}), \eprint{2303.15565}.

\bibitem[{\citenamefont{{Okumura} et~al.}(2024)\citenamefont{{Okumura}, {Taruya}, {Kurita}, and {Nishimichi}}}]{okumura+24}
\bibinfo{author}{\bibfnamefont{T.}~\bibnamefont{{Okumura}}}, \bibinfo{author}{\bibfnamefont{A.}~\bibnamefont{{Taruya}}}, \bibinfo{author}{\bibfnamefont{T.}~\bibnamefont{{Kurita}}}, \bibnamefont{and} \bibinfo{author}{\bibfnamefont{T.}~\bibnamefont{{Nishimichi}}}, \bibinfo{journal}{\prd} \textbf{\bibinfo{volume}{109}}, \bibinfo{eid}{103501} (\bibinfo{year}{2024}), \eprint{2310.07384}.

\bibitem[{\citenamefont{{Taruya} et~al.}(2024)\citenamefont{{Taruya}, {Kurita}, and {Okumura}}}]{taruya+24}
\bibinfo{author}{\bibfnamefont{A.}~\bibnamefont{{Taruya}}}, \bibinfo{author}{\bibfnamefont{T.}~\bibnamefont{{Kurita}}}, \bibnamefont{and} \bibinfo{author}{\bibfnamefont{T.}~\bibnamefont{{Okumura}}}, \bibinfo{journal}{arXiv e-prints} \bibinfo{eid}{arXiv:2409.06616} (\bibinfo{year}{2024}), \eprint{2409.06616}.

\bibitem[{\citenamefont{{Pyne} et~al.}(2022)\citenamefont{{Pyne}, {Tenneti}, and {Joachimi}}}]{pyne+22}
\bibinfo{author}{\bibfnamefont{S.}~\bibnamefont{{Pyne}}}, \bibinfo{author}{\bibfnamefont{A.}~\bibnamefont{{Tenneti}}}, \bibnamefont{and} \bibinfo{author}{\bibfnamefont{B.}~\bibnamefont{{Joachimi}}}, \bibinfo{journal}{\mnras} \textbf{\bibinfo{volume}{516}}, \bibinfo{pages}{1829} (\bibinfo{year}{2022}), \eprint{2204.10342}.

\bibitem[{\citenamefont{{Linke} et~al.}(2024)\citenamefont{{Linke}, {Pyne}, {Joachimi}, {Georgiou}, {Hoffmann}, {Mandelbaum}, and {Singh}}}]{linke+24}
\bibinfo{author}{\bibfnamefont{L.}~\bibnamefont{{Linke}}}, \bibinfo{author}{\bibfnamefont{S.}~\bibnamefont{{Pyne}}}, \bibinfo{author}{\bibfnamefont{B.}~\bibnamefont{{Joachimi}}}, \bibinfo{author}{\bibfnamefont{C.}~\bibnamefont{{Georgiou}}}, \bibinfo{author}{\bibfnamefont{K.}~\bibnamefont{{Hoffmann}}}, \bibinfo{author}{\bibfnamefont{R.}~\bibnamefont{{Mandelbaum}}}, \bibnamefont{and} \bibinfo{author}{\bibfnamefont{S.}~\bibnamefont{{Singh}}}, \bibinfo{journal}{arXiv e-prints} \bibinfo{eid}{arXiv:2406.05122} (\bibinfo{year}{2024}), \eprint{2406.05122}.

\end{thebibliography}
\end{document}